%% file: paper_practice.tex
\documentclass[final]{siamltex}

\usepackage[]{fontenc}
\usepackage[utf8]{inputenc}
\usepackage{verbatim}
\usepackage{amsmath}
\usepackage{amssymb}
\usepackage{dsfont}
\usepackage{amsfonts}
\usepackage{mathrsfs}
\usepackage{graphicx}
\usepackage{color}
\usepackage{ulem}
\usepackage{xspace,bbm,epic,eepic}
\usepackage{pgfplots}
\usepackage{setspace}
\usepackage{xspace}

\usepackage{stmaryrd}
\usepackage{graphicx}
\usepackage{lineno}
\usepackage{color}
\usepackage{url}
\usepackage{algorithm}
\usepackage{algpseudocode}

\usepackage{amsmath,amssymb,color} 
\usepackage{mathtools}

\usepackage{hyperref}

\include{macros}

\begin{document}

\title{Adaptive Reduced Multilevel Splitting}

\author{Fr\'ed\'eric C\'erou$^{*}$, Patrick H\'eas$^{*}$,   Mathias Rousset\thanks{Inria and Irmar,  University of Rennes, France. \\({\tt frederic.cerou@inria.fr,patrick.heas@inria.fr,mathias.rousset@inria.fr}) }
   }

\date{}
\maketitle

%
%
%
\begin{abstract}

	This paper considers the classical problem of sampling with Monte Carlo methods a target rare event  distribution defined by a score function that is very expensive to compute. 
	We assume we can build using evaluations of the true score, an approximate surrogate score certified with error bounds. This work proposes a fully adaptive algorithm to sequentially sample surrogate rare event distributions with increasing target levels. An essential contribution consists in sampling at each iteration the surrogate rare event at a critical level corresponding {\color{black} to a specific cost. This cost is related to importance sampling for a target for a given budget. The critical level is calculated solely from the reduced score and its error bound.}
	From a practical point of view, sampling the proposal sequence is performed by extending the framework of the popular adaptive multilevel splitting algorithm to the use of  score approximations. Numerical experiments evaluate the proposed importance sampling algorithm in terms of computational complexity versus squared error. In particular, we investigate the performance of the algorithm when simulating rare events related to the solution of a parametric PDE, which is approximated by a reduced basis.

\end{abstract}
\begin{keywords}
 Rare event simulation, reduced modeling, importance sampling, adaptive multilevel splitting, {\color{black} relative entropy}.  \vspace{-0.25cm}
\end{keywords}

\section{Introduction}

\subsection{Context and objective}

Let $\mref(\d x)$ denotes a given, easily simulable, reference probability distribution on a state space $\Space$. A random variable with distribution $\pi$ may describe for instance unknown physical parameters, uncertainty on initial/boundary conditions, or additional noise modeling either a chaotic behavior or the interaction with an environment. We also consider a real-valued {\it score} function
$
\Score:\Space \to \R
$
which is assumed to be of the form
$$
\Score(x) = \mrm{score}(\Psi^\ast(x)),
$$
where $\mrm{score}$ is an easily computable function of physical interest, and $\Psi^\ast(x)$ is the outcome of a  presumably exact numerical computation of a complex physical system parametrized by $x \in\Space$. With the previous ingredients, one can define the following family of conditional probability distributions on $\Space$:
\begin{equation}\label{eq:target}
	\eta^\ast_l( \d x) \eqdef \frac{1}{p_l^\ast} \one_{\Score(x) > l} \mref( \d x) \qquad l \in \R 
\end{equation}
defined by conditioning the distribution $\pi$ on the event $\set{\Score > l}$ which may be interpreted as a certain {\color{black} failure region/event}.
The quantity 
$p^\ast_l \eqdef \int_{\Space} \one_{\Score > l} \d \mref = \mref \p{ \set{\Score > l}}$
is the associated normalization, here the probability of the considered (possibly rare) event $\set{\Score > l}$, for $l \in \R$. 
$\eta_l^\ast$ is the probability distribution of the system conditioned to {\color{black} the failure event}. 

We consider in this paper the classical problem of designing an algorithmic Monte Carlo procedure that samples according to $\eta^\ast_{l_\mrm{max}}$ -- for one specific large $l_{\mrm{max}} \in \R$ -- and estimate the associated rare event probability $ p^\ast_{l_{\mrm{max}}}$. {\color{black} Note that depending on the problem at hand, the final event of interest might be either $\set{\Score > l_{\mrm{max}}}$ or $\set{\Score \geq l_{\mrm{max}}}$, the intermediate sets still being defined as $\set{\Score > l}$ for $l<l_{\mrm{max}}$.}

In this work, we will focus on scenarios with \textit{two} types of difficulty.
A first difficulty arises when ${l_\mrm{max}}$ is large, generating a very small (yet strictly positive) probability $p^\ast_{l_\mrm{max}}>0$. This rare event context  confronts the practitioner with a first computational challenge: the majority of the probability mass of the conditional distribution $\eta_{l_\mrm{max}}^\ast$ is concentrated in specific areas of $\Space$, which are roughly described by the maxima of $\Score$ which are usually unknown. A direct sampling with a Monte Carlo simulation of $\pi$ requires a sample size of order at least $1/p^\ast_{l_\mrm{max}}$, which is often infeasible. In that context, a popular efficient strategy to simulate a sample according to $\eta^\ast_{l_\mrm{max}}$ and to estimate the associated normalizing constant is to resort to 'Sequential Monte Carlo' (SMC) methods. Such methods start with a Monte Carlo sample of clones  (also called below 'particles') of the system distributed according to $\mref$ and then sequentially samples $\eta^\ast_l$ for increasing values of $l$. To do so, those methods resort to a combination of Importance Sampling (IS), re-sampling (selection) of weighted samples, and mutations of particles based on suitable Markov Chain Monte Carlo transitions. In this work we will focus on a specific instance called {\it Adaptive Multilevel Splitting} (AMS, see~\cite{cdfg,cerou:hal-01417241}) which moreover chooses  adaptively the sequence of levels $l$ where re-sampling is performed.

A second difficulty  appears when {the numerical evaluation of $\Psi^\ast$ is extremely expensive}, so  that the number of evaluations required in, say, a usual SMC or AMS approach becomes prohibitive. In order to circumvent this issue, we assume that the physical model $\Psi^\ast$ can be approximated by a \textit{reduced model}  denoted $\Psi$, the cost of evaluating $\Psi$  being small as compared to one evaluation of the true score with $\Psi^\ast$. The reduced model $\Psi$ we consider is assumed to come with two key features. To explain these, let us denote from now on the approximate score function by
$$
\Srom(x) = \mrm{score}\p{\Psi(x)} \in \R.
$$
The first feature is an \textit{a posteriori} pointwise error quantification associated with $\Srom$. This error quantification comes in the form of an explicit error bound denoted
$
\Err(x) = \mrm{error}\p{\Psi(x)} \in \R,
$
and satisfying the pointwise estimate:
\begin{equation}\label{eq:error_prior}
\abs{\Srom(x) - \Score(x)} \leq \Err(x), \qquad \forall x \in \Space.
\end{equation}
The second key feature assumes that we are able to {update the reduced model}, through a procedure of the following form
\begin{equation*}
 \set{(X_1,\Psi^\ast(X_1)),\ldots,(X_k,\Psi^\ast(X_k)} \xmapsto{\textrm{reduced model}}
  \Psi= \Psi^{(X_1,\ldots,X_k)}, 
\end{equation*}
which takes as an input any sequence of states in $\Space$ -- called here \textit{a sample of snapshots} -- together with the solution of the associated true model $\Psi^\ast$ evaluated for each of those snapshots. The output is the reduced model $\Psi$. To improve the readability of the document, we will use a superscript $(k)$ to specify the dependency on $(X_1,\ldots,X_k)$, or even omit this dependency if the context is unambiguous,
\[
(\Srom^{(X_1,\ldots,X_k)},E^{(X_1,\ldots,X_k)}) = (\Srom^{(k)},E^{(k)}) = (S,E).
\]


\subsection{Previous works}\label{sec:previouswork}

\paragraph{Reduced models for rare events}
The idea of using  a simpler surrogate model in Monte-Carlo simulation for target distributions is not new and there is a large amount of literature on the subject. Most references are concerned with either kriging on the one hand, and reduced basis methods on the other hand. For kriging approaches \cite{echard2011ak,moustapha2022active,razaaly2020efficient,zhang2018adaptive,balesdent2013kriging,ling2019coupled, bect2017bayesian}, the surrogate score function $\Srom:\Space \to \R $ is a {\it random} field, distributed according to the posterior of a prior Gaussian field conditioned by fitting the exact values of the evaluated snapshots. The estimate~\eqref{eq:error_prior} is however not satisfied almost surely, and one has to resort to averaged error estimates.
For {\it reduced basis} approaches  \cite{chen2013accurate,heinkenschloss2020adaptive,gallimard2019adaptive,heas2020selecting},  the surrogate score function is deterministic, and constructed by projection of the function $\Psi^\ast$ 
on a subspace where it is efficiently approximated  and characterized with precise error bounds of the form~\eqref{eq:error_prior}. This model reduction framework is known to be very efficient  for the numerical approximation of problems involving the repeated solution of parametric {\it Partial Differential Equations}  (PDEs) \cite{quarteroni2015reduced}. We note that besides  these two  main approaches,  other families of reduced models  \cite{cannamela2008controlled,elfverson2022adaptive,wagner2020multilevel,peherstorfer2018survey} or  combination of them \cite{elfverson2022adaptive,wagner2020multilevel,peherstorfer2018survey}  have been proposed for rare event simulation.

 
Arguably, the most sensitive feature of any  of these reduced model approaches is the specific collection of snapshots  $(X_k,\Psi^\ast(X_k))$, $k \geq 1$ with which the reduced model is updated along the iterations. A natural approach consists in sampling sequentially and  adaptively the collection of snapshots, the reduced model being updated after each new evaluation of the true model at iteration $k$. The updated reduced model (and the associated error estimates) can then be used to pick the next snapshot in presumably {\it interesting} areas. 
Sampling the snapshot $X_k$ rely on two ingredients: i) a probability distribution used to generate candidates, and ii) a weighting function, also called {\it learning function}, used to pick the presumably best candidate. 

The idea of ingredient ii) can be found in the seminal work of \cite{echard2011ak} where selecting snapshots is called {\it active learning}. In the latter approach, snapshots are selected by optimizing a criterion (the learning function) that will most likely reduce the uncertainty of the surrogate model. An exhaustive set of references and a very quantitative review of methods related to active learning and kriging is provided in~\cite{moustapha2022active}. Some other references, as \cite{heinkenschloss2020adaptive}, pick the snapshots $(X_k)_{k \geq 1}$ uniformly among candidates using a decreasing sequence of subsets that contain the target rare event, constructed using the iteratively updated error estimates.

However, as we will argue in the present work, ingredient i) is also of high importance. Indeed, a key general phenomenon in a rare-event and high-dimensional setting is the following: the true target distribution $\eta^\ast_{l}$ for $l$ large is extremely concentrated around very specific but unknown points in $\Space$, so that,  unless the pool of candidates among which snapshots $(X_1, \ldots, X_k)$ are picked is partly {also located} in such regions, the information given by the reduced model may eventually lack the accuracy required in order to describe the rare event. As a consequence, unless one is able to sample an extremely large amount of candidates, it is critical to select such candidates in areas that carefully track the true rare event. To make the picture clearer, consider at  iteration $k$  the following surrogates of the target conditional distributions~\eqref{eq:target}  built using the reduced score:
\begin{equation}\label{eq:prop}
	\mu_l \eqdef \frac{1}{Z_l} \one_{\Srom > l} \d \mref \qquad l \in \R,
\end{equation} 
where the normalization 
$Z_l \eqdef \int_{\Space} \one_{\Srom > l} \d \mref ,$
represents an approximation of the probability of the rare event. Then, for large $l$, the reduced rare event is also rare, hence very concentrated, and thus usually  too far away from the true target rare event.  If we want to guarantee sufficient mass  in the region of the target rare event, we propose to specify a \textit{maximum possible level $l$} below which the true conditional distribution $\eta_l^\ast$ and the reduced one $\mu_l$ do match $\eta_l^\ast \simeq \mu_l$.  This delicate issue is only rarely addressed in the literature.  Most references avoid it and  generate pools of candidates for the choice of snapshot  relying on the reference distribution $\pi$. 
To the best of our knowledge, the only related pieces of work deal with the reverse problem of selecting (in a hierarchy of pre-computed surrogates)~\cite{heas2020selecting} or refining~\cite{bect2017bayesian} the reduced score $S$, so that it is appropriate for a certain level $l$ according to some error quantification. 


 The main contribution of the present work is to use the following guidelines  to design an efficient and adaptive algorithm that synchronizes: i) the sequential Monte Carlo sampling of rare events using a reduced model (that is sampling the flow $l \mapsto \mu_l$), and ii) the sequential update of the reduced model with evaluations of the true model.

\paragraph{Reduced Monte-Carlo simulation}
 Strategies for Monte-Carlo sampling  of the reduced conditional distributions $\mu_l$ can be  grouped in two broad categories: on the one hand {\it importance sampling} (IS), usually through a parametric approach, and on the other hand  splitting methods, \ie  {\it sequential Monte Carlo}~\cite{del2006sequential} methods, which are non-parametric, and in particular  AMS~\cite{cdfg} also called  {\it subset simulation}~\cite{au2001estimation}.

Many  approaches to IS of the reduced distribution $\mu_l$ have been proposed in recent years. In particular,  {\it cross-entropy methods}~\cite{de2005tutorial} have been proposed relying on kriging~\cite{balesdent2013kriging} or reduced basis approximations~\cite{gallimard2019adaptive,heas2020selecting}.   In order to increase  the effectiveness of the cross-entropy method,  it has also been suggested to approximate the parameter variable $X$ by constraining its support in well-chosen low dimensional subspaces of $\Space$  \cite{uribe2021cross}.  Other recent contributions aim to  build specific adaptive proposals for IS~\cite{razaaly2020efficient,zhang2018adaptive} or for stratified sampling \cite{cannamela2008controlled} (that can be interpreted as a form of IS). However, the effectiveness of IS approaches depends heavily on the choice of parametric family of proposals, for which we don't necessarily have any prior knowledge. 


On the other hand, splitting methods have also been used in conjunction with  (adaptively constructed) kriging-based surrogate models~\cite{ling2019coupled, bect2017bayesian}. We refer again to the review on methods related to active learning and kriging provided in~\cite{moustapha2022active}. However,  in these works, the splitting method samples an approximation $\mu_l$ of the target probability \eqref{eq:target}, which may be biased. We note nevertheless that in \cite{elfverson2022adaptive} subset simulation is performed using nested multiresolution sets (provided by an accuracy hierarchy of the surrogate model), yielding an estimate with an associated {\it a posteriori} error.

Adapting splitting methods to reduced scores can be optionally supplemented with a {\it bridging}  procedure. The goal of this bridge  is to sample a new proposal $\mu'_{l'}$ defined with some updated reduced score ${\Srom}'$ and  level $\l'$,  starting from the samples of the current proposal $\mu_{l}$. In general, this sampling is not trivial. For example, the nesting underlying a AMS simulation  is broken, since in general updating the reduced score and the level will imply $\set{ {\Srom}' > l'} \not\subset \set{\Srom > l}$.  There is very little work in the literature on this bridging problem. Let us mention the work \cite{wagner2020multilevel}, which proposes  a bridging procedure with a constant $l'=l$ level. Specifically, the authors use the sequential Monte Carlo method with soft levels and tempering  instead of targeting distributions defined by conditioning on surrogate rare events. The authors mention that such an approach outperforms the multilevel subset simulation~\cite{ullmann2015multilevel} algorithm they previously proposed, which dealt with the nesting problem by computing correction terms and performing Markov chain Monte Carlo simulations.


\subsection{Contributions}
In this work, we propose a sequential algorithm relying on reduced scores and error bounds. The  algorithm provides an answer to the  questions and limitations listed in Section~\ref{sec:previouswork}. We  consider the family of  {proposal distributions} \eqref{eq:prop}  (built using the reduced score) for sampling the target conditional distributions~\eqref{eq:target}.  The   sketch of the proposed algorithm is the following. At each iteration~$k$, the algorithm
\begin{itemize}
	\item computes adaptively a maximum level denoted $l^{(k)}$ such that the proposal $\mu_{l^{(k)}}^{(k)}$ matches the true conditional distribution $\eta^\ast_{l^{(k)}}$, as enabled by the accuracy of the current reduced model $\Srom^{(k)}$ 
	\item simulates (with a population Monte Carlo procedure) the proposal $\mu_l^{(k)}$ from an initial level $l=l_b^{(k)}$ to the maximum level $l=l^{(k)}$; 
	\item  samples the new snapshot $(X^{(k)}, \Psi^\ast(X^{(k)}))$ according to a weighted version of the distribution $\mu_{l^{(k)}}^{(k)}$. 
	The weights (or \tit{learning function}) may favor states $x$ with large reduced score error $E^{(k)}(x)$ (see \eqref{eq:error_prior}). 
	\item updates the reduced model score $\Srom^{(k)} \to \Srom^{(k+1)}$.
	\item  computes (using records of previous iterations) the next initial level ${l_b^{(k+1)}}$ so that the nesting (inclusion) of the support of the considered proposals are guaranteed.
	\item  bridges with (population) Monte Carlo simulation records of previous iterations to the next proposal  $\mu_{l_b^{(k+1)}}^{(k+1)}$.  
\end{itemize}
The Monte Carlo aspect of the algorithm is implemented using an AMS simulation methodology, in which $\mu_l^{(k)}$ is approximated by the empirical distribution of a sample (population) of particles (clones).\medskip

%

Our {\it first main methodological contribution}  is the specific criterion used to adaptively compute the maximum level of iteration $k$ denoted $l^{(k)}$. At each iteration $k$, we will use the information given by the error quantification~\eqref{eq:error_prior} 
so that $\mu_{l^{(k)}}^{(k)}$ remains a reasonable importance distribution for sampling the true target $\eta^\ast_{l^{(k)}}$. 
The evaluation of the accuracy of an importance distribution will be done using a quantification of the  logarithmic cost of importance sampling by  relative entropy:
$$
\ln \mrm{cost} \simeq \Ent(\eta_{l^{(k)}} \mid \mu^{(k)}_{l^{(k)}})\eqdef \eta_{l^{(k)}} (\ln \frac{\eta_{l^{(k)}} }{ \mu^{(k)}_{l^{(k)}}}).
$$
This choice is discussed and justified in Appendix~\ref{app:cost}. The latter cannot be evaluated exactly, but we can leverage the existence of error estimates in order to obtain (more or less pessimistic) approximations of it.
Relying on this cost, we are able to stop the increase of rareness of the targets when the reduced modeling becomes critically inaccurate, the associated maximum level being  $l^{(k)}$. 

Snapshot sampling is then preformed using the distribution $\mu_{l^{(k)}}^{(k)}$, which we may reweight by favoring states with higher error quantification. Remark that snapshot sampling is performed at the precise level $l^{(k)}$ of the flow of reduced proposal distributions $(\mu_{l}^{(k)})_{l \in \R}$  where increasing the quality of these proposals becomes necessary: this prevents a poor sampling of a new snapshot inconsistent with the unknown target while keeping the associated level $l^{(k)}$ as large as possible.

The proposed method can also yield importance sampling estimators. If for instance the snapshots are sampled  from the start without re-weighting according to proposal $X^{k+1} \sim \mu^{(k)}_{l^{(k)}}$, it naturally leads to the following estimator: 
\begin{equation}\label{eq:estim}
 \hat{\mref}^{(k)} \eqdef \frac{1}{k} \sum_{k'=1}^{k}   \frac{\d \pi}{\d \mu^{(k'-1)}_{l^{(k'-1)}}}(X^{(k')}) \delta_{X^{(k')}},
\end{equation}
which by construction, is an \textit{unbiased}  empirical estimator of the reference distribution $\mref$ (see Section~\ref{sec:estim} for a proof), up to sampling errors of the proposal $\mu_l$, which will be done in practice using AMS.  This leads to an IS estimator \textit{e.g.} of the rare event probability 
$$\hat p =\hat{\mref}^{(k)}( \one_{\Score > l}).$$

As mentioned above, the sequence of importance distribution $\mu_{l^{(k)}}^{(k)}$ will be simulated by a population of particles and an AMS methodology  in the spirit of current rare event simulation algorithms. The only difference is that the target distribution is constructed with a reduced score that is adaptively enriched by few evaluations of the true score throughout the simulation. As remarked in Section~\ref{sec:previouswork}, updating the reduced score will in general break the nesting of the sequence of distributions. A simple way of addressing this problem is to  perform a fresh AMS simulation each time that a new snapshot is picked and that the reduced score is updated.  
 To significantly reduce the number of evaluations of the reduced model, another option is to propose a {\it bridging} procedure. This is the {\it second main methodological contribution} of this work. At iteration $k$, this procedure  searches  in the simulation history for a { well-chosen} previous proposal denoted $\mu_{l^{(k_{b})}}^{(k_{b})} $ with $k_{b} \le k$, 
  so that the updated proposal denoted $\mu_{l_{0}^{(k+1)}}^{(k+1)}$, with ${l_{0}^{(k+1)}}$  a {well-designed} level, is dominated  and sufficiently close to the former.  
The latter proposal is then simulated starting from the former one 
using an AMS  framework.\medskip

In numerical experiments, we study the proposed algorithm, called {\it Adaptive Reduced Multilevel Splitting} (ARMS) and show the empirical convergence of the ARMS algorithm on a toy model and a rare-event problem based on a realistic elliptic PDE and a \textit{reduced basis} approach for model reduction.  Our study confirms that the proposed relative entropy criterion plays a crucial role in ARMS and prevent the simulation to be taken  into spurious regions of the state-space remote from the region of interest, resulting in a large variance of the IS estimator. Moreover, by evaluating the computational complexity in relation to the squared error of the estimate, we show empirically that the bridging procedure reduces in a certain regime the simulation cost while achieving a comparable squared error. \medskip

The paper is organized as follows. The main  methodological concepts underlying the ARMS algorithm are presented in Section~\ref{sec:ingredients}. We end this section by  
 introducing an idealized version of the ARMS algorithm,  which  assumes the exact evaluation of expectations and exact sampling of distributions.  In Section~\ref{sec:practical}, we  define  ARMS, which consists in a practical implementation of this idealized algorithm using  AMS simulations. In Section~\ref{sec:num}, we numerically evaluate the proposed algorithm for various rare event problems. A final section deals with conclusions and perspectives. The appendix collects some technical proofs and details.

\section{Underlying Methodological Concepts}\label{sec:ingredients}

\subsection{Importance sampling cost}\label{sec:prop}\label{sec:imp}

Let $l_{\max}$ denote a large level of interest, and let us assume for the sake of the argument within the present section, that one is able to simulate a random variable $X$ distributed according to the proposal distribution $X \sim \mu_{l} = \frac{1}{Z_{l}} \one_{\Srom > l} \d \pi$, the normalization $Z_l$ being known. \textit{Importance sampling} of the target rare event $\set{\Score > l}$ can be performed in our context if the distribution $\mu_{l}$ dominates  the measure $\one_{\Score > l} \d \pi $, and simply amounts to remark that one can estimate the averages
$$
   Z_{l} \E \b{ \one_{\Score(X) > l} \ph(X) } = \pi\p{\ph \one_{\Score > l}  },
$$
where $\ph$ denotes a generic test function. For instance taking $\ph =\one$, the above yields an unbiased estimator of the normalization probability $p^\ast_{l}$. \medskip

The domination requirement, denoted $\mu_{l} \gg\eta_{l}^\ast, $ holds true by definition if and only if one has the inclusion $\set{\Score > l} \subseteq \set{\Srom > l}$. This is not guaranteed here as $\Srom$ is an approximation of the true score $\Score$, and since $\Score$ is not computed for each sample states, one cannot check it in a systematic fashion. Moreover, even if the above domination condition holds, importance sampling becomes quickly infeasible if the target distribution diverges too widely from the proposal, a typical phenomenon in high dimension. 
As  discussed in Appendix~\ref{app:cost}, it has been thoroughly argued in~\cite{ChaDia18} (see also~\cite{cerou2022entropy}) that the cost of importance sampling -- in terms of the required sample size $K$-- is quite generically and roughly given by the exponential of the relative entropy  between $\eta^\ast_{l}$ and $\mu_l$:
\begin{equation}\label{eq:cost}
K \simeq \e^{\Ent(\eta^\ast_{l} \mid \mu_l)}.
\end{equation}
This fact will justify the choice of various heuristics that will be found in this paper. We will  call $\Ent(\eta \mid \mu)$ the \textit{log cost of importance sampling} of the target $\eta$ by the proposal $\mu$. In the present rare event case, note that relative entropy is simply the logarithm of small probability ratios:
\[
\Ent(\eta^\ast_{l} \mid \mu_l) = \ln \frac{Z_l}{p^\ast_{l}}
\]

Note that domination between measures is implied by finite relative entropy, and will rather use throughout the paper the condition
$ \Ent(\eta \mid \mu) < +\infty $ 
in order to check the domination condition $\mu \gg \eta$. 

\subsection{First non-achievable level}\label{sec:prop_temp}

In this section, we describe how to choose  the target level $l^{(k)}$, which is suited to the current score function $\Srom^{(k)}$ at iteration $k$ of the  algorithm -- $k$ being the number of sampled snapshots of the true model. The idea is to tune $l^{(k)}$ with a criteria evaluating that surpassing this level should require, in some sense, a better surrogate model, and thus new snapshots. The adjusted proposal $\mu^{(k)}_{l^{(k)}}$ will then be used to sample a new snapshot, and obtain an updated score function $\Srom^{(k+1)}$ for the next iteration $k+1$. In short, we are going to develop a heuristic criterion on levels to detect when a new snapshot is needed. \medskip 

The general insight enabling to define $l^{(k)}$ is the following. Consider the following  problem: one is given a family of rare event target distributions $\p{\eta^\ast_l}_{l \in \R}$ in the form of~\eqref{eq:target}, a given computational budget, and a family of importance sampling proposal distributions $(\mu_{l}^{(k)})_{l\in \mathbb{R}}$ in the form of~\eqref{eq:prop}. One wants to perform some sequential algorithm to sample the distributions $\mu_l$ along an increasing sequence of levels $l$ as high as possible, and then use importance sampling. However, we are given limited resources preventing sampling for levels that are unreasonably high. As an idealized criteria formalizing achievable levels, we propose the importance sampling cost~\eqref{eq:cost} of the true target. The (ideal) first non-achievable (hence critical) level is thus given by the smallest level for which this cost matches a certain given numerical threshold $\Cte$, that is:
$$\text{Find first} \, l : \quad \Ent(\eta^\ast_{l} \mid \mu^{(k)}_{l})  \geq \Cte.$$
  By doing so, we ensure that the snapshot sampled with respect to $\mu_{l^{(k)}}^{(k)}$ will not diverge too much from the target distribution $\eta^\ast_{l^{(k)}}$. As an example, if, as it is the case in the present context, $\mu_l^{(k)}$ has been constructed using a reduced model based on $k$ snapshots $(X_1,\ldots,X_k)$, the first non-achievable level $l^{(k)}$ will typically be of order $\max(\Score(X_1),\ldots,\Score(X_k))$, as the quality of the approximation $\Score-\Srom^{(k)}$ will deteriorate in areas with higher scores. \medskip

In a practical context, the log cost $\Ent(\eta^\ast_{l} \mid \mu_{l}^{(k)})$  has to be approximated using some prior information on $\eta^\ast_{l}$, for instance using some form of error quantification. We propose in this work the following method, which is by no means unique. We consider a \textit{pessimistic} (respectively an \textit{optimistic}) surrogate  of $\eta^\ast_l$, denoted by  $\check{\eta}_{l}^{(k)}$ (respectively  by $\hat{\eta}_{l}^{(k)}$), constructed  with the reduced score $\Srom^{(k)}$ and its error quantification $\Err^{(k)}$, and typically associated with a much lower (respectively higher) rare event probability than $\eta^\ast_l$. In the present work, we will choose \eqref{eq:pessTarget}-\eqref{eq:optTarget} below which satisfy:
\[
\check{\eta}_{l}^{(k)} \ll \eta^\ast_l \ll\hat{\eta}_{l}^{(k)} .
\]
The first non-achievable level will be estimated using the \textit{pessimistic} surrogate by searching the  closest level such that the log cost $\Ent\p{\check{\eta}_{l}^{(k)} \mid \mu_{l}^{(k)} } $  matches the threshold $\Cte$. 
Given an initial level $l_b$, this  yields  the following rigorous definition of  the  (critical) first non-achievable level:
\begin{equation}\label{eq:level}
 l^{(k)}(\Cte,l_b)= \min \{ l\geq l_b : \; \Ent\p{\check{\eta}_{l}^{(k)} \mid \mu_{l}^{(k)} }  \geq  \Cte \;  \;  \text{or}  \;  \; \Ent\p{\hat{\eta}_{l_{\mrm{max}}}^{(k)} \mid \mu_{l}^{(k)} }   = +\infty\}.
\end{equation}
$l^{(k)}(c,l_b)$ is the smallest level above $l_b$ such that, in the (unfortunate) case where the true target is given by the pessimistic choice $\check{\eta}_{l}^{(k)}$,  a log cost of $\Cte$   is necessary in order to perform importance sampling. For convenience, the dependency on $c$ and $l_b$ of $l^{(k)}$ will be removed if the context makes it clear.
Note that the additional inequality constraint $\Ent\p{\hat{\eta}_{l_{\mrm{max}}}^{(k)} \mid \mu_{l}^{(k)}} < + \infty$ in \eqref{eq:level} ensures that the support of $\mu_{l}^{(k)}$ contains the support of the final target ${\eta}^\ast_{l_{\mrm{max}}}$. 


\medskip

In the present rare event case,  the importance distribution is \eqref{eq:prop}, the pointwise error bound is~\eqref{eq:error_prior} and we propose the pessimistic and optimistic  surrogate of the  target  \eqref{eq:target}  
\begin{equation}\label{eq:pessTarget}
\check{\eta}_l^{(k)}  \eqdef \frac{1}{\check Z_l^{(k)}} \one_{\Srom^{(k)}-\Err^{(k)} > l} \d \mref \quad \textrm{with} \quad \check Z_l^{(k)} \eqdef \mref(\one_{\Srom^{(k)}-\Err^{(k)} > l}),
\end{equation}
and 
\begin{equation}\label{eq:optTarget}
\hat{\eta}_l^{(k)}  \eqdef \frac{1}{\check Z_l^{(k)}} \one_{\Srom^{(k)}+\Err^{(k)} > l} \d \mref \quad \textrm{with} \quad\hat Z_l^{(k)} \eqdef  \mref( \one_{\Srom^{(k)}+\Err^{(k)} > l} ).
\end{equation}

 \medskip

\begin{Rem}
Using the surrogate \eqref{eq:pessTarget},  the particularization of  the worst-case log cost in \eqref{eq:level} is 
$ \Ent\p{\check{\eta}_{l}^{(k)} \mid \mu_{l}^{(k)} } =\ln \frac{Z_{l}^{(k)}}{\check Z_{l}^{(k)}},$
 thus the logarithm of the ratio of  the approximate probability ${ \mref  \p{ \set{\Srom^{(k)}  > {l}}}}$ over the smaller pessimistic probability ${ \mref  \p{ \set{\Srom^{(k)} -\Err^{(k)} > {l}}}}$.\medskip
\end{Rem}
\begin{Rem}
It can easily be  checked that the critical level $l^{(k)}(\Cte,l_b)$ tends to infinity when the error $E^{(k)}$ vanishes pointwise.\medskip
\end{Rem}

%
%

\subsection{Snapshot sampling and learning function}\label{sec:learn}

Given the importance distribution  $\mu^{(k)}_{l^{(k)}}$, we now discuss the procedure used to sample a new snapshot. A possible choice used in the present work is the penalization of the importance distribution as follows:
\[
X \sim \frac{1}{\mu^{(k)}_{l^{(k)}}(\e^{\tau^{(k)} \Err^{(k)}})} \e^{\tau^{(k)} \Err^{(k)}} \mu^{(k)}_{l^{(k)}},
\]
with $\tau^{(k)}\ge 0$.
 The above penalization is similar to the so-called \textit{learning function} in the literature on 'active learning'~\cite{moustapha2022active}. In this work, we simply choose an exponential function of the error $\Cte$ with a penalty parameter  $\tau^{(k)}$  adaptively chosen at  iteration $k$ of the algorithm. In particular, $\tau^{(k)}$ will be chosen to be high in the beginning of the algorithm and then be reduced to $0$ at some point. The idea here is that $\tau^{(k)}$ enables to find a compromise between two objectives: first i) the snapshots are used to refine the reduced model $\Srom^{(k)}$, and ii) the snapshots are also  used in estimation by importance sampling. A large $\tau^{(k)}$ is associated with the objective i). Indeed, favoring candidates with high errors in a $\mu^{(k)}_{l^{(k)}}$-distributed sample enables a faster learning of the true score $\Score$. A small or vanishing $\tau^{(k)}$ is associated with objective ii). Indeed, the distribution $\mu^{(k)}_{l^{(k)}} $  is constructed to be a robust importance sampling distribution for the target $\eta_{l^{(k)}}^\ast$ so that states with smaller error are needed for estimation, ensuring a limited variance for importance sampling. \medskip

In the numerical experiments of Section~\ref{sec:num}, we will consider a very large $\tau^{(k)}$  in a first phase of the algorithm (until a given number of snapshots are taken in the rare event $\set{\Score > l_{\mrm{max}}}$ of interest), and then, because estimation is then possible, we will take $\tau^{(k)}=0$.

No attempt is made to optimize the learning function and the choice of the penalty parameter; those studies are left for future work, taking into account the thoroughful studies on active learning methods as found in the survey~\cite{moustapha2022active}.

\subsection{Bridging two distributions with updated score approximation (optional)}\label{sec:bridging}

We will now focus on what happens after the update of the reduced model that follows the sampling of a new snapshot. We are given a certain level $l^{(k)}$ and an associated importance distribution $\mu_{l^{(k)}}^{(k)}$. The problem is to find a new initial level which we will denote $l_{b}(k+1)$ and then sample the new importance distribution $\mu_{l_{b}(k+1)}^{(k+1)}$ associated with the updated score approximation $\Srom^{(k+1)}$. 

We emphasize that this bridging methodology is only optional. It is perfectly possible (and we will provide some numerical comparisons) to set $l_b(k+1)= - \infty$ which amounts to start at the next iteration with an importance distribution $mu_{l_{b}(k+1)}^{(k+1)} = \pi$ given by the reference distribution.
  
This initial level $l_b(k+1)$ will then be used in the subsequent $k+1$ iteration to determine (with definition \eqref{eq:level})  the critical level at iteration $k+1$:  $l^{(k+1)}(\Cte,l_{b}(k+1) )$ for  $\Srom^{(k+1)}$. This will complete the main loop of the algorithm.
\medskip

However, if one tries to sample $\mu_{l_{b}(k+1)}^{(k+1)}$ starting from $\mu_{l^{(k)}}^{(k)}$, using some form of importance or sequential sampling, there is an issue with inclusion. For instance, simply setting the initial level  ${l_{b}(k+1)}$ to $l^{(k)}$, the inclusion $\{\Srom^{(k+1)} > l^{(k)} \} \subseteq   \{\Srom^{(k)} > l^{(k)}\}$ will generally not be valid, making the task infeasible.

In order to circumvent this issue, we propose to keep in memory the importance samples $\mu^{(k')}_{l^{(k')}}$ for $k' \leq k$. We then set a quantile $\theta =\frac{M}{N} \in (0,1)$ associated with an effective sample size $N-M \in \N$, and find an index $$\text{Find:} \quad k_{b}\le k$$ and a level $$\text{Find:} \quad l_{b}(k+1)$$ 
verifying:
\begin{equation}\label{eq:BridgeQuantile}
	\Ent\p{ \mu_{l_{b}(k+1)}^{(k+1)} \mid \mu_{l^{(k_{b})}}^{(k_{b})} } \le -\ln(1-\theta).
\end{equation}
This condition ensures that a relatively straightforward importance sampling procedure enables to obtain the initial importance sampling distribution $\mu_{l_{b}(k+1)}^{(k+1)} $ of the forthcoming iteration, from a previous importance distribution $\mu_{l^{(k_{b})}}^{(k_{b})}$. In particular, the inclusion condition $\{\Srom^{(k+1)} \ge {l_{b}(k+1)} \} \subseteq   \{\Srom^{(k_{b})} > l^{(k_{b})}\}$ will hold true.

 \medskip
 
 In addition of the condition~\eqref{eq:BridgeQuantile}, we also want to  ensure that the two conditions in \eqref{eq:level} used to define the maximal achievable level $l^{(k+1)}$ hold for this initial level ${l_{b}(k+1)}$. This is necessary to ensure formally that ${l_{b}(k+1)} \leq l^{(k+1)}$. Explicitly:  
   \begin{align}
    & \Ent(  \check \eta^{(k+1)}_{l_{b}(k+1)} \mid  \mu^{(k+1)}_{l_{b}(k+1)}              ) \leq \Cte , \label{eq: wlcCondition}\\
  & \Ent(  \hat \eta^{(k+1)}_{l_{\mrm{max}}} \mid  \mu^{(k+1)}_{l_{b}(k+1)}  ) < +\infty.\label{eq: domCondition}
 \end{align}

\medskip
A pair $(k_{b},{l_{b}(k+1)})\in \llbracket 0, k \rrbracket \times \R$ satisfying 
 \eqref{eq:BridgeQuantile}, \eqref{eq: wlcCondition} and \eqref{eq: domCondition}  determine two consecutive proposal distributions $\mu_{l^{(k_{b})}}^{(k_{b})} $ and $\mu^{(k+1)}_{l_{b}(k+1)} $  ``bridging'' the score approximations $\Srom^{(k_{b})}$ and  $\Srom^{(k+1)}$. This pair can be determined according to various strategies. 
 \begin{itemize}
 \item A  trivial feasible pair is to set $k_{b}=0$ such that $l^{(k_{b})}= l_{b}(k+1)=-\infty$. As explained in Section~\ref{sec:practical}, this will account in practice to restart the AMS simulation with the new score $\Srom^{(k+1)}$ from the beginning until it reaches the critical level
$
l^{(k+1)}(\Cte,-\infty). 
$
However, this option can lead to a significant computational overhead if the complexity of the reduced score evaluation is not negligible.\medskip

\item A non trivial feasible pair of interest is given by 
the largest (lexicographic) pair $(k_{b},{l_{b}(k+1)}) \in \llbracket 0, k \rrbracket \times \R$ 
 satisfying the three conditions \eqref{eq:BridgeQuantile}, \eqref{eq: wlcCondition} and \eqref{eq: domCondition}. This means first selecting the importance sampling distribution $\mu_{l^{( k_{b})}}^{( k_{b})}$, among the already computed distributions, from the most recent to the oldest, and then to search the largest level ${l_{b}(k+1)}$  
such that the three conditions hold.
We will see  in Section~\ref{sec:practical} that in practice, this solution will lead to an important computation saving, but will require nevertheless to save the history of the previous AMS simulations.  

\end{itemize}

\subsection{Stopping criteria}\label{sec:stop}
In practical situations, a maximum level $l_{\mrm{max}}$ is setting the range of the computation of interest: one is interested in computing the rare event probability $p^\ast_{l_{\mrm{max}}}$.
The level $l^{(k)}(\Cte,l_b)$ defined in~\eqref{eq:level} is constrained to stay below $l_{\mrm{max}}$, and the definition of $l^{(k)}(\Cte,l_b)$ is modified as the minimum between $l_{\mrm{max}}$ and the critical level given by~\eqref{eq:level}. The algorithm can thus continue as long as one wishes, for instance until enough snapshot samples are obtained. \medskip
 
In addition, in order to avoid unnecessarily enriching an already accurate reduced score and obtain computational savings, it is possible to stop updating the reduced score if the worst-case log cost at the critical level $l^{(k)}$ is almost zero and if this critical level is just below $l_{\max}$. More precisely we define the update stopping index\footnote{ We call a stopping index any exit criteria of the algorithm that is independent of the future snapshots. } as the first index $k$ such that
\begin{equation}\label{eq:stopUpdate}
 \Ent\p{\check{\eta}_{l^{(k)}}^{(k)} \mid \mu_{l^{(k)}}^{(k)} }  \le \epsilon \quad \& \quad l^{(k)}\simeq l_{\max},
\end{equation}
where  $\epsilon>0$ is a given threshold parameter close to the machine precision. This stop-update rule must however be implemented only when using a non-trivial bridging procedure. We will indeed see in Section~\ref{sec:practical} that using such a stopping of update of the reduced score, the normalization constant is no longer updated resulting in a sustained variance while new snapshots are sampled. Yet this sustained variance always occurs when using a non trivial bridging procedure. We will hence see in Section~\ref{sec:practical} that using the latter stop-update rule with a non-trivial bridge usually largely compensate additional variance with computational (and memory) savings. 

\subsection{The estimator}\label{sec:estim}
We are now ready to discuss the estimation of the probability $p^\ast_{l_{\mrm{max}}}$. It is  more convenient to work with the un-normalized version of the target distribution $\eta^\ast_{l_{\mrm{max}}}$ defined in~\eqref{eq:target}, which will be denoted from now on by $\gamma^\ast_{l_{\mrm{max}}}$ and defined by
$$
\gamma^\ast_{l_{\mrm{max}}}(\ph) \eqdef \pi(\ph \one_{\Score > {l_{\mrm{max}}}}), 
$$
for the level of interest ${l_{\mrm{max}}} \in \R$ and some test function $\ph$. Note that $p^\ast_{l_{\mrm{max}}} = \gamma^\ast_{l_{\mrm{max}}}(\one)$. \medskip

The importance sampling estimator $\hat \gamma_{l_{\max}}^H$ of $ \gamma^\ast_{l_{\max}}$ is defined using a random iteration index which we will denote by $K_{j_0}$. Only the snapshots sampled at and after iteration $K_{j_0}+1$ will be used to build our importance sampling estimator. More precisely, $K_{j_0}$ must be of  {\it optional type}, in the sense that $K_{j_0}$ is defined as the first iteration index $k$ for which a certain property depending on the snapshots $X_1, \ldots , X_k$ holds true.

We propose the following option: $K_{j_0}$ is the first iteration index $k$ such that the score of the past snapshots have reached the level $l_{\max}$ (the rare event) $j_0$ times, for some prescribed $j_0\ge 1$. At iteration $K_{j_0}+H$ (for some $H> 0$) of the algorithm, the  estimator of $\gamma^\ast_{l_{\mrm{max}}}$ is then given by:
\begin{equation}\label{eq:estim2}
 \hat{\gamma}^{H}_{l_{\mrm{max}}} \eqdef \frac{1}{H} \sum_{k'=K_{j_0}+1}^{K_{j_0}+H} f_{l_{\mrm{max}}}^{(k')}(X_{k'+1}) \delta_{X_{k'+1}}  ,
\end{equation}
 where, if   we choose  for $k'\ge K_{j_0}$ to set the {\color{black} learning parameter as $\tau=0$}, the IS density ratio can be simplified as
$$
f_{l_{\mrm{max}}}^{(k')}(x) \eqdef   \one_{\Score\ge {l_{\mrm{max}}} }(x)\frac{\d \pi}{ d \mu_{l^{(k')}}^{(k')}}(x).
$$
 Setting a learning parameter $\tau>0$  in the early stages of the algorithm  is crucial as the proposal law has a low probability mass over the set of rare events, which implies that snapshots taken so far are still insufficient to learn an accurate  reduced model  in this rare event regime.  However, once $j_0$ snapshots have already been sampled from the set of rare events, the choice of $\tau=0$ implies that the snapshot ${X_{k'+1}}$ is drawn according to the proposition ${ d \mu_{l^{(k')}}^{(k')}}$, thus favoring the robustness of the IS estimator rather than the accuracy of an already refined reduced model.
Note that : i) the entropic criterion in \eqref{eq:level} and \eqref{eq: domCondition}  ensures that $\eta_{l_{\mrm{max}}}^\ast \ll \mu_{l^{(k)}}^{(k)}$, and ii) $f_{l_{\mrm{max}}}^{(k)}(x)$ can be evaluated pointwise as long as we have an estimate of the normalizations constant of the proposal $\mu^{(k)}_{l^{(k)}}$.

As shown in Appendix~\ref{app:martingal} with a martingale argument, because of the optional property of $K_{j_0}$, this  estimator is unbiased
$$
\E \, \hat{\gamma}_{l_{\mrm{max}}}^{H}  = \gamma_{l_{\mrm{max}}}^\ast,
$$
and its variance decreases in $\bigO(H^{-1})$. Note that samples $X_{k}$ in \eqref{eq:estim2}  are assumed to be  drawn exactly from the sequence of importance distributions ${\mu_{l^{(k-1)}}^{(k-1)}}$. In practice,  this assumption will not hold exactly because of the adaptive features used in the sampling of the  proposals by the Monte Carlo routine, yielding a small bias of order the inverse of the sample size.  This issue is discussed in  Section~\ref{sec:practical}.

\subsection{The idealized importance sampling algorithm}\label{sec:ideal}

Based on the concepts introduced in Sections~\ref{sec:imp} to \ref{sec:estim}, we can now describe in this section the algorithm that will lead to an estimation of $\gamma^\ast_{l_{\mrm{max}}}$ based on importance sampling. \medskip

{\bf Algorithm 1 (Idealized ARMS) }
{\small 
\begin{algorithmic}[1]
 \Require level of interest $l_{\mrm{max}}$, initial reduced score $\Srom^{(1)}$ and error $\Err^{(1)}$ functions, budget of snapshots $K$, worst-case log cost threshold $\Cte$, reduced score precision $\epsilon \ll e$, number of hits of rare event before estimation $j_0$, quantile $\theta = M/N \in (0,1)$ for the bridge between proposals, boolean $restart$ variable for restarting or bridging after a reduced model update, initial learning parameter $\tau_0$ 

\Comment{{(\color{blue} initialization)}}
\State Set  $l^{(0)} = l_{b}(1) = -\infty$, $H=0$, $j=0$, $k=0$, $\tau=\tau_0$, ${Z_l}^{(1)}=1$ 
\While{$K>0$}
\State{$k \leftarrow k+1$}

\Comment{{(\color{blue} searching  the critical level, see Section~\ref{sec:prop_temp})}}
\State Simulate the flow of proposal distributions $\mu^{(k)}_l$ for $l\leq l^{(k)}$. The maximal level $l^{(k)}$ must be a level $l\in [l_{b}(k),l_{\mrm{max}}]$
satisfying both of the following two conditions:
\begin{align*}
&\Ent(  \check \eta^{(k)}_l \mid  \mu^{(k)}_l ) \leq \Cte \\
& \Ent(  \hat \eta^{(k)}_{l_{\mrm{max}}} \mid  \mu^{(k)}_l  ) < +\infty.
\end{align*}
$l^{(k)}$ is defined as the smallest $l \in [l_{b}(k),l_{\mrm{max}}]$  (in practice $l^{(k)} = l-\delta$  for some $\delta>0$ very small)  violating one the above two conditions; if search fails, then $l^{(k)} = l_{\mrm{max}}$. 
\State Compute\footnote{With a Sequential Monte Carlo approach for instance, see next sections.} the associated normalisation $Z^{(k)}_{l^{(k)}}$.
%
%

\Comment{{\color{blue} (snapshot sampling \&  score update, see Section~\ref{sec:learn})}}
\State Draw 
$$ X \sim \frac{ \e^{\tau E^{(k)} }}{\mu_{l^{(k)}}^{(k)}(e^{\tau \Err^{(k)} })} \mu_{l^{(k)}}^{(k)}  $$
\State{Compute  the snapshot  $\Score(X)=\mrm{score}( \Psi(X))$}
\If{ $ \Ent(  \check \eta^{(k)}_{l^{(k)}} \mid  \mu^{(k)}_{l^{(k)}} ) >\epsilon$ or  $l^{(k)} < l_{\max}$}
\State Update  reduced model: {build $(\Srom^{(k+1)} ,E^{(k+1)} )$ by enriching the previous surrogate model  $(\Srom^{(k)} ,E^{(k)} )$ with   $(X,\Score(X))$}.
\Else
\State Don't update the reduced model: set $(\Srom^{(k+1)} ,E^{(k+1)} ) \leftarrow (\Srom^{(k)} ,E^{(k)} )$.
\EndIf
\State {\bf if} {$\Score(X)> l_{\max} $}  {\bf then}  $j \leftarrow j+1$  {\bf end if}
\State{{\bf if} $j< j_0$ {\bf then} update  the learning function parameter $\tau$  {\bf else}  set  $\tau=0$  {\bf end if} } 

 
\Comment{{(\color{blue} importance sampling estimation, see Section~\ref{sec:estim})}}
\If{ $j>j_0$}
\State{$H\leftarrow H+1$}
\State {$$\hat \gamma^{H}_{l_{\max}} \leftarrow \frac{1}{H} \left( (H-1)\hat \gamma^{H-1}_{l_{\max}} +  Z^{{(k)}}_{l^{(k)}} \one_{\Score(X)> l_{\max}}  \delta_{X}\right)$$}
  \EndIf
  \State Set $K\leftarrow K-1$
 \If{$(\Srom^{(k+1)} ,E^{(k+1)} ) \neq (\Srom^{(k)} ,E^{(k)})$ }
 
 \Comment{{(\color{blue} bridging, see Section~\ref{sec:bridging})}}
   \If{$! restart$} 
 \State Find the largest (lexicographic) pair $(k',l') \in \llbracket 0, k \rrbracket \times \R$ 
 satisfying the following three conditions:
 \begin{align*}
  & \Ent( \mu^{(k+1)}_{l'} \mid  \mu^{(k')}_{l^{(k')}}              ) \leq - \ln (1-\theta)  \\
  & \Ent(  \check \eta^{(k+1)}_{l'} \mid  \mu^{(k+1)}_{l'}              ) \leq \Cte , \\
  & \Ent(  \hat \eta^{(k+1)}_{l_{\mrm{max}}} \mid  \mu^{(k+1)}_{l'}  ) < +\infty,
 \end{align*}
 and denote  the  solution pair by 
 $
 (k_{b}(k+1),l_{b}(k+1) ) \leftarrow (k',l').
 $

 \State Simulate the proposal $\mu^{(k+1)}_{l_{b}(k+1)}$ using $\mu^{(k_{b}(k+1))}_{l^{k_{b}(k+1)}}$; then compute the associated normalization $Z^{(k+1)}_{l_{b}(k+1)}$. 
 \Else 
  \State Set $l_{b}(k+1) = -\infty$
 \State Set  $Z^{(k+1)}= 1$  and reinitialize $\mu^{(k+1)}_{l_{b}(k+1)}$ with $\pi$
  \EndIf 
 \Else
 \State Set $l_{b}(k+1) = l^{(k)}$
 \State Set  $Z^{(k+1)}= Z^{(k)}$  and $\mu^{(k+1)}_{l_{b}(k+1)}=\mu^{(k)}_{l^{(k)}}$
 \EndIf
%
%

\EndWhile

\State \textbf{Return}  Estimate $\hat \gamma^{H}_{l_{\max}}$
\end{algorithmic}
}
\medskip

This algorithm is an idealized concept and,  in practice, it is not directly computable. Indeed it assumes exact evaluation of expectations and exact sampling according to the proposal and the surrogate distributions which are defined up to a normalization constant.

\section{Practical Implementation: AMS Simulations}\label{sec:practical}
 In this  section, we  propose a practical implementation of the idealized algorithm introduced in Section~\ref{sec:ideal} using an AMS methodology. The latter will simulate the flow of proposals for growing level parameters $l$ by providing Monte Carlo samples with approximately correct distributions for large sample sizes.
 
\subsection{AMS}\label{sec:AMS}

We wish to draw a sample of size $N$ according to the importance sampling distribution $\mu^{(k)}_{l^{(k)}}$ in the form~\eqref{eq:prop}, where, for ease of presentation, we temporarily assume that the target level ${l^{(k)}}$ is known.  We suppose we start with a given sample of size $N$ aproximately distributed  according to the law $\mu^{(k)}_{l_b}$  with $l_b< {l^{(k)}}$, as well as with a given estimator of the associated normalization $Z^{(k)}_{l_b}$.  The AMS algorithm~\cite{cdfg} will be used here to generate a finite sequence of $N$-samples approximately distributed according to a sequence of nested distributions $\{\mu^{(k)}_{l_i}\}_{i \geq 1}$, as well as estimators of the associated normalizations $ \set{ Z^{(k)}_{l_i} }_{i \geq 1}$, \ie such that $l_b<l_1<\ldots < l_m = l^{(k)}$. At iteration $i \geq 0$, the  $N$-sample of particles (clones) will be denoted 
\[(\Xi^{(i)}_{1}, \ldots, \Xi^{(i)}_{N})
\] 
and we omit in notation the dependence in $(k)$. The associated empirical distribution will be denoted \[ {\mu^{N,{(k)}}_{l_i}} := \frac1N \sum_{n=1}^N \delta_{\Xi^{(i)}_{n}} \simeq  \mu_{l_i}^{(k)}, \]
and the estimator of the normalization constant  $ Z^{(k)}_{l_m} $  will be denoted $ {Z^{N,{(k)}}_{l_m}}$. As a consequence, AMS will yield an approximation of the reference distribution $\pi$ restricted to the current reduced score  $\Srom^{(k)}$ as follows: 
$$
Z^{N,{(k)}}_{l_m} {\mu^{N,{(k)}}_{l_m}}(\ph\one_{\Srom^{(k)} > l^{(k)}} ) \simeq  \pi(\ph {\one_{\Srom^{(k)} > l^{(k)}}}),
$$
 for $\ph$ any given test function.   \medskip

Let us describe briefly the AMS algorithm. At the $i$-th iteration, the algorithm performs sucessively:  {\it i)} a weighting and selection step, and {\it ii)} a mutation step. The weighting step {\it i)} assigns weights to the empirical distribution ${\mu^{N,{(k)}}_{l_i}}$ so that it approximates $\mu_{l_{i+1}}^{(k)}$, up to a normalization constant given by an empirical approximation of the normalization ratios. More precisely, the empirical distribution at level $l_{i+1}$ and the associated normalizations are obtained by weighting the  current $N$-sample:  
$$ \frac{Z^{N,{(k)}}_{l_{i+1}}}{Z^{N,{(k)}}_{l_{i}}} {  {\mu_{l_{i+1}}^{N,(k)}}} := \frac1N  
\sum_{j=1}^N\one_{\Srom^{(k)} > l_{i+1}} \delta_{\Xi^{(i)}_{j}},$$ 
where the  equality is well-defined because $l_i<l_{i+1}$. This empirical distribution of course requires the determination (or a choice) of $l_{i+1}$. AMS performs an adaptive choice of $l_{i+1}$ as follows. AMS starts by sorting the $N$ particles according to their scores in ascending order. AMS then considers an adaptive target distribution defined by a fixed number $M < N$ of particles. The latter are assigned a weight equal to zero, and the level ${l_{i+1}}$ is defined by the $M$-th smallest particle score
$${l_{i+1}}=\Srom^{(k)}(\Xi^{(i)}_{(M)}),$$
in which $(1), \ldots ,(N)$ is the usual shorthand notation for increasing order statistics (based on scores, here).  
In other words, this corresponds to set ${l_{i+1}}$ as the highest level $l$ in a set given by a maximal relative entropy
 \[
l_{i+1} := \sup \set{l \in [l_i,l_{\mrm{max}}]: \, \Ent(\mu_l^{(k)} \mid \mu^{(k)}_{l_i}) \simeq - \ln \mu^{N,(k)}_{l_i}(\one_{\Srom^{(k)} > l}) <  -\ln\p{\frac{N-M}{N}}},
\]
with  the quantile (proportion of killed particles) satisfies  $M =  \lfloor \theta N\rfloor $. 

In the selection step, the first $M$ particles with zero-weights are removed and  surviving particles are duplicated. The $M$ resampled particles are chosen uniformly among the surviving particles. The AMS approximation of the normalization constant thus follows the recursion\footnote{To ease the presentation we  have assumed  here that particles have different scores. 
However,  in the case $\exists \mathcal{J} \subset  \mathbb{N}$ such that $ \Srom^{(k)}(\Xi^{(i)}_{(M)})= \Srom^{(k)}(\Xi^{(i)}_{(M+j)})$ for $j\in \mathcal{J}$, a good practice  is to  kill in addition the  set of particles $\{\Srom^{(k)}(\Xi^{(i)}_{(M+j)})\}_{j\in\mathcal{J}}$, and locally change the value of $M$ in \eqref{eq:updateZ} to $M+\textrm{card}(\mathcal{J})$ accordingly.
}  
\begin{equation}\label{eq:updateZ}
{Z^{N,{(k)}}_{l_{i+1}}}={Z^{N,{(k)}}_{l_{i}}} {\mu_{l_{i}}^{N,(k)}}(\one_{\Srom^{(k)} > l_{i+1}} )={Z^{N,{(k)}}_{l_{i}}} (1-{M}/{N}),
\end{equation}
yielding 
  $${Z^{N,{(k)}}_{l^{(k)}}} = {Z^{N,{(k)}}_{l_m}}={Z^{N,{(k)}}_{l_b}}(1-{M}/{N})^m.$$  

The mutation step {\it ii)} then consists of a Markovian transition that leaves $\mu_{l_{i+1}}^{(k)}$ invariant and is applied individually and independently a prescribed number of times to the $M$ particles after resampling, the goal being to bring diversity in the $N$-sample to counterbalance the effect of selection. In practice, Markovian transitions can be performed using a Metropolis-Hasting algorithm or other techniques studied in the Markov Chain Monte Carlo (MCMC) literature. We mention that the Markovian kernel is often chosen to be adaptive in order to maintain a given Metropolis acceptance rate~\cite{garthwaite2016adaptive}.  \medskip 

Note that AMS typically involves many evaluations of the score function underlying the  sequence of targeted  distributions. This paper precisely focuses on cases where the true score $\Score$  too expensive for so many evaluation, is substituted by but a reduced score $\Srom^{(k)}$, much cheaper to evaluate. \medskip

\subsection{Reaching the critical level with AMS}\label{sec:increaseAMS}

Let us now describe the main Monte Carlo sampling routine of the algorithm. We recall that after $k$ iterations, one is given with a current reduced score $\Srom^{(k)}$, an associated pointwise error estimate $\Err^{(k)}$, an initial level $l_{b}(k) $  and a sample of size $N$ approximately distributed according to $\mu_{l_{b}(k)}^{(k)}$, that we denote using a sample-size superscript: $\mu_{l_{b}(k)}^{N,(k)}$ .  One want to perform an AMS simulation targeting the distribution $\mu_{l^{(k)}}^{(k)}$, where we recall that ${l^{(k)}}$ is the critical, non-achievable level the closest to $l_{b}(k)$. The latter is formally defined by \eqref{eq:level} (for $N=+\infty$). When using sample-based empirical distributions, one will need to adapt the latter conditions.

The AMS simulation will proceed as explained in Section~\ref{sec:AMS}, creating a sequence $(l_i)_{i\geq 0}$ of increasing levels starting from $l_b={l_{b}(k)}$, creating a sequence of empirical distributions ${\mu^{N,{(k)}}_{l_i}}$ for $i \geq 0$. We then check for each proposed level increase from $l_i$ to $l_{i+1}$ that the entropic conditions \eqref{eq:level} using the empirical distribution  ${\mu^{N,{(k)}}_{l_i}}$ to approximate the latter. More precisely, we remark that
\begin{align}
	\Ent\p{\check{\eta}_{l_{i+1}}^{(k)} \mid \mu_{l_{i+1}}^{(k)} }   & = \ln \frac{ {\mu^{{(k)}}_{l_{i}}}(\one_{\Srom^{(k)}  > l_{i+1}}) }{ {\mu^{{(k)}}_{l_{i}}}(\one_{\Srom^{(k)}  - \Err^{(k)}  > l_{i+1}}) }
	 \simeq \ln \frac{ {\mu^{N,{(k)}}_{l_{i}}}(\one_{\Srom^{(k)}  > l_{i+1}}) }{ {\mu^{N,{(k)}}_{l_{i}}}(\one_{\Srom^{(k)}  - \Err^{(k)}  > l_{i+1}}) } , \label{eq:logCostAMS} \\
	\Ent\p{\hat{\eta}_{l_{\max}}^{(k)} \mid \mu_{l_{i+1}}^{(k)} } & = \ln \frac{ {\mu^{{(k)}}_{l_{i}}}(\one_{\Srom^{(k)}  > l_{i+1}}) }{ {\mu^{{(k)}}_{l_{i}}}(\one_{\Srom^{(k)}  + \Err^{(k)}  > {l_{\max}}}) } 
	 \simeq \ln \frac{ {\mu^{N,{(k)}}_{l_{i}}}(\one_{\Srom^{(k)}  > l_{i+1}}) }{ {\mu^{N,{(k)}}_{l_{i}}}(\one_{\Srom^{(k)}  + \Err^{(k)}  > {l_{\max}}}) },\label{eq:dominationAMS}
\end{align}
so that \eqref{eq:level} is  re-written as:
\begin{align*}
	& l^{(k)} =l_m := \\
	& \min\{ l_i, \,   i \geq 0  : \;   \ln \frac{ {\mu^{N,{(k)}}_{l_{i}}}(\one_{\Srom^{(k)}  > l_{i+1}}) }{ {\mu^{N,{(k)}}_{l_{i}}}(\one_{\Srom^{(k)}  - \Err^{(k)}  > l_{i+1}}) } \geq \Cte \;  \; \text{or} \;  \;  \ln \frac{ {\mu^{N,{(k)}}_{l_{i}}}(\one_{\Srom^{(k)}  > l_{i+1}}) }{ {\mu^{N,{(k)}}_{l_{i}}}(\one_{\Srom^{(k)}  + \Err^{(k)}  > {l_{\max}}}) }   =+\infty\}.
\end{align*}
The above means that when the AMS simulation proposes to increment the level $l_i$ to $l_{i+1}$ and that at least one of these conditions is violated, then the critical level set to $ l^{(k)}=l_i$ is reached.\medskip

\begin{Rem}\label{rem:1} If the logarithm  of the ratio in  \eqref{eq:dominationAMS} is finite when $N \to \infty$ for all  intermediate levels $l_{j}$ from $-\infty$ to $l_i$
 then we can verify recursively that, at a formal level,
  $$\lim_{N\to \infty}\ln \frac{ {\mu^{N,{(k)}}_{l_{i}}}(\one_{\Srom^{(k)}  > l_{i+1}}) }{ {\mu^{N,{(k)}}_{l_{i}}}(\one_{\Srom^{(k)}  + \Err^{(k)}  > {l_{\max}}}) } <+\infty  \Leftrightarrow  \Ent\p{\hat{\eta}_{l_{\max}}^{(k)} \mid \mu_{l_{i+1}}^{(k)} }   <+\infty ,$$
  which precisely means that the rare event of interest is a subset of the support of the current proposal distribution $\mu_{l_{i+1}}^{(k)}$.\medskip
\end{Rem}

The  procedure to reach the critical level in practice is summarized  by the following algorithm.\medskip

{\small 
{\bf Algorithm 2 (AMS simulation up to the critical level)}
\begin{algorithmic}[1]
 \Require  level of interest $l_{\mrm{max}}$, initial distribution  $\mu^{N,(k)}_{l_b(k)}$ and normalization $Z^{N,(k)}_{l_b(k)}$, related reduced score $\Srom^{(k)}$ and error $\Err^{(k)}$,  quantile $\theta \in [1/N,1[$,  worst-case log cost threshold $\Cte$.  \\
\State Set   $i=0$, $c_{try}=0$
\While{  $l_i < l_{\mrm{max}}$ \&  $c_{try} \leq  \Cte$ \&  $d_{try} < +\infty $

}
    \State {Order scores $\Srom^{(k)}$ relative to particles of ${\mu^{N,{(k)}}_{l_i}}$} 
    \State {Set $l_{i+1}=\min(l_{M},l_{\mrm{max}})$ with  $l_{M}$ the $M$-th smallest score}
    
    \State{Compute $$c_{try}= \ln \frac{ {\mu^{N,{(k)}}_{l_{i}}}(\one_{\Srom^{(k)}  > l_{i+1}}) }{ {\mu^{N,{(k)}}_{l_{i}}}(\one_{\Srom^{(k)}  - \Err^{(k)}  > l_{i+1}}) }  \simeq \Ent\p{\check{\eta}_{l_{i+1}}^{(k)} \mid \mu_{l_{i+1}}^{(k)} }$$
    	as well as 
    	$$ d_{try} = \ln \frac{ {\mu^{N,{(k)}}_{l_{i}}}(\one_{\Srom^{(k)}  > l_{i+1}}) }{ {\mu^{N,{(k)}}_{l_{i}}}(\one_{\Srom^{(k)}  + \Err^{(k)}> l_{i+1}} )} \simeq \Ent\p{\hat{\eta}_{l_{\max}}^{(k)} \mid \mu_{l_{i+1}}^{(k)} }$$ as defined in \eqref{eq:logCostAMS} and \eqref{eq:dominationAMS}}

    \If{ $l_{i+1} < l_{\mrm{max}}$ \&  $c_{try} \leq  \Cte $ \&  $d_{try} < +\infty $}
       \State{Kill the $M\ge \lfloor \theta N\rfloor$ particles with the lowest score or with score equal to $l_{i+1}$ }
	\State{Update  ${Z^{N,{(k)}}_{l_{i+1}}} $ with recursion  \eqref{eq:updateZ}}
     \State{Perform $ M$ duplications (uniformly among the survivors)}
	 \State Mutate the duplicates using a MCMC leaving target $\mu_{l_{i+1}}^{(k)}$ invariant. 
	 \State  Denote the resulting sample ${\mu_{l_{i+1}}^{N,(k)}}$.

	\Else
	\State {Set $l^{(k)}=l_{i}$}
   \EndIf
   	\State {Set $i  \leftarrow{i+1}$}

\EndWhile
\State \textbf{Return}  Empirical measure  ${\mu_{l_{l^{(k)}}}^{N,(k)}}$ and normalizing constant ${Z^{N,{(k)}}_{l^{(k)}}}$
 \end{algorithmic}
   }

The following remark is useful in practice, as it provides some prior information on the worst-case log cost threshold parameter $\Cte$.\medskip

 \begin{Rem}[Worst-case log cost: a lower bound in practice]
 	 Consider, in Algorithm~$2$ above, the empirical distribution ${\mu_{l_{i}}^{N,(k)}}$. By construction of AMS, $l_{i+1}$ is the $M$-th smallest score among  the $N$ scores. Set $c_{\min}(M,N) \eqdef  \ln \frac{N-M}{N-M-1}$. 
As long as $ \Err^{(k)}$ is non zero, at least one particle of the sample, the $M$-th particle, lies outside the level set obtained by subtracting the error to the score. $ c_{\min}(M,N)$ is thus the lower bound of the entropy between the empirical measures with and without the error term, that is 
$$ c_{\min}(M,N) \le 
\ln \frac{\mu_{l_{i}}^{N,(k)}(\one_{\Srom^{(k)}  > l_{i+1}}) }{\mu_{l_{i}}^{N,(k)}(\one_{\Srom^{(k)} - \Err^{(k)} > l_{i+1}}) }.$$
Setting the numerical threshold to the minimal possible value $\Cte= c_{\min}(M,N)$ can be considered a safe strategy, leading to a slower but safer convergence of the algorithm.
\end{Rem}

\subsection{Bridging with an AMS step}\label{sec:bridge}


As exposed in Section~\ref{sec:bridging}, the bridging procedure at the end of iteration $k$ consist in selecting a past index $=k_{b}:=k_{b}(k)\le k$ and a new initial level  ${l_{b}(k+1)}$ which will be used in the AMS simulation with the new score approximation $\Srom^{(k+1)}$ starting from a past (appropriately recorded) empirical distribution $\mu^{N,{(k_{b})}}_{l^{(k_{b})}}$. As defined previously,  bridging consists in choosing, using a lexicographic search, a feasible pair $(k_{b},{l_{b}(k+1)}) \in \llbracket 0, k \rrbracket \times \R $, where feasibility means satisfying the quantile condition \eqref{eq:BridgeQuantile} and the two entropic conditions ~\eqref{eq: wlcCondition} and \eqref{eq: domCondition}. It is now necessary to define empirical counterparts to the latter conditions. 
As previously, the two  entropic conditions are approximated by 
\begin{eqnarray}
\Ent\p{\check{\eta}_{l'}^{(k+1)} \mid\mu_{l'}^{(k+1)} }  \simeq 
\ln \frac{ {\mu^{N,{(k')}}_{l^{(k')}}}(\one_{\Srom^{(k+1)}  > {l'}}) }{ {\mu^{N,{(k')}}_{l^{(k')}}}(\one_{\Srom^{(k+1)}  - \Err^{(k+1)}  > {l'}}) } \le \Cte ,\label{eq:wlcConditionAMS}\\
 \Ent\p{\hat{\eta}_{l_{\max}}^{(k+1)} \mid \mu_{l'}^{(k+1)}} \simeq 
  \ln \frac{ {\mu^{N,{(k')}}_{l^{(k')}}}(\one_{\Srom^{(k+1)}  > {l'}}) }{ {\mu^{N,{(k')}}_{l^{(k')}}}(\one_{\Srom^{(k+1)}  + \Err^{(k+1)}  > {l_{\max}}}) } < +\infty,\label{eq:domConditionAMS}
\end{eqnarray}
 where in the above $(k',l')$ denote feasible pairs that are candidates for the solution $(k_{b},l_{b}(k+1))$. Note that the well-posedness of \eqref{eq:domConditionAMS} can be justified by Remark \ref{rem:1}.
Concerning the quantile condition, we remark that, given some $k'$ and ${l'}$, and assuming that the domination relation stands
\begin{equation}\label{eq:NestedAMS}
{\mu_{l^{(k')}}^{(k')}} \gg \mu_{l'}^{(k+1)} \Leftrightarrow \set{ \Srom^{(k+1)} > {l'} } \subset  \set{ \Srom^{(k')} > l^{(k')}},
\end{equation}
 an empirical approximation of the quantile condition  \eqref{eq:BridgeQuantile} is    
\begin{equation}\label{eq:quantileCOnditionAMS}
{\mu_{l'}^{N,(k')}}(\one_{\Srom^{(k+1)} \leq {l'} } )\le \theta.
\end{equation}
In order to verify \eqref{eq:NestedAMS}, we propose to browse the history of past AMS simulations and check the inclusion of the support of $ \mu_{l'}^{(k+1)} $ in the one of ${\mu_{l^{(k')}}^{(k')}} $ using all the recorded $N$-samples targeting the past proposal distributions  ${\mu_{l^{(\tilde k)}}^{N,(\tilde k)}}$ with $\tilde k < k_{b}$, whose samples are scattered between the reference distribution and the current distribution. This check must be carried out for all $\tilde k < k_{b}$   and is given by 
\begin{equation}\label{eq:const1}
\mu_{l^{(\tilde k)}}^{N,(\tilde k)} \p{ \one_{\Srom^{(k')} > l^{(k')}}\one_{\Srom^{(k+1)} > l'} +\one_{\Srom^{(k+1)} \le l'}}=1.
\end{equation}
 
To sum up, bridging consist in selecting some feasible pair, \ie some pair $(k_{b},{l_{b}(k+1)})$ satisfying  \eqref{eq:wlcConditionAMS}, \eqref{eq:domConditionAMS},   \eqref{eq:quantileCOnditionAMS} and \eqref{eq:const1}.  
Note that once the pair has been determined,  the  normalization estimate can be updated as
\begin{equation}\label{eq:inclusionZRefined}
{Z^{N,{(k+1)}}_{l_{b}(k+1)}}= {Z^{N,{({ k_{b}})}}_{l^{( k_{b})}}} {\mu_{l^{( k_{b})}}^{N,{({ k_{b}})}}}(\one_{\Srom^{(k+1)} > {l_{b}(k+1)}} ).
\end{equation}
Then the $\tilde M( k_{b} )$ 
particles (at max given by $\tilde M( k_{b} )= \lfloor \theta N\rfloor $ when~\eqref{eq:quantileCOnditionAMS} is saturated) with updated scores  below or equal to ${l_{b}(k+1)}$ are replaced by duplicating surviving particles and performing mutation targeting $\mu_{l_{b}(k+1)}^{(k+1)}$  (in an AMS line of thinking).  The next AMS simulation with $\Srom^{(k+1)}$ can then be started from the level ${l_{b}(k+1)}$. The following iteration then starts up to the critical level $ l^{(k+1)}(\Cte,{l_{b}(k+1)})$, an so on.

  As mentioned in Section~\ref{sec:bridging}, among the feasible set, we focus on either the trivial pair  $(k_{b},{l_{b}(k+1)})=(0, -\infty)$, or the largest (lexicographic) pair $(k_{b},{l_{b}(k+1)}) \in \llbracket 0, k \rrbracket \times \R$.  
A simple way to solve this  maximisation problem is to begin with the highest possible index $k':=k$ and then to decrease $k'$ by a unit if no feasible level $l'$ has been found for this index. To search for the largest feasible level $l'$ at a given index $k'$, one can proceed as follows. Begin by setting $l'$ to the highest  level satisfying \eqref{eq:quantileCOnditionAMS},  then   lower this level until  \eqref{eq:wlcConditionAMS} and \eqref{eq:domConditionAMS}  hold and finally   perform  for all $\tilde k<k_{b}$ the check \eqref{eq:const1}. 

The non-trivial  bridging  procedure is summarized by the following algorithm.\medskip

{\small 
  {\bf Algorithm 3 (Bridging)}
\begin{algorithmic}[1]
\Require  level of interest $l_{\mrm{max}}$, current index $k$, for all $k'\le k$, particles of distribution  $\mu^{N,(k')}_{l^{( k')}}$, normalization $Z^{N,( k')}_{l^{(k')}}$ and related reduced score $\Srom^{(k')}$ and error $\Err^{(k')}$,  reduced score $\Srom^{(k+1)}$, the quantile  $\theta \in [1/N,1[$,  worst-case log cost threshold $\Cte$.\\

\State {Set $k'=k$ and  $feasible=false$}
\Comment{{(\color{blue}   find largest feasible pair)}}
\While{$!feasible$ \& $k'>0$}
\State {Compute $\Srom^{k+1}$ for particles in $\mu_{l^{(k')}}^{(k')}$ and sort their  scores}
\State{Set $\tilde M= \lfloor N \theta \rfloor$}
\While{$!feasible$ \& $\tilde M>0$}

\State{Set $l'=\min(l_{\tilde M},l_{\mrm{max}})$ with  $l_{\tilde M}$ the $\tilde M$-th smallest score }
\State{Compute~\eqref{eq:wlcConditionAMS}:
	 $c_{try}= \ln \frac{ {\mu^{N,{(k')}}_{l^{(k')}}}(\one_{\Srom^{(k+1)}  > {l'}}) }{ {\mu^{N,{(k')}}_{l^{(k')}}}(\one_{\Srom^{(k+1)}  - \Err^{(k+1)}  > {l'}}) }  \simeq  \Ent\p{\check{\eta}_{l'}^{(k+1)} \mid\mu_{l'}^{(k+1)} } 
	 $ }
\State{Compute~\eqref{eq:domConditionAMS}:
	$d_{try} =\ln \frac{ {\mu^{N,{(k')}}_{l^{(k')}}}(\one_{\Srom^{(k+1)}  > {l'}}) }{ {\mu^{N,{(k')}}_{l^{(k')}}}(\one_{\Srom^{(k+1)}  + \Err^{(k+1)}  > {l_{\max}}}) } \simeq \Ent\p{\hat{\eta}_{l_{\max}}^{(k+1)} \mid \mu_{l'}^{(k+1)}} $}
\State{{\bf if}  $c_{try} \le \Cte \, \& \,  d_{try}  < +\infty$ {\bf then} }
\State{\quad\quad Set $feasible=true$}
\State{\quad\quad {\bf for} { $\tilde k \in \llbracket 0, k' -1\rrbracket $} {\bf do} }
 	\State{\quad\quad \quad\quad  Compute $\Srom^{(k+1)}$ and $\Srom^{(k')}$ for particles in $\mu_{l^{(\tilde k)}}^{(\tilde k)}$   }
  	\State{\quad\quad \quad\quad  {\bf if} ${\mu_{l^{(\tilde k)}}^{N,(\tilde k)}} (\one_{\Srom^{( k')} > l^{(k')}}\one_{\Srom^{(k+1)} > l'}+\one_{\Srom^{(k+1)} < l'})\neq1$ }
   	\State{\quad\quad \quad\quad \quad\quad Set $feasible=false$}
\State{\quad\quad \quad\quad  {\bf end if}}
\State{\quad\quad {\bf end for}}
\State{Set $\tilde M \leftarrow \tilde M-1$}
\EndWhile
\State{Set $k' \leftarrow k'-1$}
\EndWhile
\If{$feasible$}  \Comment{{(\color{blue} $k'>0$: perform an AMS step)}}
\State{Set the largest pair as $(k_{b},l_{b}(k+1) )= (k',l')$  }

 	 	\State{Compute  ${Z^{N,{(k+1)}}_{l_{b}(k+1)}}$ using  \eqref{eq:inclusionZRefined} }
		\State{Kill the $\tilde M$ particles in $\mu_{l^{(k_{b})}}^{N, ( k_{b})}$ of smallest score $\Srom^{(k+1)}$  or with score equal to $l_{b}(k+1)$}
 		\State{Perform duplications (uniformly among the survivors) to replace the killed particles}
 	\State{Simulate $\mu_{l_{b}(k+1)}^{N,(k+1)}$ by mutating the duplicates to target $\mu_{l_{b}(k+1)}^{(k+1)}$ }
\Else{} \Comment{{(\color{blue} $k'=0$: initialize a fresh AMS)}}
\State{Set $l_{b}(k+1) = -\infty$  }
 \State{Set ${Z^{N,{(k+1)}}_{l_{b}(k+1)}}=1$}
\State {Simulate  $\mu_{l_{b}(k+1)}^{N,(k+1)}$ (by sampling the reference distribution $\pi$) }

\EndIf
\State \textbf{Return}  Empirical measure  $\mu_{l_{b}(k+1)}^{N,(k+1)}$ and normalizing constant ${Z^{N,{(k+1)}}_{l_{b}(k+1)}}$
\end{algorithmic}
}

\subsection{AMS-based importance sampling estimator}\label{sec:ISAMS}

 Using the sequence of empirical distributions $\{ \mu_{{l^{(k)}}}^{{N,(k)}}\}_{k=1}^K$ generated by AMS simulations, we can write a closed-form approximation of the importance sampling estimator $\hat \gamma_{l_{\max}}^{H}$ of  $\gamma^\ast_{l_{\mrm{max}}}$  defined by~\eqref{eq:estim2}.   The AMS approximation of the estimator $\hat \gamma_{l_{\max}}^{H}$ is defined  by 
\begin{equation}\label{eq:ISpractice}
\hat \gamma^{H,N}_{l_{\max}}(\varphi) \eqdef \frac{1}{H} \sum_{k=K_{j_0}+1}^{K_{j_0}+H}  {Z_{l^{(k)}}^{N,(k)}} \varphi(X_{k+1}){ \one_{\Score> l_{\max} }(X_{k+1})} ,  
\end{equation}
In particular, the AMS approximation of the   estimator  $\hat \gamma^H_{l_{\max}}(\one)$ of $p^\ast= \gamma^\ast_{l_{\max}}(\one)$ is
 \begin{equation}\label{eq:ISpractice2}
\hat p_{IS} \eqdef \hat \gamma^{H,N}_{l_{\max}}(\one),  
\end{equation}
AMS  also provides as by-products a sequence of approximate estimators of $\pi(\{\Srom^{(k)}\ge {l_{\max}}\})$ given by
\begin{equation}\label{eq:AMSpractice}
\hat p_{AMS} \eqdef
 {Z_{l^{(k)}}^{N,(k)}}\mu_{l^{(k)}}^{N,(k)}(\one_{\Srom^{(k)} > {l_{\max}}}),
\end{equation}
This estimator can be an increasingly accurate approximation of $p^\ast$, but as our numerical experiments show, it can also be highly biased, even in the case where the reduced score $\Srom^{(k)}$ converges uniformly to $\Score$ as $k$ increases. 

\subsection{The ARMS algorithm}\label{sec:algo}
The complete methodology proposed in this work is summarized in the complete algorithm presented below named adaptive reduced multilevel splitting (ARMS). ARMS is a practical implementation of the idealized importance sampling estimator (Algorithm 1), built on the AMS simulation framework described above.   \medskip

{\small 
{\bf Algorithm 4 (ARMS) }
\begin{algorithmic}[1]
\Require  level of interest $l_{\mrm{max}}$, initial reduced score $\Srom^{(1)}$ and error $\Err^{(1)}$,  budget of snapshots $K$, the quantile $\theta \in (0,1)$ proposing the bridge level and  level increases in AMS,   number of hits of rare event target $j_0$, number of particles $N$ for AMS, worst-case log cost threshold $\Cte$, reduced score precision $\epsilon \ll e$, boolean $restart$ variable for restarting or bridging after a reduced model update, initial learning parameter $\tau_0$.

\Comment{{(\color{blue} initialization)}}
\State Set  $l^{(0)} = l_{b}(1) = -\infty$, $H=0$, $j=0$, $k=0$, $\tau=\tau_0$, $Z_{l^{(0)}}^{N,(0)}=1$ 
\State {Set  $\mu_{ l_{b}(1)}^{N,(1)}$  by drawing a $N$-sample of the reference distribution $\pi$ }
\While{$K>0$}
\State{$k \leftarrow k+1$}

\Comment{{(\color{blue} AMS simulation to  critical level)}}
\State{ Use {\bf Algorithm 2} (see	 Section~\ref{sec:increaseAMS}) from  $l_b= l_{b}(k)$ to:
\begin{itemize}
\item[]   \quad\quad $\bullet$  estimate the critical level  $l^{(k)} \in [l_{b}(k),l_{\mrm{max}}]$ 
\item[]   \quad\quad $\bullet$  simulate $\mu_{l^{(k)}}^{N,(k)}$ starting from $\mu_{l_{b}(k)}^{N,(k)}$
\item[]   \quad\quad $\bullet$  compute the associated normalisation $Z^{N,(k)}_{l^{(k)}}$.
\end{itemize}}

\Comment{{\color{blue} (snapshot sampling  \&   score update)}}
\State Draw $$ X \sim   \frac{{\mu^{N,{(k)}}_{l^{(k)}}} e^{\tau E^{(k)}}}{{\mu^{N,{(k)}}_{l^{(k)}}}(e^{\tau E^{(k)}}) }  $$ 
\State{Compute  the snapshot  $\Score(X)$}
\If{ $l^{(k)} < l_{\max}$ or $ \Ent(  \check \eta^{(k)}_{l^{(k)}} \mid  \mu^{(k)}_{l^{(k)}} ) \simeq - \ln \mu^{N,{(k)}}_{l^{(k)}}( \one_{\Srom^{(k)}  - \Err^{(k)}  > {l^{(k)}}} )  > \epsilon$} 
\State {Build $(\Srom^{(k+1)} ,E^{(k+1)} )$ by enriching $(\Srom^{(k)} ,E^{(k)} )$ with   $(X,\Score(X))$}.
\Else
\State Set $(\Srom^{(k+1)} ,E^{(k+1)} ) \leftarrow (\Srom^{(k)} ,E^{(k)} )$.
\EndIf
\State {\bf if} {$\Score(X)> l_{\max} $}  {\bf then}  $j \leftarrow j+1$   {\bf end if} 
\State{{\bf if} $j< j_0$ {\bf then} update  the learning function parameter $\tau$  {\bf else}  set  $\tau=0$  {\bf end if} }

\Comment{{(\color{blue} importance sampling)}}

\If{ $j>j_0$}
\State{$H\leftarrow H+1$}
\State {Update estimate (see Section~\ref{sec:ISAMS}) $$\hat \gamma^{H,N}_{l_{\max}} \leftarrow \frac{1}{H} \left((H-1)\hat \gamma^{H-1,N}_{l_{\max}} +  Z^{N,{(k)}}_{l^{(k)}}  \one_{\Score(X)> l_{\max}} \delta_X\right)$$}
  \EndIf
  \State Set $K\leftarrow K-1$
 \If{$(\Srom^{(k+1)} ,E^{(k+1)} ) \neq (\Srom^{(k)} ,E^{(k)})$ }
 \If{$! restart$}
 \Comment{{(\color{blue} AMS step for bridging)}}
 \State Use {\bf Algorithm 3} (see Section~\ref{sec:bridge})  to: 
\begin{itemize}
 \item[]   \quad\quad\quad\quad $\bullet$ find the largest feasible pair $(k_{b},l_{b}(k+1) )$  
\item[]   \quad\quad\quad\quad $\bullet$  simulate $\mu^{N,(k+1)}_{l_{b}(k+1)}$ from $\mu^{N,(k_{b})}_{l^{(k_{b})}}$
\item[]   \quad\quad\quad\quad $\bullet$  compute the associated normalisation $Z^{N,(k+1)}_{l_{b}(k+1)}$.
\end{itemize}
 \Else  \Comment{{(\color{blue} or  AMS reinitialization)}}
\State{Set $l_{b}(k+1) = -\infty$  }
 \State{Set ${Z^{N,{(k+1)}}_{l_{b}(k+1)}}=1$}
\State {Set  $\mu_{l_{b}(k+1)}^{N,(k+1)}$ by drawing a $N$-sample of the reference distribution $\pi$ } 
 \EndIf
 \Else
 \State Set $l_{b}(k+1) = l^{(k)}$ 
 \State Set $Z^{N,(k+1)}_{l_{b}(k+1)}= Z^{N,(k)}_{l^{(k)}}$, $\mu^{N,(k+1)}_{l_{b}(k+1)}=\mu^{N,(k)}_{l^{(k)}}$
 \EndIf

\EndWhile
\State \textbf{Return}  Estimate $\hat \gamma^{H,N}_{l_{\max}}$
\end{algorithmic}
}
~\\
We comment hereafter the ARMS algorithm. The simulation starts from an i.i.d. sample with the reference distribution.  ARMS then iterates through four blocks until the computing budget allocated to snapshots is exhausted.\medskip

The \emph{first block} (lines 4 to 5) consists in  increasing the level until reaching the first non-achievable level presented in Section~\ref{sec:prop_temp} and implemented in practice by an AMS routine in Section~\ref{sec:increaseAMS}. In short, it simulates the distribution flow $ l \mapsto \mu^{N,(k)}_{l}$ with $l\in[{l_{b}(k)}, l^{(k)}]$, where  $l^{(k)}$ is the  critical level where IS of the true target $ \eta^{\ast}_{l^{(k)}}$ by the current reduced model $\Srom^{(k)}$ is expected to be too expensive. \medskip

The \emph{second block} (lines 6 to 14) samples a new sample according to  the proposal $\mu^{N,(k)}_{l^{(k)}}$, penalized by some learning function in the early iterations. It then computes the related snapshot, updates the reduced score and finally updates the learning function.  This block follows the guideline presented in Section~\ref{sec:learn}. \medskip

The  \emph{third block} (lines 15 to 19) is the estimation phase beginning when the algorithm has reached the rare event in some sense. This block performs importance sampling with the proposal $\mu^{N,(k)}_{l^{(k)}}$ and updates the IS estimate \eqref{eq:ISpractice}. \medskip

The \emph{fourth and final block} (lines 20 to 31)
 is the restart or  bridge routine presented in Section~\ref{sec:bridging} and implemented in practice by an AMS step in Section~\ref{sec:bridge}. In short, if the restart option is untriggered  ($restart=false$), then the bridging procedure searches in past simulations to find a certain pair $(k_{b},l_{b}(k+1) )$ which enables to simulate from $\mu^{N,(k_{b})}_{l^{(k_{b})}}$ towards  $\mu^{N,(k+1)}_{l_{b}(k+1)}$  in a single AMS step. Otherwise (in the case $restart=true$),  a fresh AMS simulation is initialized at each score approximation update. 
 We mention that the bridging routine is avoided in the case where the stopping criterion \eqref{eq:stopUpdate} is reached. Indeed, in this case, no bridging is needed, since the reduced score is not updated anymore and the next initial level can simply be set to $l^{(k)}$. 

\section{Numerical Simulations}\label{sec:num}

We first consider a toy example given by an explicit one dimensional true score. Reduced modeling is performed by cubic spline interpolation. The rare event simulation is in this case one-dimensional but nonetheless challenging, as the score presents several maxima and moreover splines may produce highly oscillating approximations. We then focus our attention on a  more realistic scenario in which the rare event is defined through the high-dimensional solution of an elliptic  PDE  with random inputs. The PDE solution is approximated by a reduced basis approach which yields the reduced models.   All our numerical simulations are carried out using an implementation of the ARMS algorithm in the Julia language, available at \url{https://gitlab.inria.fr/pheas/arms}.  

\subsection{Examples}

\begin{figure}[h!] \label{fig:0}
\begin{center}
\begin{tabular}{cc}
\includegraphics[width=0.5\textwidth]{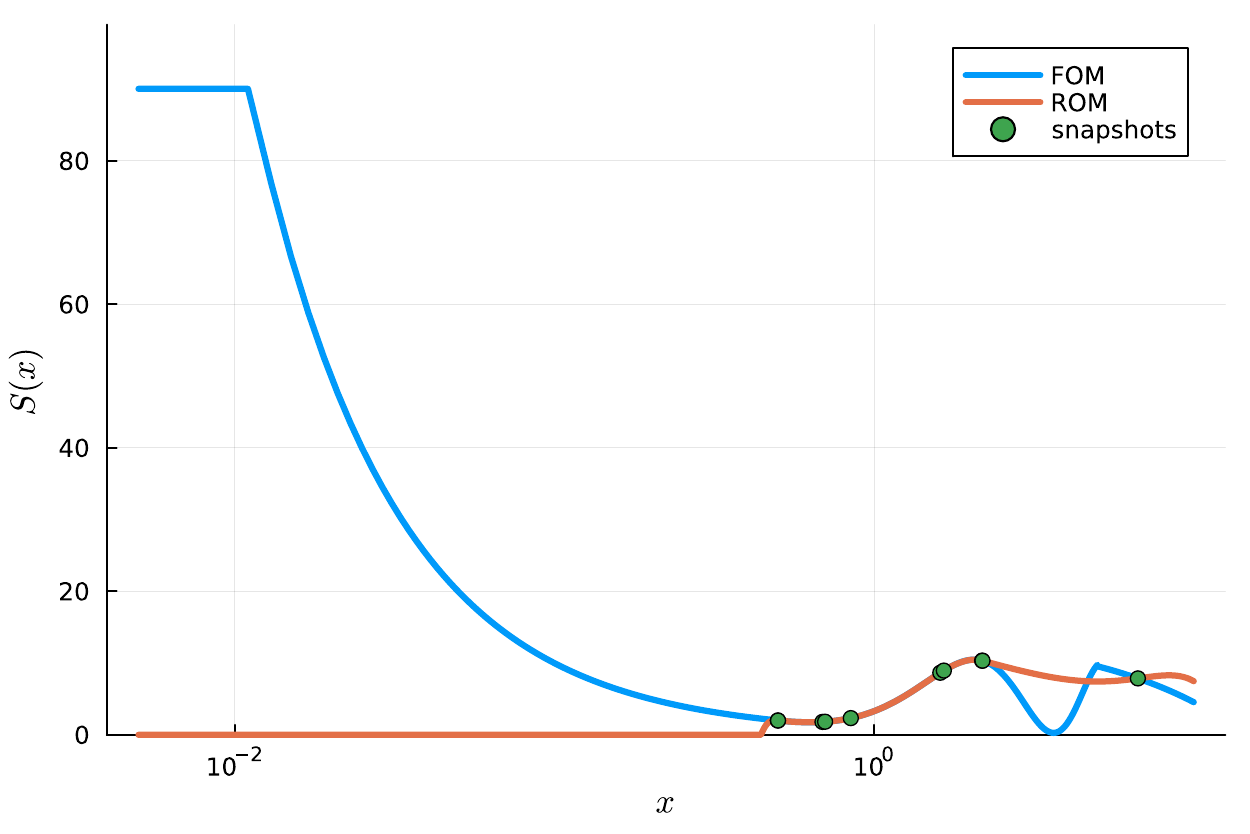}&
\includegraphics[width=0.5\textwidth]{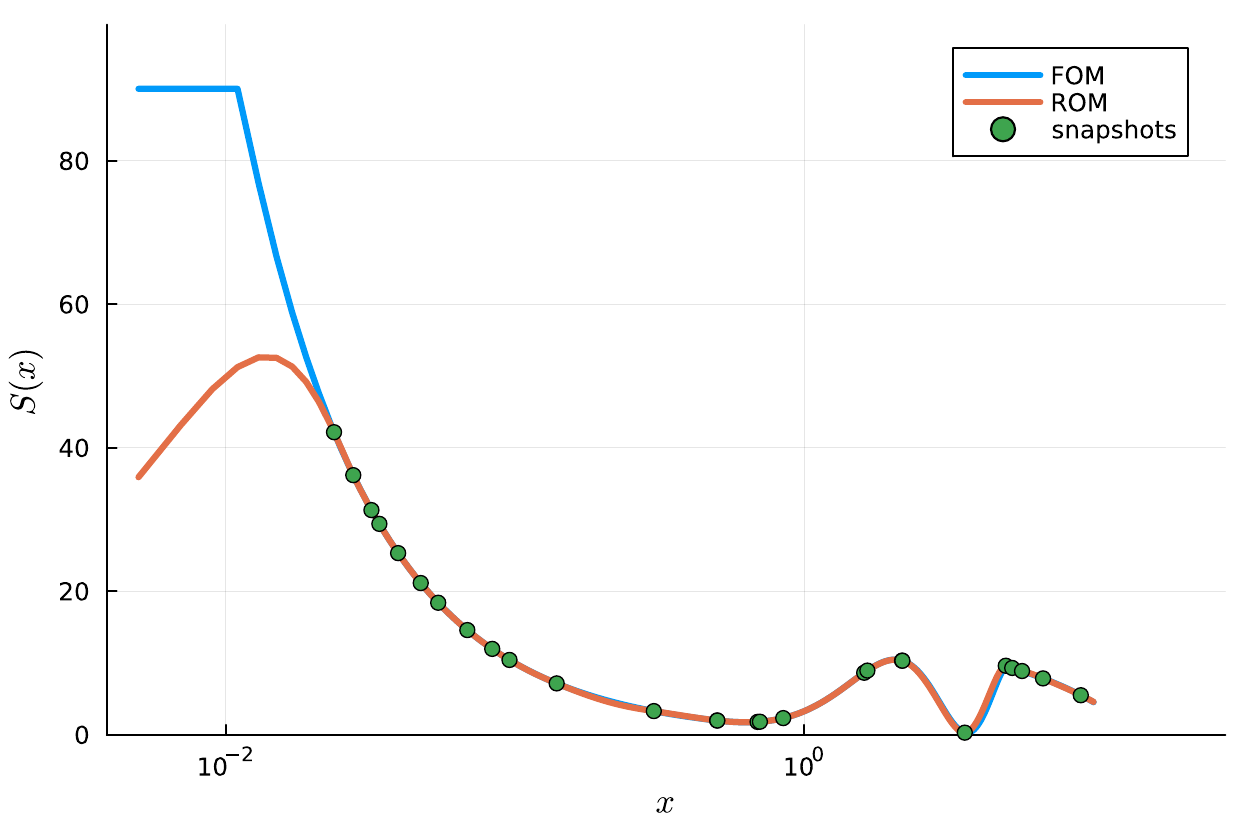}\\
\includegraphics[width=0.5\textwidth]{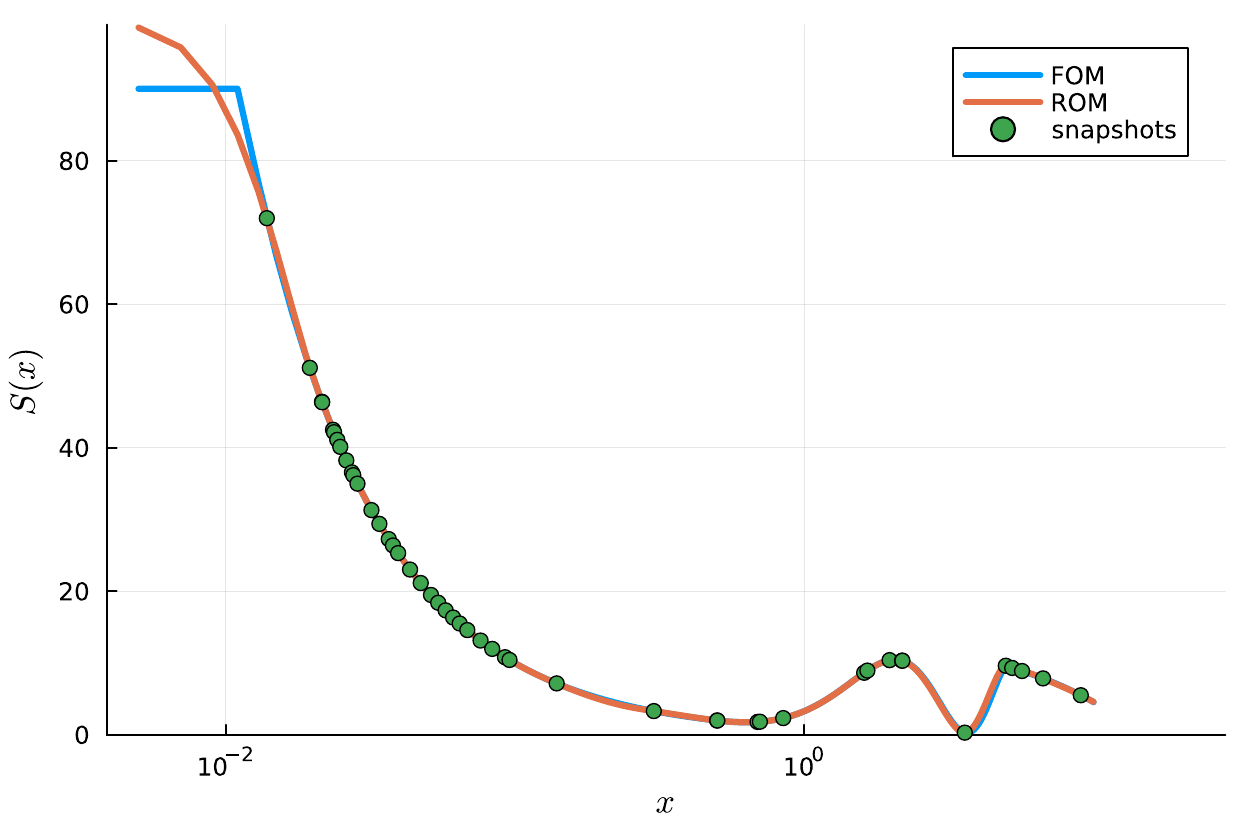}&
\includegraphics[width=0.5\textwidth]{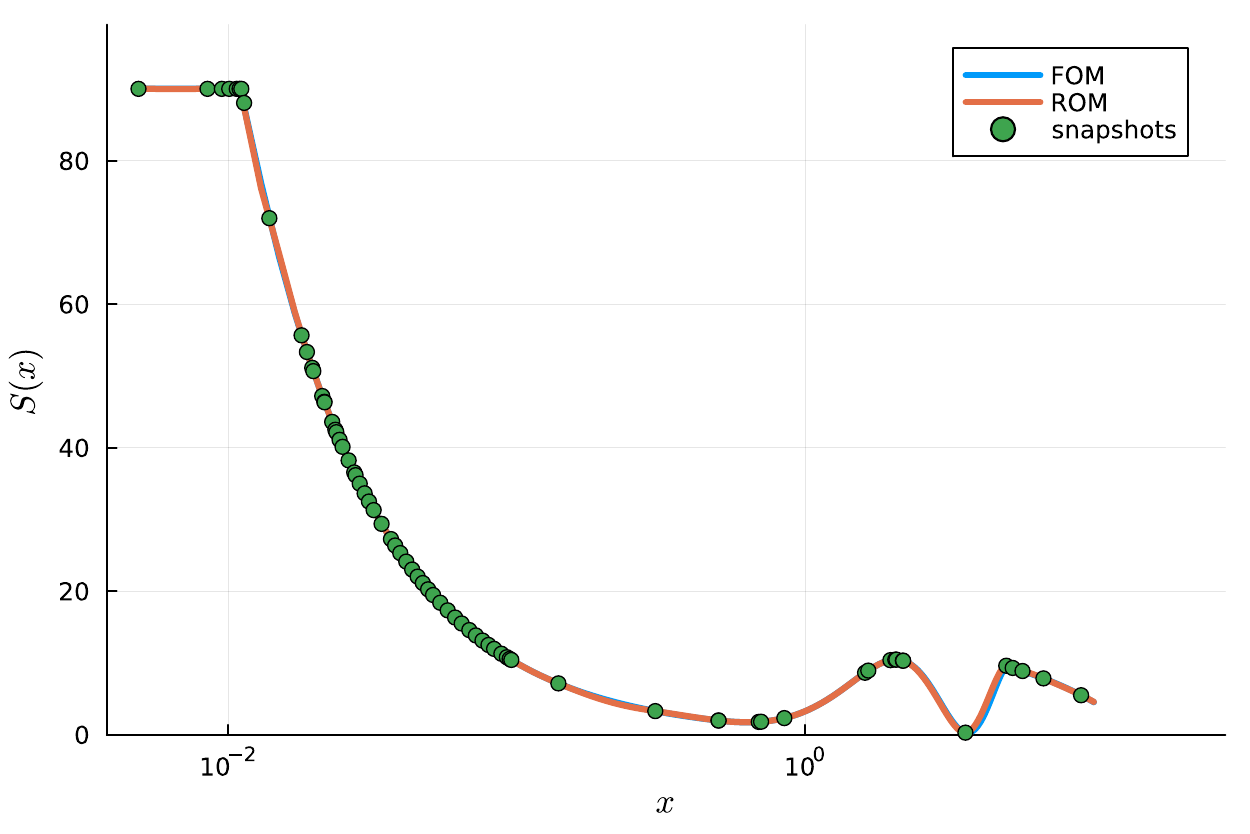} \vspace{-0.5cm}
\end{tabular}
	\caption{{\footnotesize {\bf Illustration of  the algorithm adaptive behavior (example \#1).} The four plots illustrate the algorithm at iteration $k=1,20,40,60$. They display in blue the score function $\Score$ called full order model (FOM) (the rare event being $\Score(X) \ge 90$), in red the  reduced score approximation  $\Srom^{(k)}$ called reduced order model (ROM), in green the snapshots  $(X_{1},\ldots, X_{k})$  used to build the reduced score.   }}
		\end{center}\vspace{-0.cm}
\end{figure}

We will consider the problem of  estimation of $p^\ast=\pi(\one_{\Score > l_{\max}})$ with $ \Score(X)=\mrm{score}(\Psi^\ast(X))$, for various  models $\Psi^\ast$ and reduced models $\Psi$, various dimensions $q=\dim(X)$ of the parameter space,  various model dimensions  $h=\dim(\Psi^\ast(X))$ and different score functions. \medskip  

\hspace{-0.5cm}\begin{tabular}{c|ccccccc}
&$\Psi^\ast$&$\Psi$&$\dim(X)$&$\dim(\Psi^\ast(X))$&  $\mrm{score}$& $p^\ast$&$l_{\max}$\\
\hline
example~\#1 & analytical  & spline&1&1& $| \cdot |$& 2.18e-08 &90 \\
example~\#2a& PDE  & RB  &4&5101& $L^1$& 6.31e-04 & 0.5 \\
example~\#2b& PDE  &RB& 4&5101& $L^\infty$& 1.47e-04& 3.0  \\
example~\#2c& PDE  &RB & 25&5101& $L^1$& 2.48e-14 &0.5   \\
\end{tabular}\medskip

We detail hereafter these examples.

\subsubsection{Example \#1: spline approximation of a closed-form score}
In a first numerical experiment, we design a one-dimensional analytic toy model. Cubic spline data interpolation is used to build the surrogate $\Srom(x)$ from the snapshots  and the error bound is set  to two times the absolute value of the true error $E(x)=2|\Srom(x)-\Score(x)|$.

Using spline interpolation is not recommended in practice, as it easily creates wide oscillations that can artificially create spurious high score areas that mislead the sampling routines. It is nonetheless an interesting benchmark for difficult situations where the reduced modeling produces very sub-optimal approximations in some areas.

The score function, depicted in Figure~\ref{fig:0}, is defined for $x\in\mathcal{X}=\mathbb{R}_+$ by $\Score(x)=|\Psi^\ast(x) |$ with
\begin{equation*}
\Psi^\ast(x)=
  \left\{\begin{aligned}
  l_{\max}\quad  &\textrm{if} \quad x \le 1/l_{\max} \\
1/x+f(x) \quad &\textrm{else} 
\end{aligned}\right. ,
\end{equation*}
where 
$f(x)=\alpha \left( \one_{x_0\le x <x_1}  \sin(x-x_0)^2 + \one_{x_1 \le x} ( \sin(x_1-x_0)^2 -\beta(x-x_1))\right),$
given parameters $x_1>x_0>1/l_{\max} $ and $\alpha,\beta>0$.  
The reference distribution $\pi$ is a log-normal distribution such that $\ln(x) \sim \mathcal{N}(1.5,1.5)$. {\color{black} Note that here the rare event is $\{S^*\ge l_{\rm{max}}\}$.}
The parameter are set to  $x_0=0.5$, $x_1=5$, $\alpha=15$, $\beta=0.1$ and $l_{\max}=90$,  corresponding to the rare event probability (given in closed-form) and given in the table above.

The initial surrogate is built using the 10 snapshots drawn according to the reference density.

 Figure~\ref{fig:0} shows a typical evolution of the spline-based score approximation as well as the selected snapshots  over the iterations of the algorithm.

\subsubsection{Example \#2: reduced basis approximation of a PDE score}\label{sec:PDERB}
In our second numerical experiment, the score function is defined as $\Score(x)=\|\Psi^\ast(x)\|_p$ with $p=1$ or $p=\infty$,   $\Psi^\ast(x)$ being the high-fidelity numerical approximation of the solution of a  bi-dimensional PDE parametrized by  a  $q$-dimensional  vector $x=(x_1\cdots x_q)^\intercal\in  \mathcal{X}=\R_+^q$. The chosen PDE corresponds to the well-studied  {\it thermal block problem}, which models heat diffusion  in an heterogeneous media.    The  reference distribution for $x$ will be a log-normal distribution with independent components with $\ln(x_i) \sim \mathcal{N}(1.5,1.5)$, for $i=1,\ldots,q$. The probability of interest will correspond to the occurrence of the temperature exceeding  the given threshold $l_{\max}$ either in average (examples~\#1, \#2a \#2c)  or locally (example~\#2b) over the spatial domain  $\Omega=[0,1]^2$.  PDE solutions will  be approximated using the popular reduced basis (RB) method.  \medskip

We begin by describing the parametric PDE and the high-fidelity numerical approximation of its solution. The temperature function $\Psi$ over  $\Omega$ is ruled by the elliptic problem
\begin{equation*}
 \left\{\begin{aligned}
-\nabla \cdot(\kappa(\cdot;x) \nabla \Psi)=&1\quad  \textrm{in}\quad \Omega\\
\Psi=&0\quad  \textrm{on}\quad \partial \Omega
\end{aligned}\right.,
\end{equation*}
where the function $\kappa (\cdot;x):\Omega \to \mathbb{R}_+$ represents a parametric diffusion coefficient function. This PDE models the stationary temperature field $\Psi$ of an heterogeneous medium which is uniformly heated in the domain and thermostatted at the boundary. 

The   high-fidelity approximation $\Psi^\ast(x)$ of the solution of this diffusion problem entails the solution of a variational problem.  Precisely, let $\mathcal{V}$ denote the Hilbert space $\mathcal{H}^1_0(\Omega)$ and $\|\nu\|_\mathcal{V}=\| \nabla \nu \|_{L^2}$ the norm induced by the inner product defined over $\mathcal{V}$. For any $x \in \mathcal{X}$, we define the solution $\Psi^\star(x)\in \mathcal{V}$  of the variational problem  
$a(\Psi(x),\nu;x)=\int_\Omega  \nu \d\Omega,\quad\forall \nu \in \mathcal{V},$
where  $a: \mathcal{V} \times \mathcal{V} \to \mathbb{R}$ is the bilinear form
$a(u,\nu;x)=\int_\Omega \kappa(\cdot;x) \nabla u \cdot \nabla \nu d\Omega$. We consider the piecewise constant  diffusion coefficient $\kappa(\cdot;x)=\sum_{i=1}^{q} \one_{\Omega_i}(\cdot) x_i$  with the partition $\Omega=\cup_{i=1}^{q}\Omega_i$ and $\Omega_i$'s being non-overlapping squares. 
Setting the parameter space $\mathcal{X}= (\mathbb{R}^*_+)^q$ implies that   the  bilinear form is  strongly coercive, and the  Lax-Milgram lemma  ensures the variational problem has a unique solution.

 Consider next a finite $h$-dimensional subspace $\mathcal{V}_h \subset \mathcal{V}$.  The  high-fidelity approximation $\Psi^\ast(x)\in \mathcal{V}_h$ of the solution $\Psi^\star(x)$  is defined for any $x\in \mathcal{X}$ by 
\begin{align*}
\Psi^\ast(x)=\arg\min_{\nu \in \mathcal{V}_h} \| \Psi^\star(x)-\nu\|^2,
\end{align*}
with the energy norm induced by the bilinear form $\| \nu \|=\sqrt{a(\nu,\nu;x)}$. The  high-fidelity approximation $\Psi^\ast(x)$ is obtained in practice by solving a  large  system of $h$ linear equations.~\cite{quarteroni2015reduced}. For the function $score$  defined as the $L^p$-norm, the rare event \{$\Score(X)=\|\Psi^\ast(X)\|_{L^p} > l_{\max}\}$  corresponds to reaching a critical temperature $l_{\max}$, \textit{e.g.} a melting temperature, on average  in the domain ($p=1$) or in some point of the domain ($p=\infty$).
  
  We are interested in the construction of an inexpensive approximation to the set of high-fidelity solutions  
$$
\mathcal{M}=\{\Psi^\ast(x)\in \mathcal{V}_h : x \in \mathcal{X} \},
$$
by evaluating only some of its elements.  By selecting a well-chosen set of $K$ snapshots $\{\Psi^\ast(x_i)\}_{i=1}^K$, the RB  framework  approximates $\mathcal{M}$
in a well-chosen linear subspace $ \mathcal{V}_K \subset  \mathcal{V}_h$ of much lower dimension $K \ll h$. 
Using this framework yields    {\it i)} an approximation  $\Psi$ of  the high-fidelity solution $\Psi^\ast$ which is used to compute the approximation $\Srom(x)=\|\Psi(x)\|_{L^p}$ of the score $\Score(x)$,  {\it ii)} a so-called  {\it a posteriori} error bound on $\| \Psi^\ast(x)-\Psi(x)\|_{\mathcal{V}_h}$ denoted by $\Delta_\Psi(x)$. The high-fidelity solution, the RB approximation and the a posteriori error bounds are computed via  the {\it pymor}  Python\textsuperscript{\textregistered} toolbox~\cite{milk2016pymor} available at \url{https://github.com/pymor/pymor}.   

To use the ARMS algorithm, we need to design an error bound $\Err(x)$  majoring the absolute error on the score $|\Score(x)-\Srom(x)|$, relying on the {\it a posteriori} error bound  on $\| \Psi^\ast(x)-\Psi(x)\|_{\mathcal{V}_h}$.  
In the case of $p=1$, the error bound satisfies   
\begin{align*}
|\Score(x)-\Srom(x)| &=  |\| \Psi^\ast(x)\|_{L^1}-\|\Psi(x)\|_{L^1}| \\
&\le   \| \Psi^\ast(x)-\Psi(x)\|_{L^1}\le   \| \Psi^\ast(x)-\Psi(x)\|_{L^2}  \le \pi^{-1}\| \Psi^\ast(x)-\Psi(x)\|_{\mathcal{V}_h}\\
& \le \pi^{-1}\Delta_\Psi(x) =E(x),
\end{align*}
where we have used the triangular inequality, the Cauchy-Schwarz, the Poincaré inequality with the constante $\pi^{-1}$   on the unit square domain and the a posteriori error bound. As the finite element basis is an interpolating one, we will assume  $h$  sufficiently large so that we can compute the $L^1$-norm as  the  average of the finite element coefficients. In the case of $p=\infty$, except in the one-dimensional case, one can not in general rigorously upper bound $ \| \Psi^\ast(x)-\Psi(x)\|_{L^\infty}$  by  $\| \Psi^\ast(x)-\Psi(x)\|_{\mathcal{V}_h}$. We will nevertheless use for this experiment  the rough approximation $|\Score(x)-\Srom(x)| \approx \| \Psi^\ast(x)-\Psi(x)\|_{\mathcal{V}_h} \le \Delta_\Psi(x) = E(x)$,  possibly violating  assumption \eqref{eq:error_prior}. 


Let us comment on two important features of the RB approximation in regards to  our rare event simulation framework. 
First, we emphasize that our algorithm enriches the RB approximation space using a generalization of the  popular {\it weak greedy} algorithm. Indeed, the latter selects at each iteration the snapshot in the solution set $\mathcal{M}$ which possess the worst a posteriori error bound for the current RB approximation. This setup corresponding in our algorithm to the degenerated case where the learning parameter is set to $\tau=+\infty$. At the opposite, the setup where $\tau=0$ will correspond to sampling according to the proposal, independently of the approximation error. In the general setup where $0< \tau < \infty$, snapshot selection will balance these two cases, as  the RB will be enriched with a sample drawn according to a weighted version of the proposal distribution, with weights proportional to  the  function $e^{\tau E^{(k)}}$. 

Second, as in standard RB approaches, we here extensively rely on the affine parametric dependence of the RB approximation by splitting the assembly of the reduced matrices and vectors in an {\it off-line } and {\it on-line} phases. The former, which is performed once for all at each RB update, entails the computation of all the $h$-dependent and $x$-independent structures. It requires $\mathcal{O}(Kh^2+K^2h)$ operations. Besides, one must add the evaluation of a snapshot {\it per se}, which is $\mathcal{O}(h^3)$. In the latter, on-line phase, for a given parameter $x\in \mathcal{X}$, we assemble and solve the RB system and provide the error bound  in $\mathcal{O}(K^3)$,  with a cost depending only on $K$.  For further details, we refer the reader to the introductory book on RB~\cite{quarteroni2015reduced}.

As there is no analytic value for the rare event probabilities sought, we perform standard AMS calculations with a score based on this high-fidelity approximation $\Psi^\ast(x)$ and using $N=1e3$ particles. Averaging 10 runs, we obtain, after intensive computation, an empirical Monte Carlo approximation for $p^\ast=\pi(\one_{\Score > l_{\max}})$ which is presented in the table above.


%
%
\begin{figure}[h!]
\begin{center}
\begin{tabular}{ccccc}
\hspace{-0.5cm}\rotatebox{90}{$\quad\quad k=1$} &\hspace{-0.35cm}\includegraphics[height=0.15\textwidth]{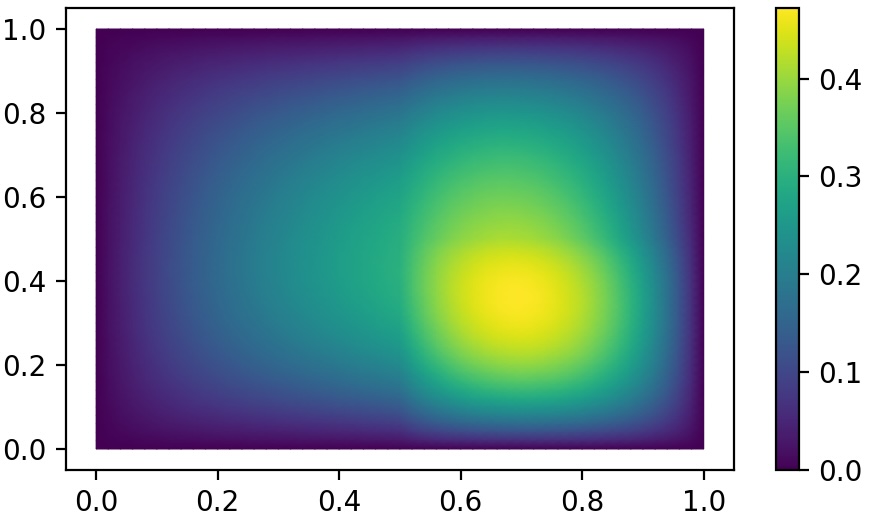}&\hspace{-0.35cm}\includegraphics[height=0.15\textwidth]{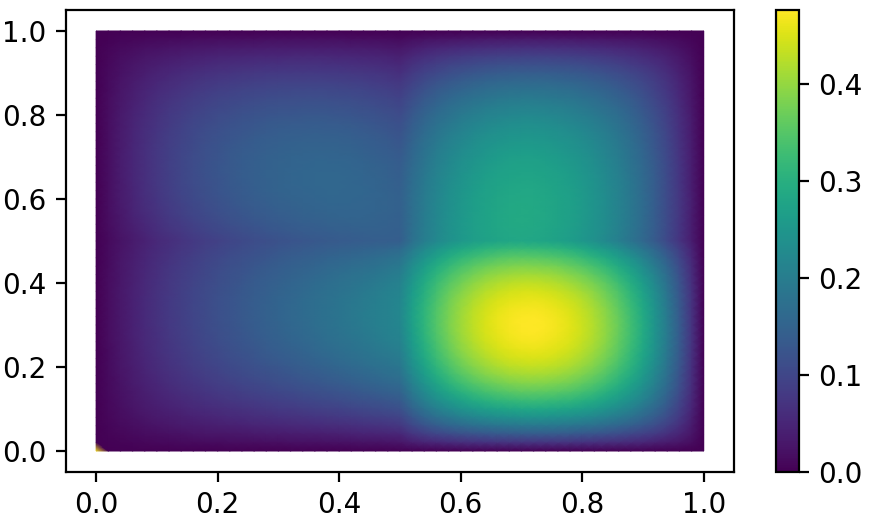}&\hspace{-0.35cm}\includegraphics[height=0.15\textwidth]{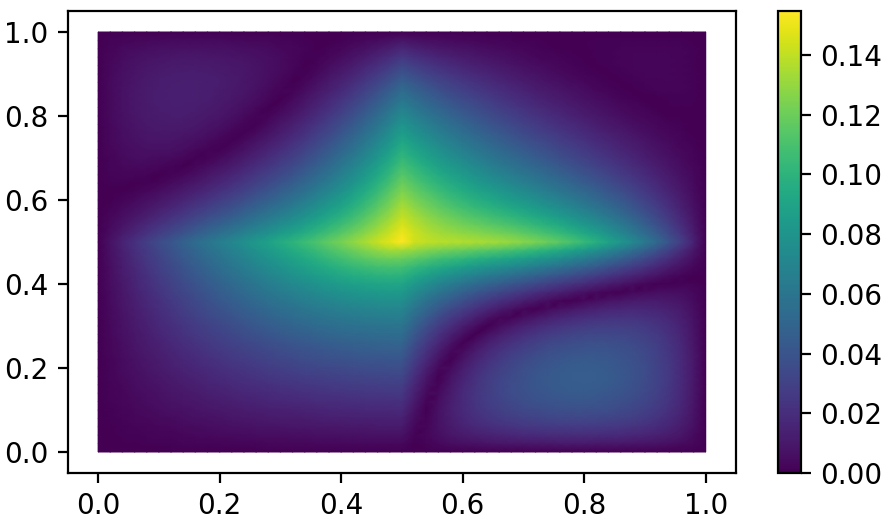}&\hspace{-0.25cm}\includegraphics[height=0.15\textwidth,width=0.2\textwidth]{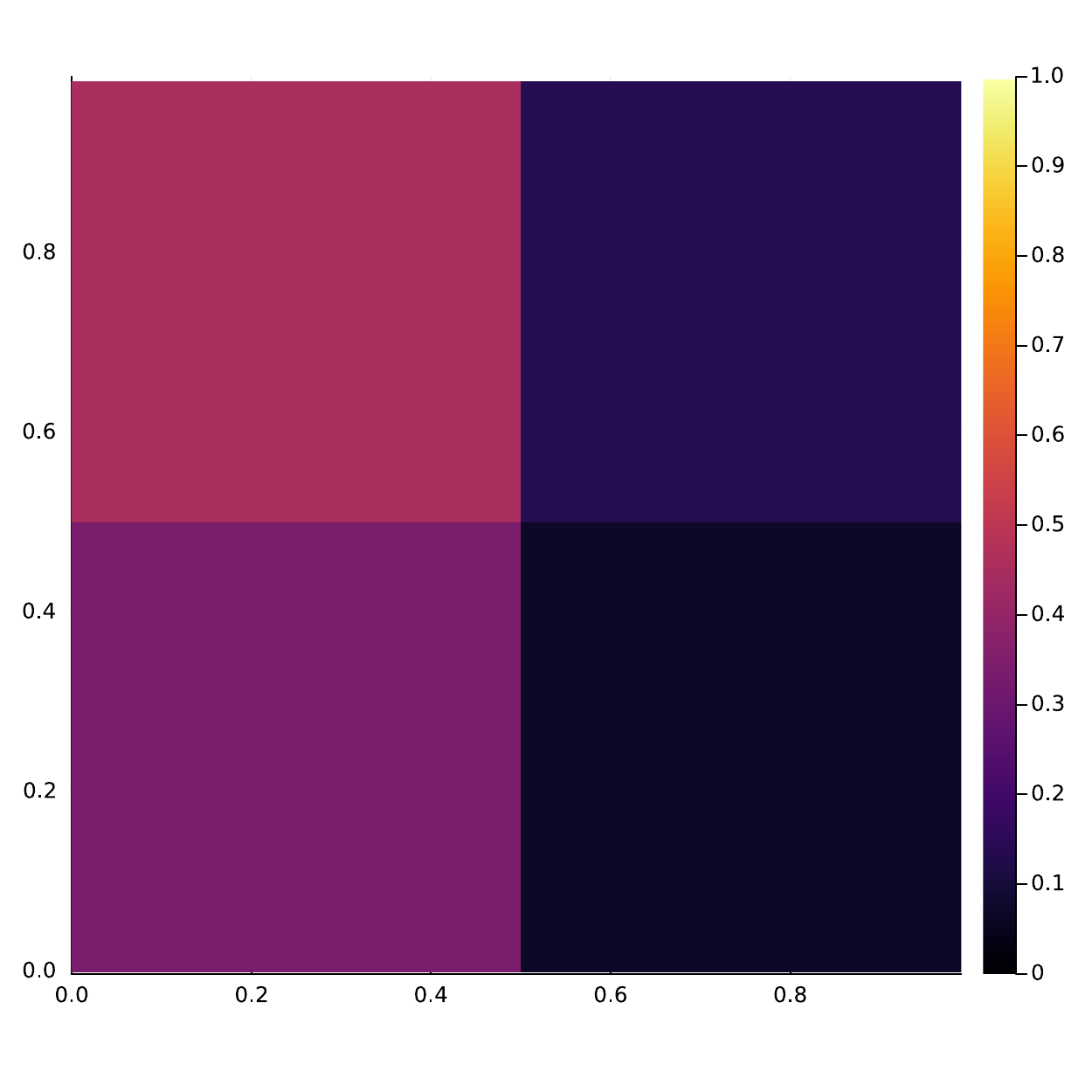}\\
\hspace{-0.5cm}\rotatebox{90}{$\quad\quad k=12$} &\hspace{-0.35cm}\includegraphics[height=0.15\textwidth]{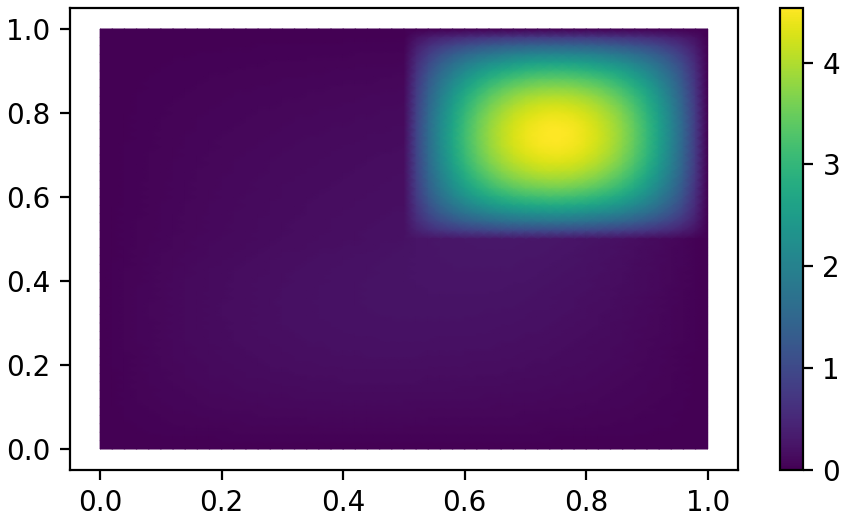}&\hspace{-0.35cm}\includegraphics[height=0.15\textwidth]{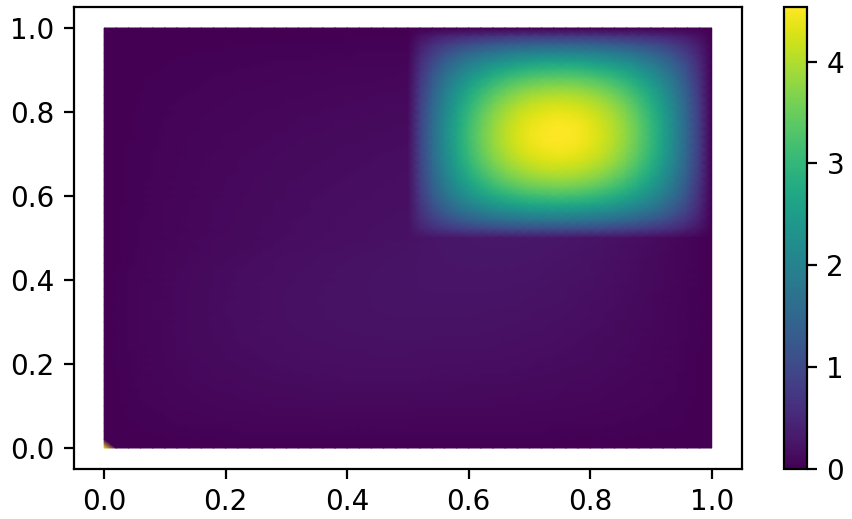}&\hspace{-0.35cm}\includegraphics[height=0.15\textwidth]{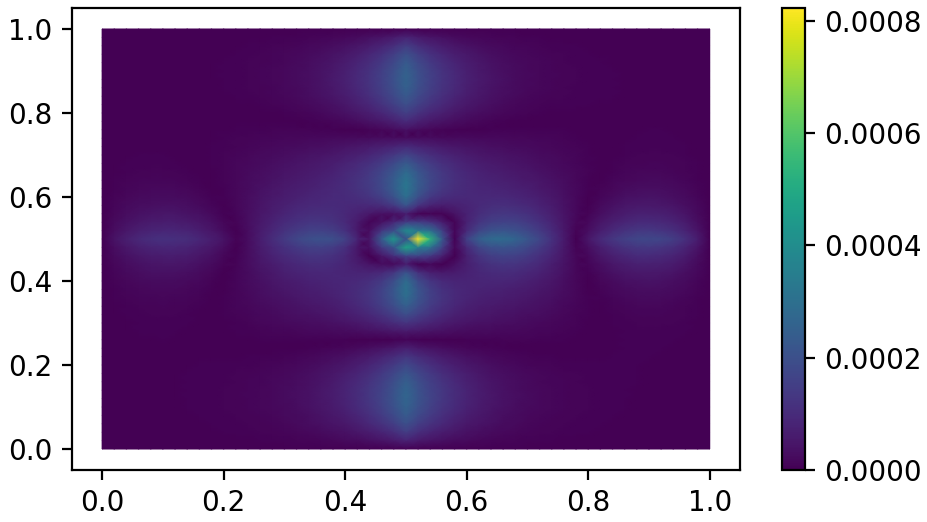}&\hspace{-0.25cm}\includegraphics[height=0.15\textwidth,width=0.2\textwidth]{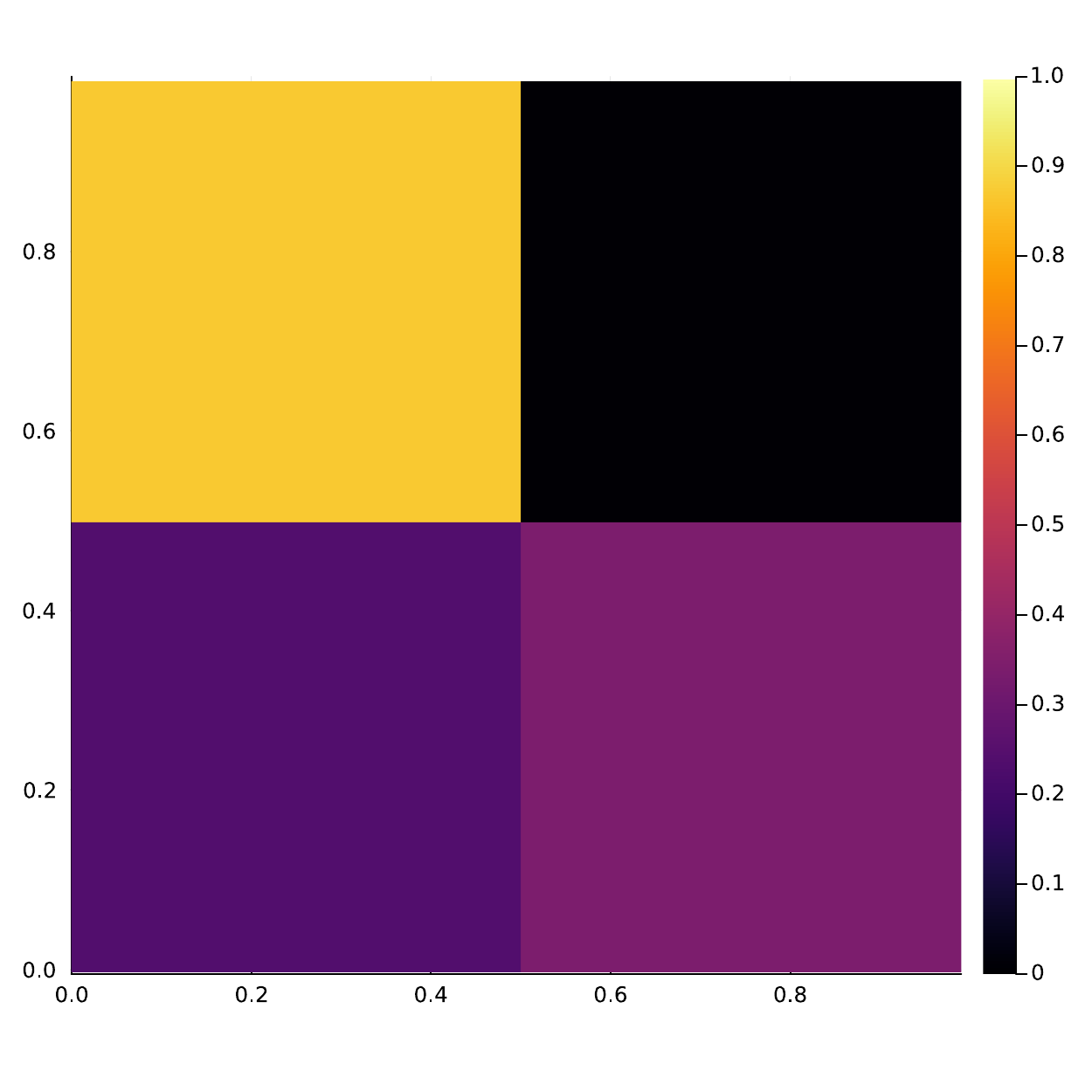}\\
\hline 
\hspace{-0.5cm}&\hspace{-0.5cm}\footnotesize{ $\Psi^\ast(X^{k+1})$}&\hspace{-0.5cm}\footnotesize{  $\Psi^{(k)}(X^{k+1})$}&\hspace{-0.5cm}\footnotesize{ error }&\hspace{-0.5cm}\footnotesize{$X^{k+1}$ }\\
\end{tabular}
	\caption{{\footnotesize {\bf  Illustration of  the algorithm adaptive behavior (example \#2a).}  The figure displays  the high-fidelity solution  of the PDE (left), its reduced basis approximation (middle left) and the related absolute error (middle right), evaluated at the snapshot  $X^{k+1}$ (right)  sampled at iteration $k$  by our ARMS algorithm.
	 {  \label{fig:4bis}}}}
		\end{center}\vspace{-0.cm}
\end{figure}

\begin{figure}[h!]
\begin{center}
\begin{tabular}{ccccc}
\hspace{-0.5cm}\rotatebox{90}{$\quad\quad k=1$} &\hspace{-0.35cm}\includegraphics[height=0.15\textwidth]{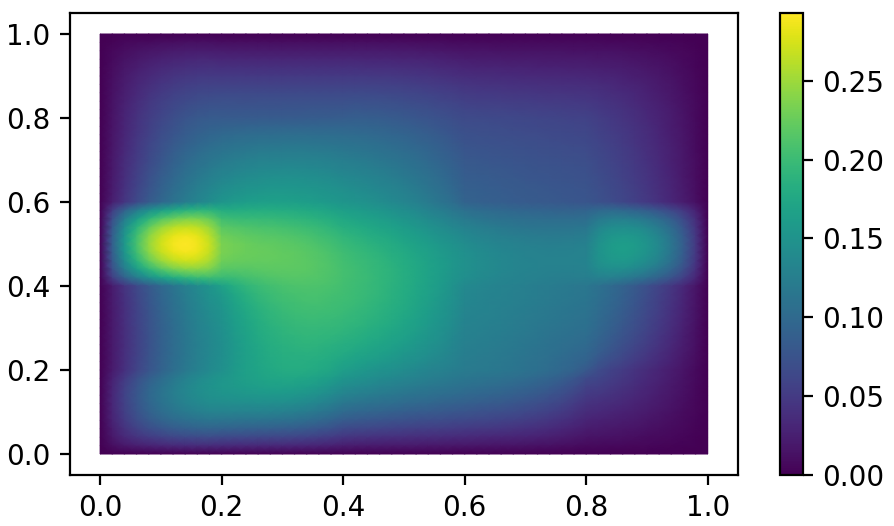}&\hspace{-0.35cm}\includegraphics[height=0.15\textwidth]{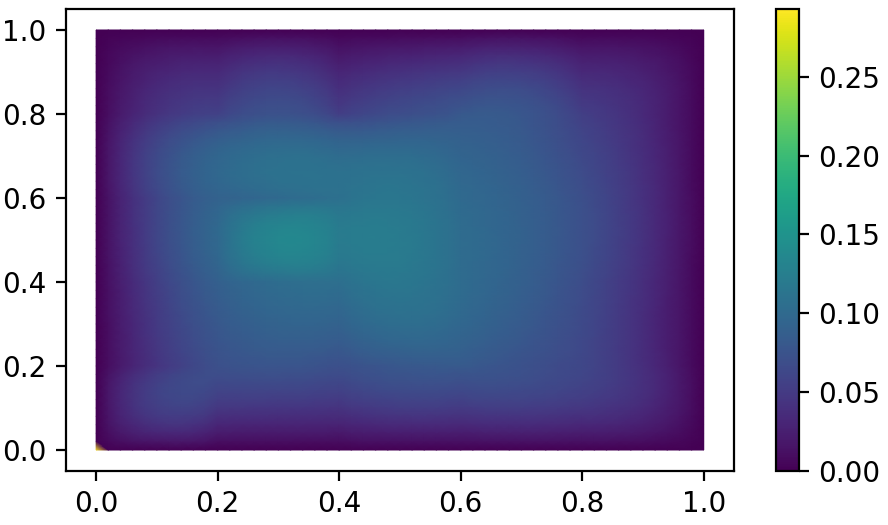}&\hspace{-0.35cm}\includegraphics[height=0.15\textwidth]{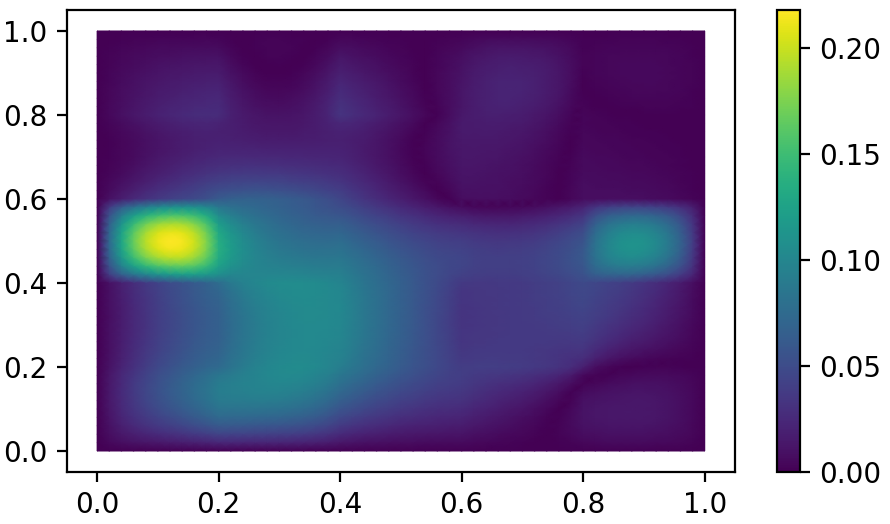}&\hspace{-0.25cm}\includegraphics[height=0.15\textwidth,width=0.2\textwidth]{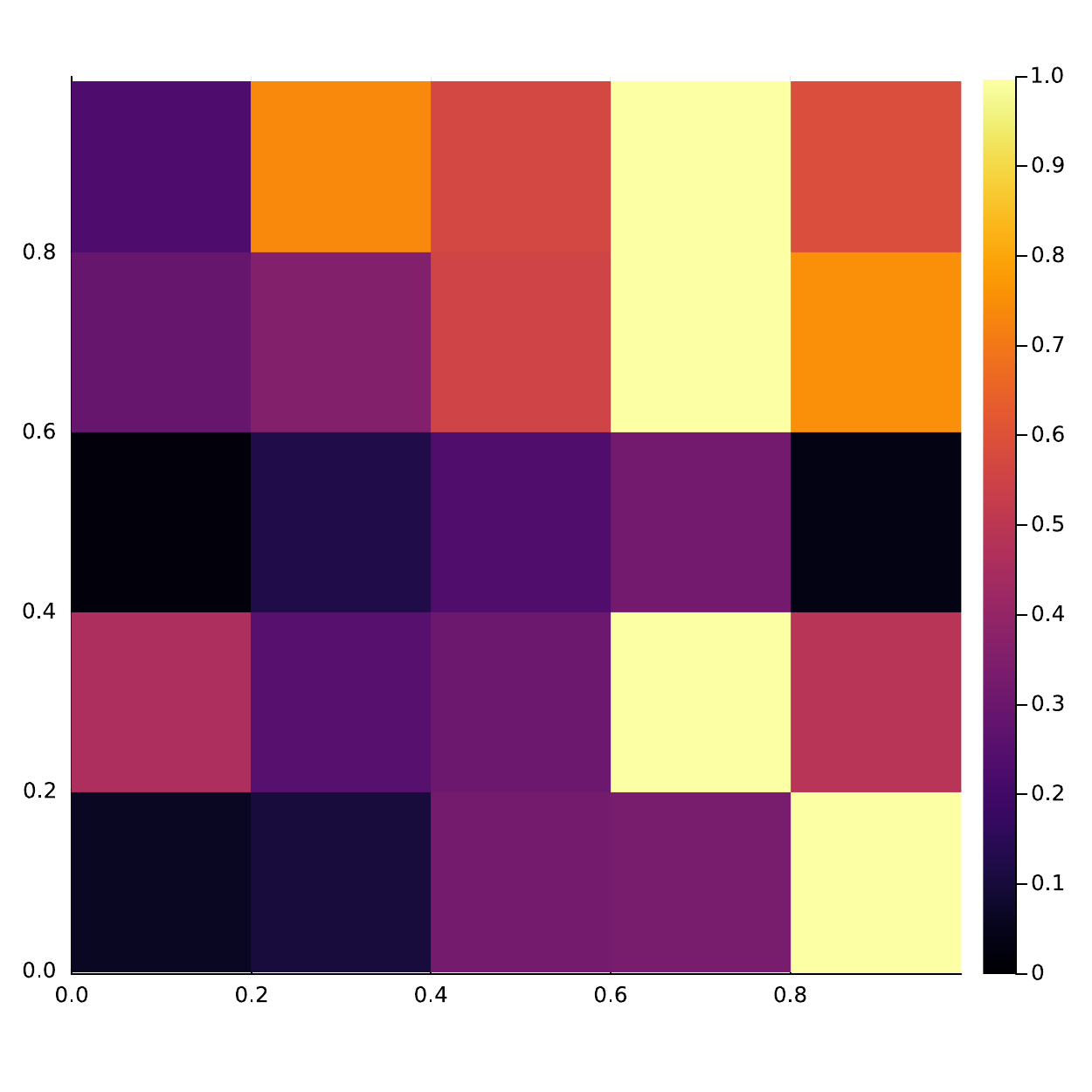}\\
\hspace{-0.5cm}\rotatebox{90}{$\quad\quad k=60$} &\hspace{-0.35cm}\includegraphics[height=0.15\textwidth]{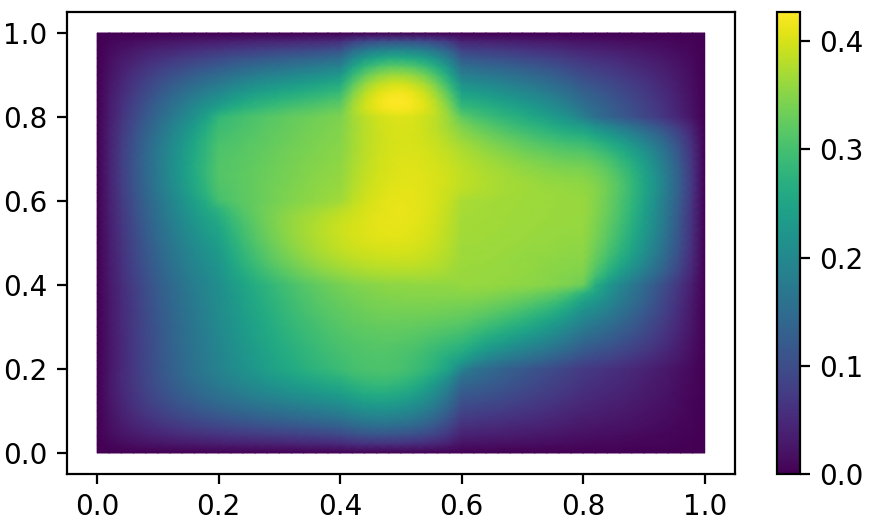}&\hspace{-0.35cm}\includegraphics[height=0.15\textwidth]{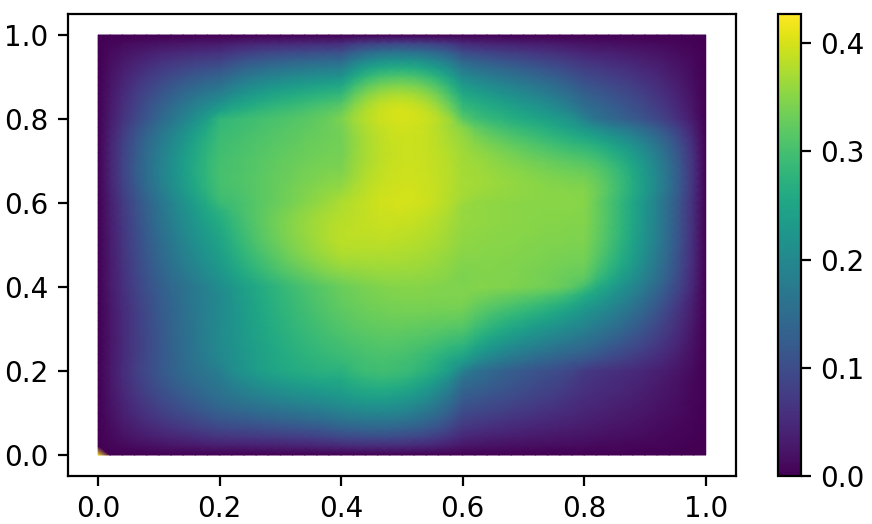}&\hspace{-0.35cm}\includegraphics[height=0.15\textwidth]{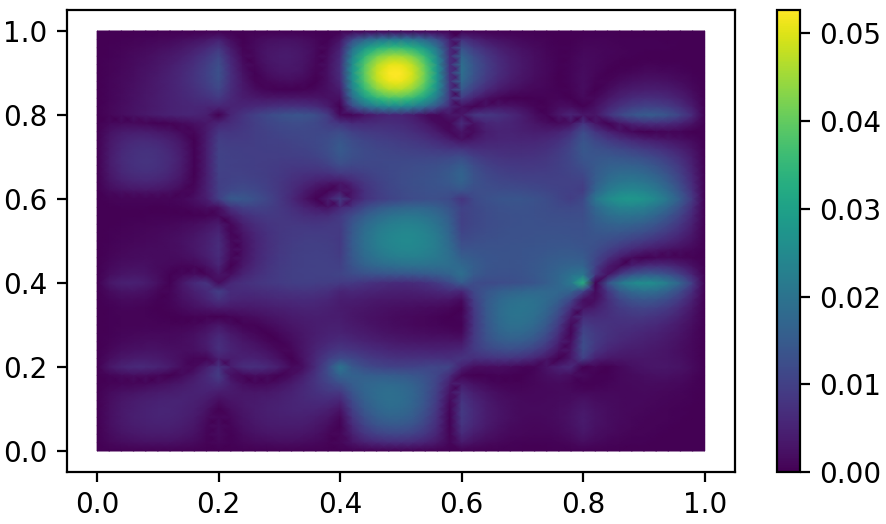}&\hspace{-0.25cm}\includegraphics[height=0.15\textwidth,width=0.2\textwidth]{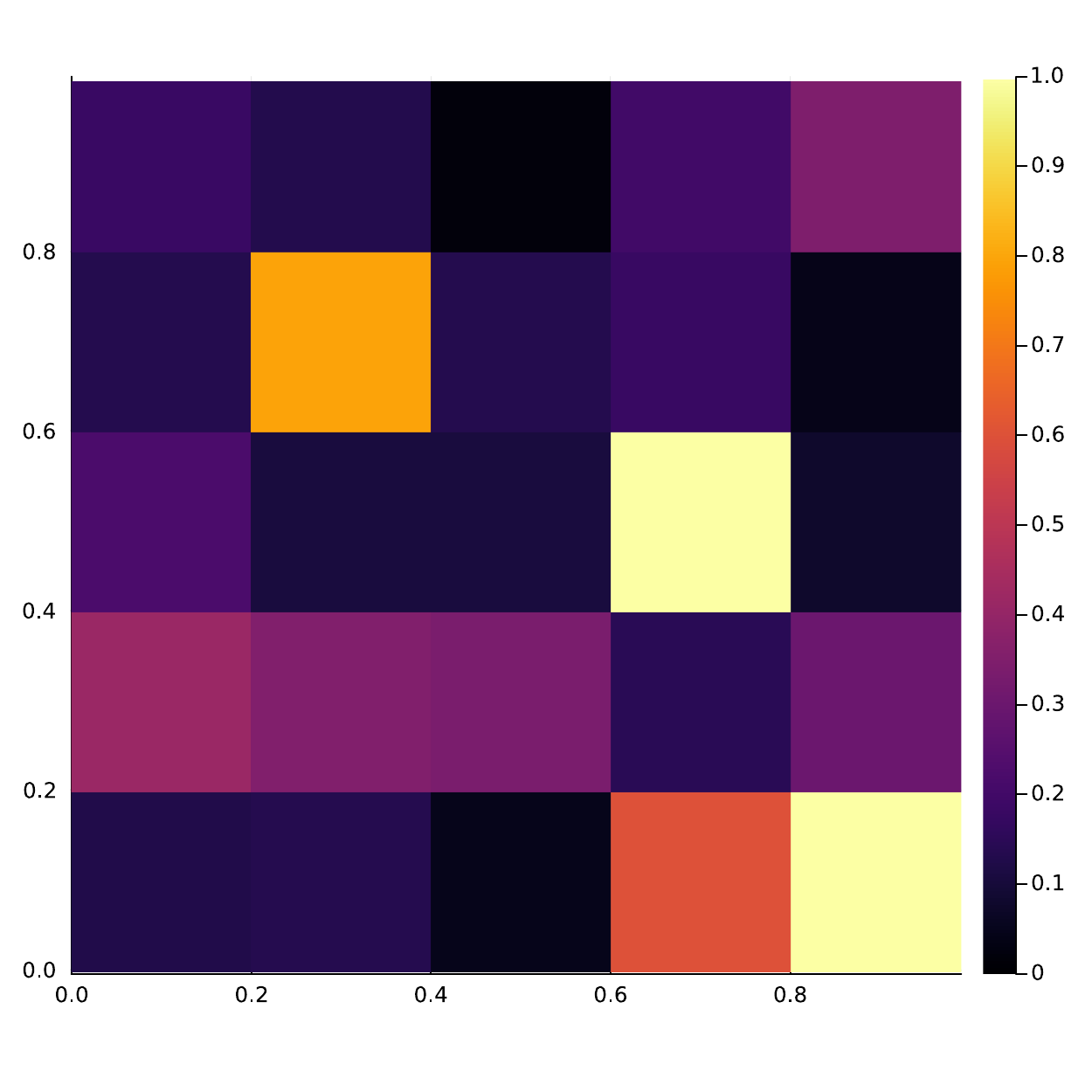}\\
\hspace{-0.5cm}\rotatebox{90}{$\quad\quad k=120$} &\hspace{-0.35cm}\includegraphics[height=0.15\textwidth]{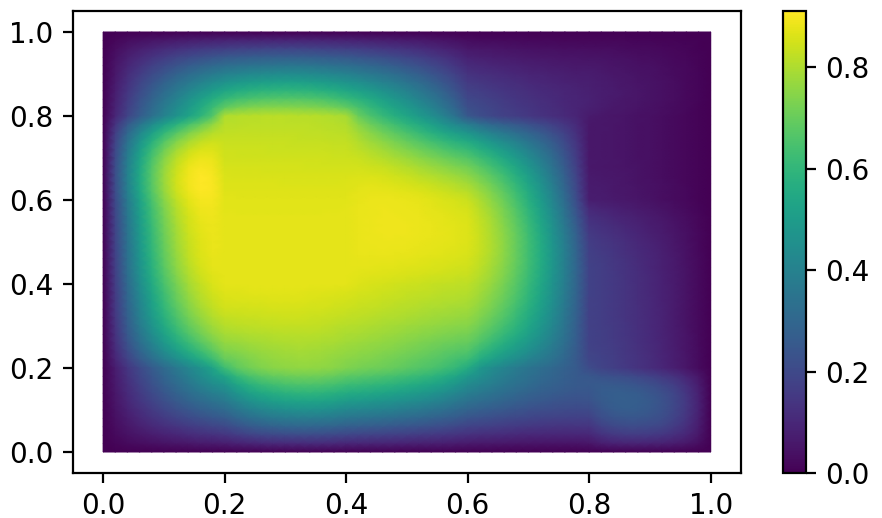}&\hspace{-0.35cm}\includegraphics[height=0.15\textwidth]{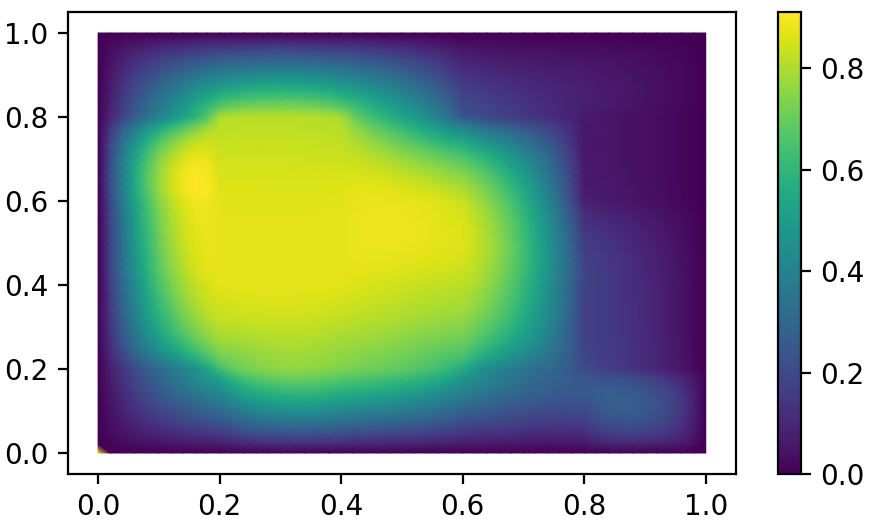}&\hspace{-0.35cm}\includegraphics[height=0.15\textwidth]{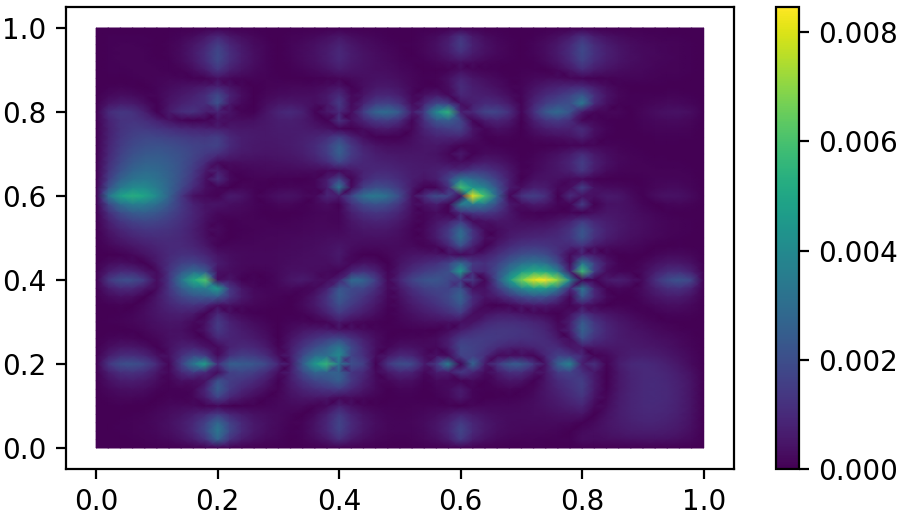}&\hspace{-0.25cm}\includegraphics[height=0.15\textwidth,width=0.2\textwidth]{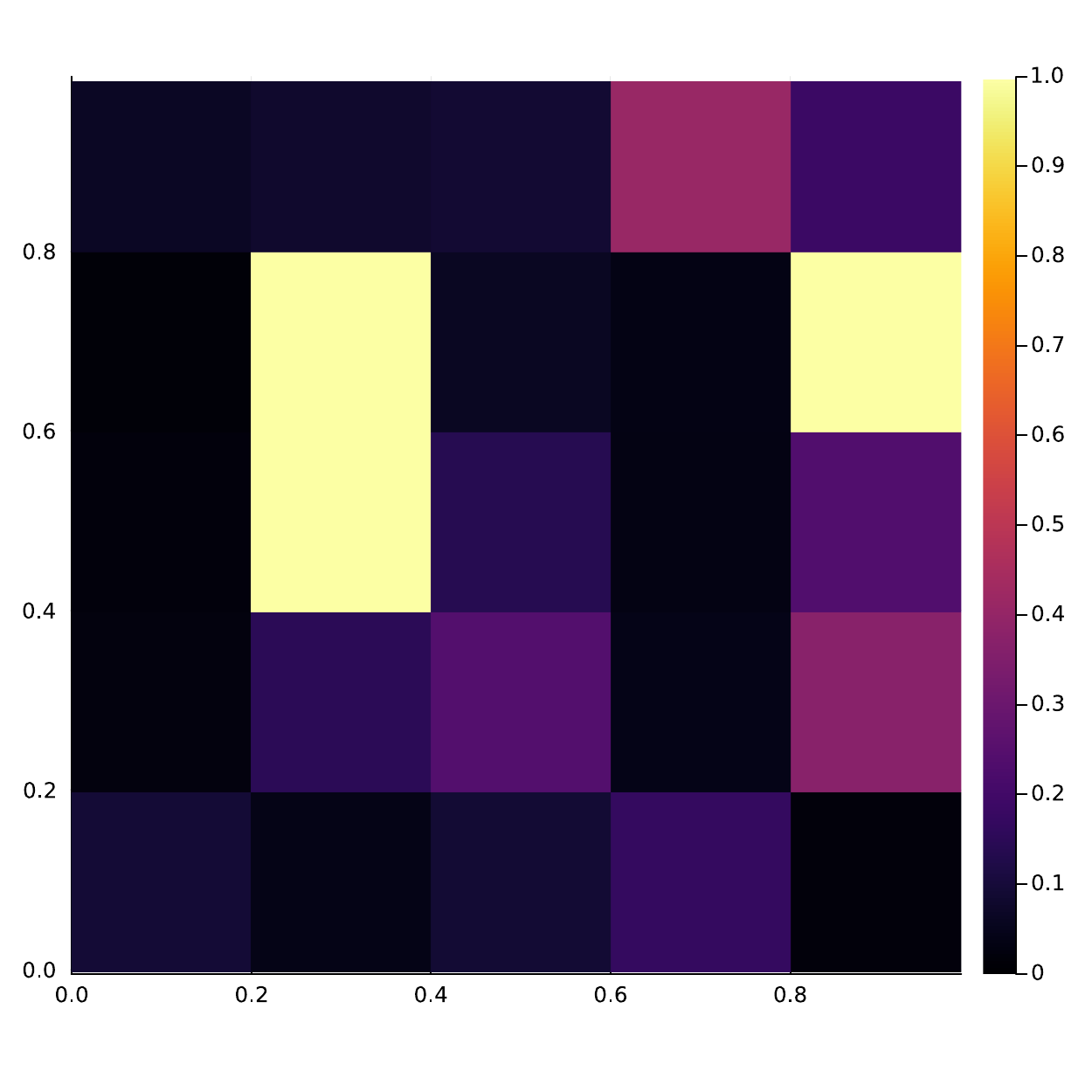}\\
\hspace{-0.5cm}\rotatebox{90}{$\quad\quad k=185$} &\hspace{-0.35cm}\includegraphics[height=0.15\textwidth]{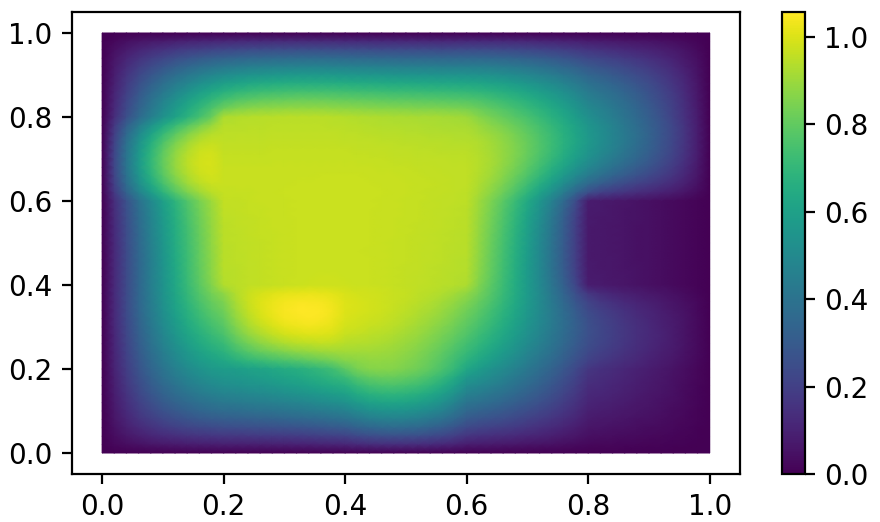}&\hspace{-0.35cm}\includegraphics[height=0.15\textwidth]{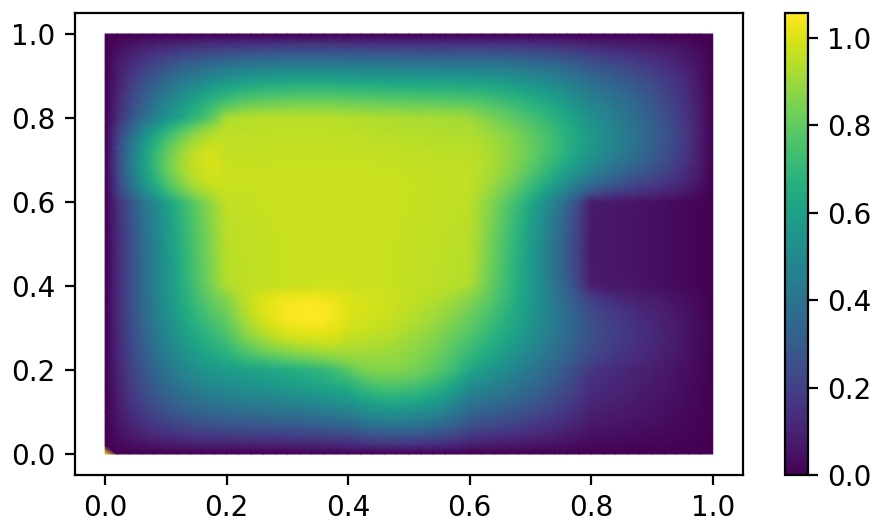}&\hspace{-0.35cm}\includegraphics[height=0.15\textwidth]{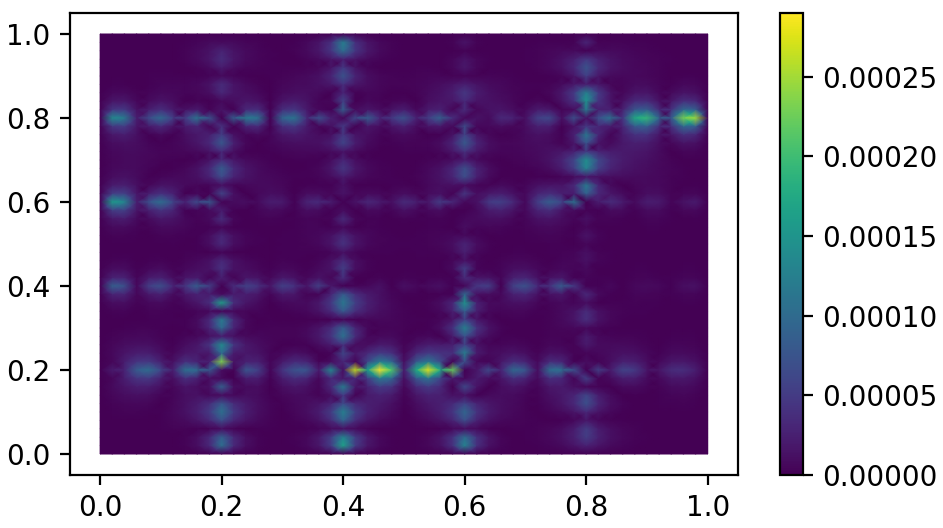}&\hspace{-0.25cm}\includegraphics[height=0.15\textwidth,width=0.2\textwidth]{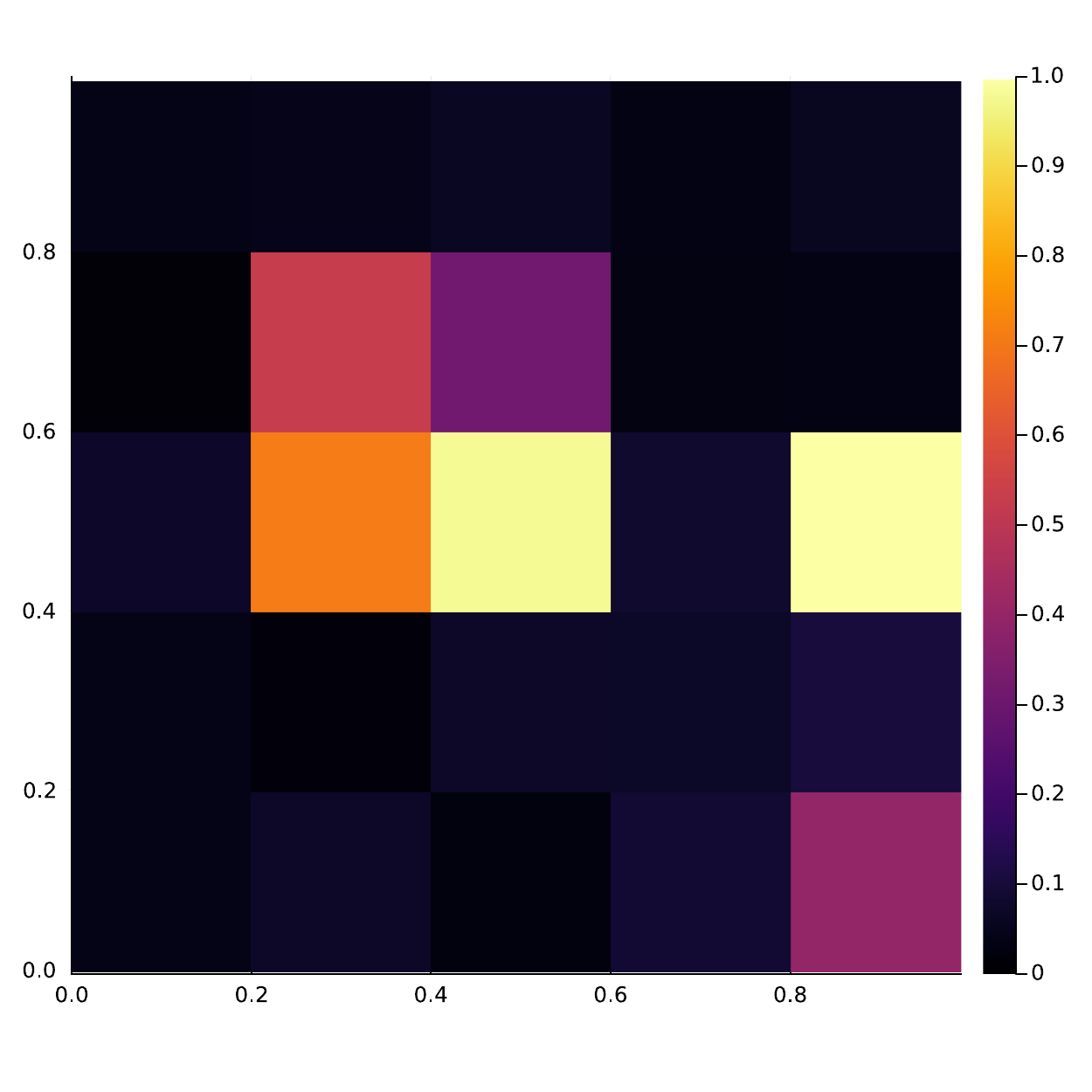}\\
\hline 
\hspace{-0.5cm}&\hspace{-0.5cm}\footnotesize{ $\Psi^\ast(X^{k+1})$}&\hspace{-0.5cm}\footnotesize{  $\Psi^{(k)}(X^{k+1})$}&\hspace{-0.5cm}\footnotesize{ error }&\hspace{-0.5cm}\footnotesize{$X^{k+1}$ }\\
\end{tabular}
	\caption{{\footnotesize {\bf  Illustration of  the algorithm adaptive behavior (example \#2c).}   The figure displays  the high-fidelity solution  of the PDE (left), its reduced basis approximation (middle left) and the related absolute error (middle right), evaluated at the snapshot  $X^{k+1}$ (right)  sampled at iteration $k$  by our ARMS algorithm.
	 {  \label{fig:4tres}}}}
		\end{center}\vspace{-0.cm}
\end{figure}

Figure~\ref{fig:4bis} and \ref{fig:4tres}  show typical evolutions along  the ARMS algorithm iterations for example \#2a and \#2c. We observe that  in the first iterations of the algorithm,  the  RB approximation of the  high-fidelity solution of the PDE is too rough to characterize the rare event of interest. While the number of iterations increase,  RB gain in accuracy, enabling importance sampling of the rare event. 

In the context of example \#2a, a typical rare event sampled by ARMS corresponds to a diffusion coefficient one decade lower than other diffusion coefficients. To provide a physical interpretation of this rare event, note that each of the $4$ subdomains touches the zero border condition, so that no coefficient configuration can isolate a subdomain.  Consequently, increasing the average temperature can only be achieved by slowing down the heat (contributed by the homogeneous source) lost by the system, which corresponds to configurations with low diffusion coefficients. As these configurations lie in the tail of the reference distribution, such events are rare, and of these events, the least rare corresponds to a decrease in one of the $4$ parameters. 

In the context of example \#2c, a typical rare event sampled by ARMS corresponds to setting the value of most diffusion coefficients  touching the zero boundary condition, one decade lower than the value of other diffusion coefficients.  The physical interpretation of this rare event is clear: isolating the domain boundary slows the loss of heat  contributed by the source, and therefore induces an increase in the average temperature within the domain.  Note that such a configuration is extremely rare (on the order of $1e-14$) according to the reference distribution, as 
$11$ coefficients out of $25$ must be sampled in the tail of the reference distribution.  

 
\subsection{ARMS setup \& evaluation criteria}

We now detail the configuration of the ARMS algorithm. The mutation Markov Chain Monte Carlo transition is a Metropolis accept/reject applied on an Ornstein-Uhlenbeck proposal, with an adaptive variance keeping the average acceptance rate over particles in a reasonable range (see \textit{e.g.} ~\cite{garthwaite2016adaptive}). 

 We use a simple scheme to initialize and  update the learning parameter $\tau$: in the early iterations of our algorithm ($k\le K_{j_0}$) we set the learning parameter to $\tau^{(k)}=+\infty$ (which corresponds to the weak greedy algorithm setting in the RB framework); once the sampled snapshots have hit $j_0$ times the  rare event set, we  set this parameter to $\tau^{(k)}=0$ for $k> K_{j_0}$ (which corresponds to sampling the next snapshots according to the current proposals). 

If not explicitly specified in the text, ARMS parameters is configured in accordance with  the table below.

\begin{tabular}{c|cccccccccccc}
&$N$& $M/N$& $K$& $T$& $R$& $j_0$& $c$\\
\hline
example~\#1&  [2.0e2, 8.0e4]& 0.3& 300 & 1e2 & 100 & 5 &5.0e-2  \\
example~\#2a &  [2.5e2, 4.0e2] & 0.1& 300 & 3e1 & 50 & 2 & 3.0 \\
example~\#2b &   [1.0e2, 4.0e3]& 0.3& 200 & 3e1 & 100 & 5 & 5.0e-2  \\
example~\#2c & [1.0e2, 2.0e2]  & 0.1& 300 & 3e1 & 50 &  2 & 3.0 \\
\end{tabular}\medskip

In the table above, $N$ is the number of particles, $M$ is the number of particles killed at each selection step, $K$ is the snapshot budget, $T$ is the number of Markov transitions at each mutation step, $R$ is the number of independent ARMS runs for each problem, $j_0$ is the number of hits before the estimator is triggered and $c$ is the log worst-cost threshold.

 We  plot the median and  quantiles related to the IS estimate $\hat p_{IS}$ and to the AMS estimate  $\hat p_{AMS}$  as a function of the number of snapshots $k$. ARMS performance 
is quantified by evaluating  the expected cost  as a function of  the relative root expected  square error. More specifically, after $k$  snapshots, the    expected relative square error is
 approximated over runs by the empirical average
\begin{equation}\label{eq:expError}
\frac{{\mathbb{E}( \hat p -p^\ast)^2}}{ {p^\ast}^2} \approx \frac{1}{R}\sum_{r=1}^R\frac{( \hat p_{r}-{p^\ast)}^2}{  {p^\ast}^2},
\end{equation}
where $ \hat p_{r}$ denotes, at the $r$-th run and  after $k$ snapshots, either AMS estimate $\hat p_{AMS}$ given in \eqref{eq:AMSpractice}, or the importance sampling estimate $\hat p_{IS}$ given in \eqref{eq:ISpractice2} (with the convention that  $\hat p_{IS}=0$  if   $H= 0$).
After $k$ snapshots, the expected cost  (in units of the cost of evaluating $\Score$) is approximated over the runs by the empirical average 
\begin{equation}\label{eq:expCost}
\textrm{cost(k)} \simeq k+ \frac{1}{R}\sum_{r=1,k'=1}^{R,k}N^{(k')}_r \textrm{gain}^{(k')}
\end{equation}
  with $ \textrm{gain}^{(k')}$ representing the ratio of the cost of evaluating the reduced score $\Srom^{(k')}$ to the cost of the true score $\Score$, and where $N^{(k')}_r$ represents the number of evaluations of the reduced score at the $k'$-th iteration and $r$-th execution of the ARMS algorithm.
  
\begin{figure}[t!]
\begin{center}
\begin{tabular}{c}
\includegraphics[width=0.5\textwidth]{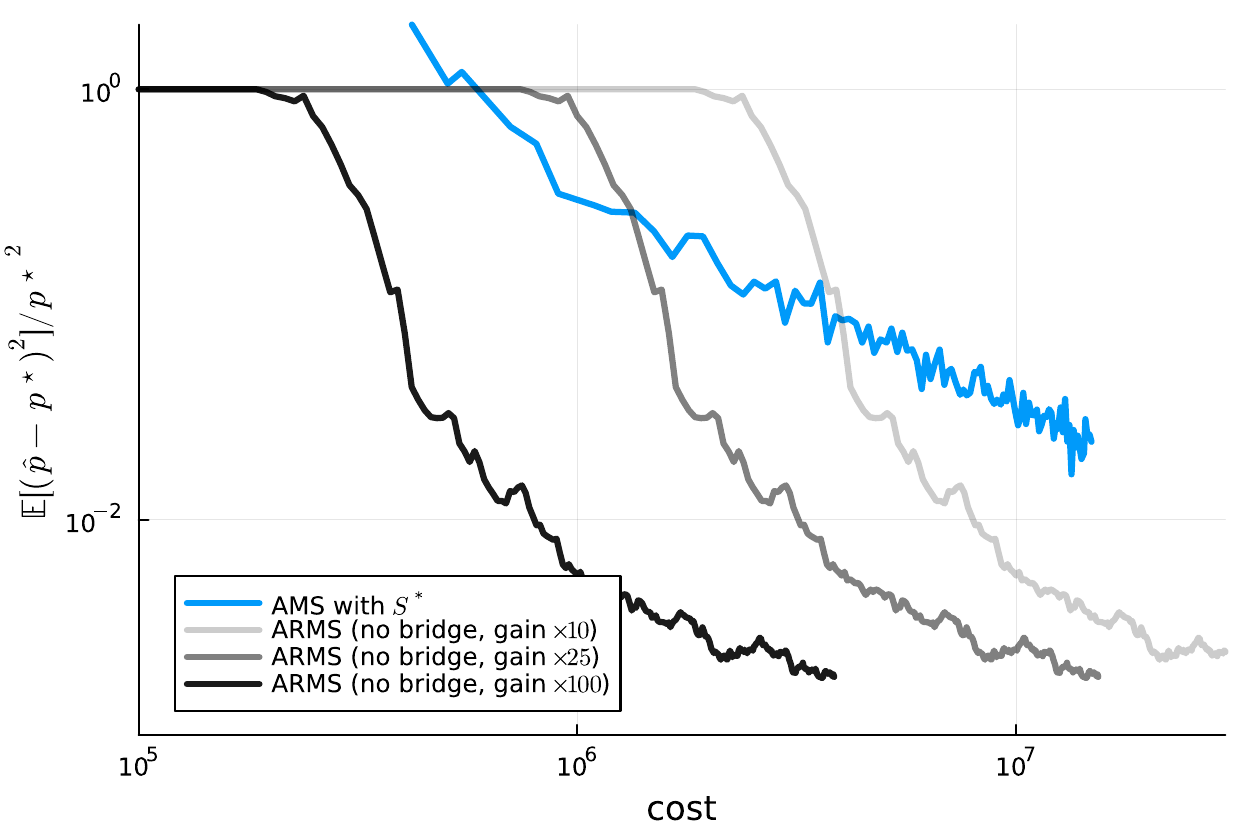}\vspace{-0.35cm}
\end{tabular}
	\caption{{\footnotesize  {{\bf Influence of  the reduced score gain (example \#1).} Expected cost \eqref{eq:expCost} as a function of the relative expected squared error \eqref{eq:expError} for the IS estimate \eqref{eq:ISpractice2}, for different values of the ratio between the cost of evaluating $\Srom^{(k)}$ and $\Score$ (grey curves) and for $N=1e3$  and the threshold $\Cte=28 \times c_{\min}$. Comparison to the performance of standard AMS with $\Score$ and $N\in [1e2, 1e4]$ (blue curve).  \label{fig:3}}}}
		\end{center}\vspace{-0.cm}
\end{figure}

  Please note that for the first example, $ \textrm{gain}^{(k')}$ will be set artificially  to emulate  realistic cases in which the cost of evaluating the surrogate is indeed cheaper than that of the true model (see Figure~\ref{fig:3}). 

We will compare the performance of ARMS with the empirical average of the estimates obtained by performing $R$ standard AMS with the true score $\Score$.
\begin{figure}[t!]
\begin{center}
\begin{tabular}{cc}
\hspace{-1.cm}{\footnotesize $\Cte=42\times c_{\min}$}&\hspace{-0.5cm}{\footnotesize $\Cte= c_{\min}$}\\
\hspace{-1.cm}\includegraphics[width=0.5\textwidth]{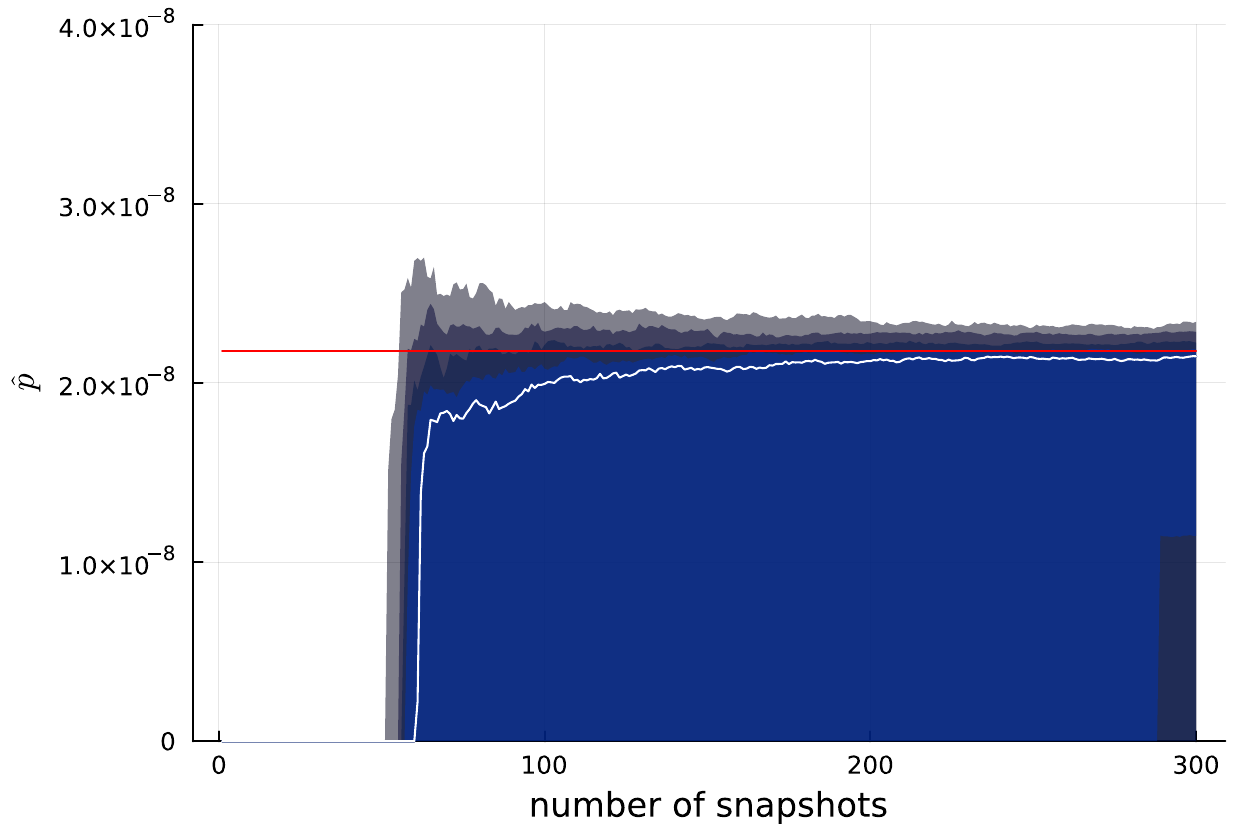}&\hspace{-0.5cm}\includegraphics[width=0.5\textwidth]{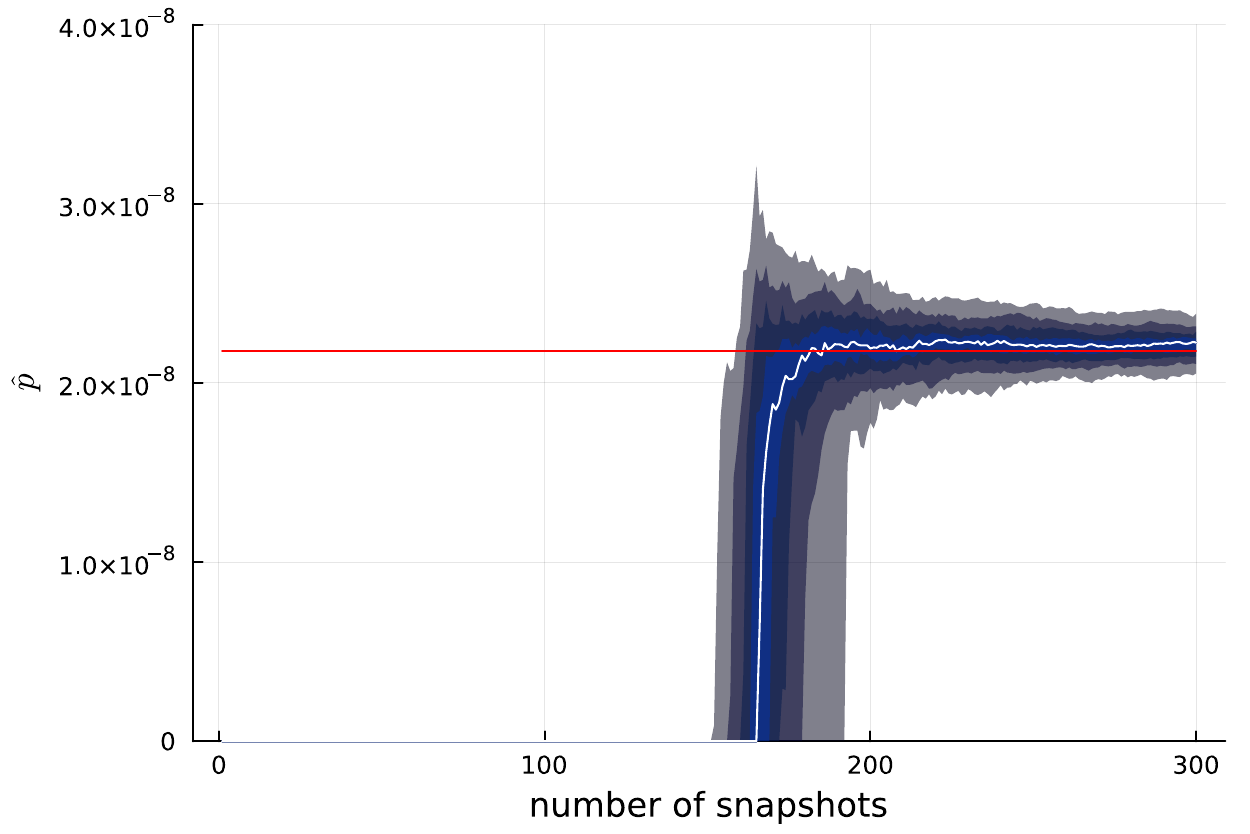}\\
\hspace{-1.cm}{\footnotesize $\Cte=28\times c_{\min}$}&\\
\hspace{-1.cm}\includegraphics[width=0.5\textwidth]{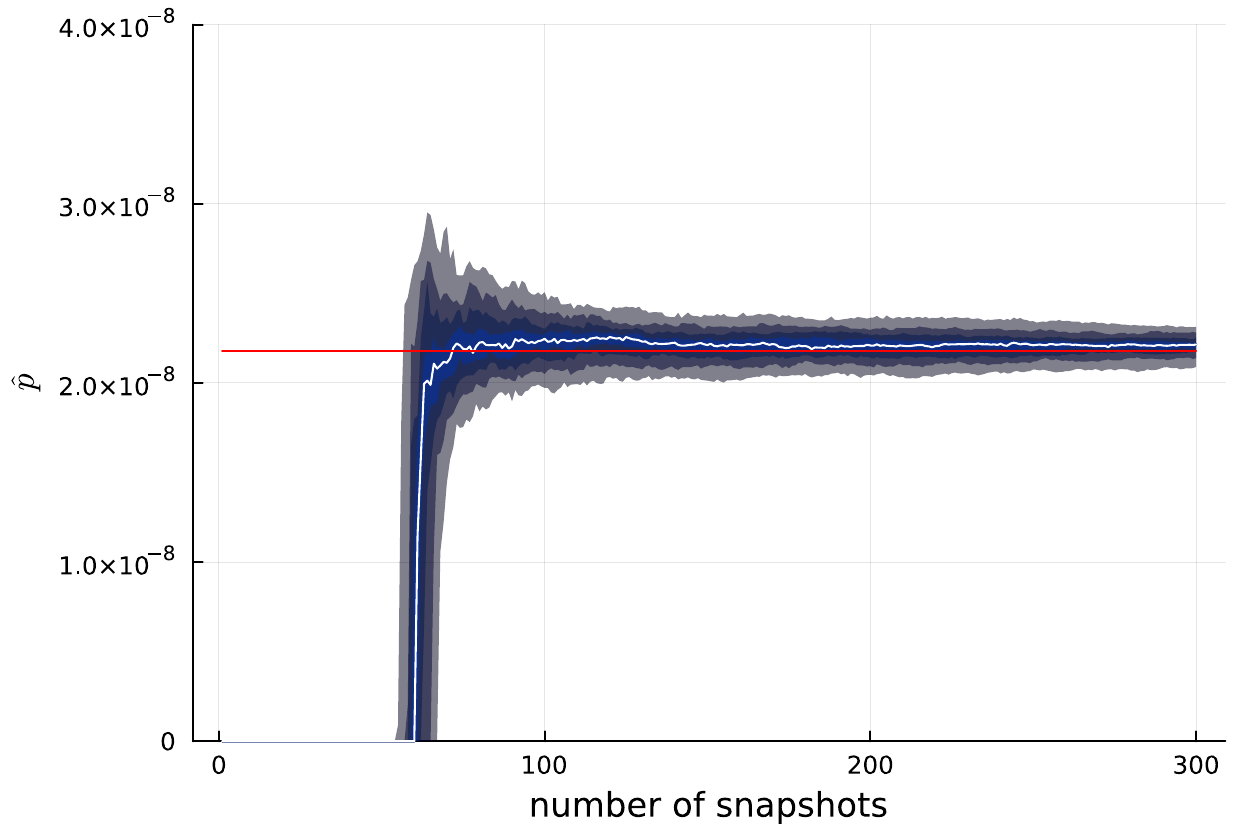}&\hspace{0.5cm}\includegraphics[width=0.5\textwidth]{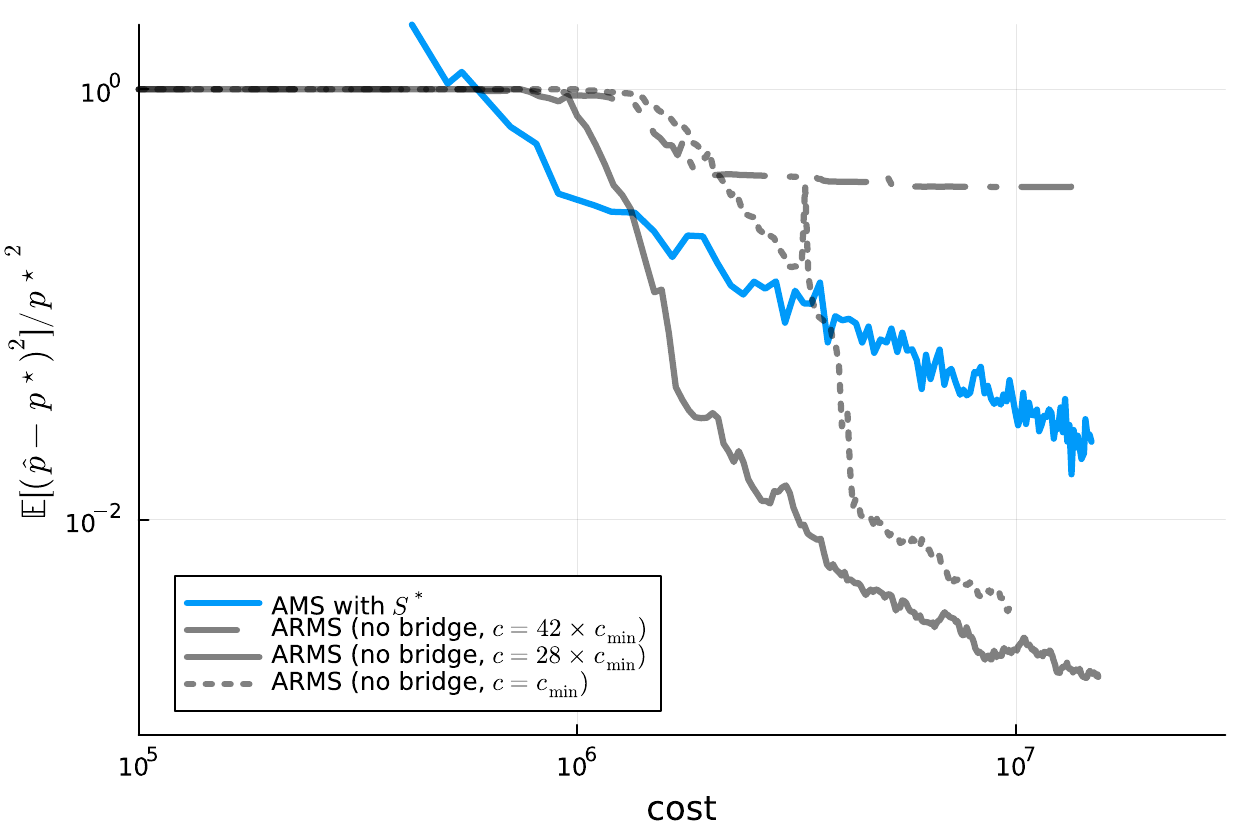}\vspace{-0.35cm}
\end{tabular}
	\caption{{\footnotesize  { {\bf Influence of the entropic criterion threshold $\Cte$ (example \#1).} Upper and lower left plots: median (in white) and  quantiles (in shades of blue-grey) related to the IS estimate \eqref{eq:ISpractice2} as a function of the number of snapshots (equal to the number of iteration $k$ of the algorithm) for $N=1e3$, a reduced model gain of  $\times 25$, and $3$ different values of the worst log cost threshold $\Cte$.  The red line denotes the true rare event probability. Lower right plot:    expected  relative square error \eqref{eq:expError}   as a function of the  expected cost \eqref{eq:expCost} (grey curves) for the IS estimate  \eqref{eq:ISpractice2}. Comparison to the performance of standard AMS with $\Score$ and $N\in [1e2, 1e4]$ (blue curve). \label{fig:1}}}}
		\end{center}\vspace{-0.cm}
\end{figure}

\begin{figure}[t!]
\begin{center}
\begin{tabular}{c}
\includegraphics[width=0.5\textwidth]{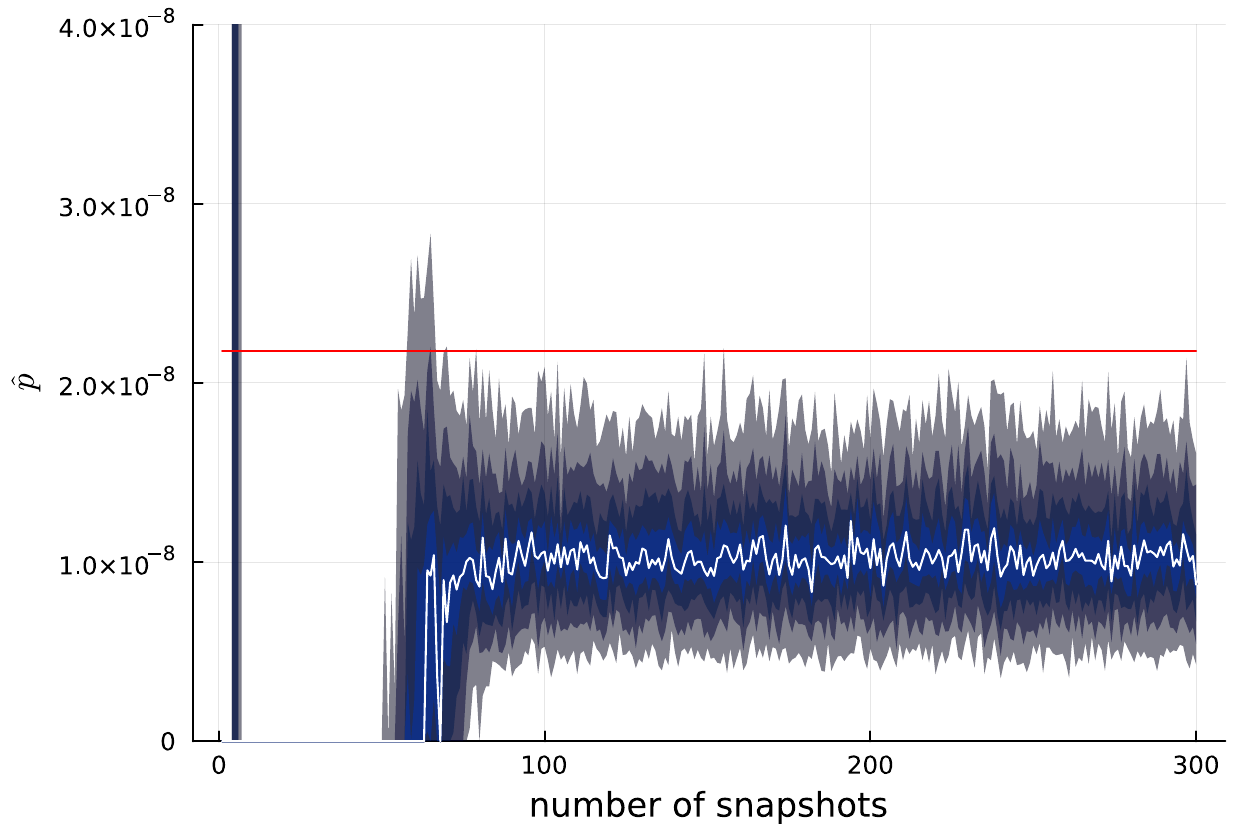}
\end{tabular}
	\caption{{\footnotesize  { {\bf Bias of the reduced score based estimate  (example \#1).}  Median (in white) and  quantiles (in shades of blue-grey) related to the AMS estimate \eqref{eq:AMSpractice} as a function of the number of snapshots; the red line denotes the true rare event probability.  Number of samples $N=1e3$, entropic criterion set to $\Cte=28 \times e_{\mrm{min}}$,   no bridging, never  stopping  the reduced score update ($\epsilon<0$) and gain between surrogate and exact is assumed to be $25$. \label{fig:2cinq}}}}
		\end{center}\vspace{-0.cm}
\end{figure}

\begin{figure}[t!]
\begin{center}
\begin{tabular}{c|c}
\hspace{-0.5cm}{\footnotesize no bridge}&\hspace{-0.5cm}{\footnotesize bridge}\\
\hline \\
\hspace{-0.5cm}\includegraphics[width=0.5\textwidth]{./Images/Restart1xEminN1e3_plotDistributionIS}&\hspace{-0.cm}\includegraphics[width=0.5\textwidth]{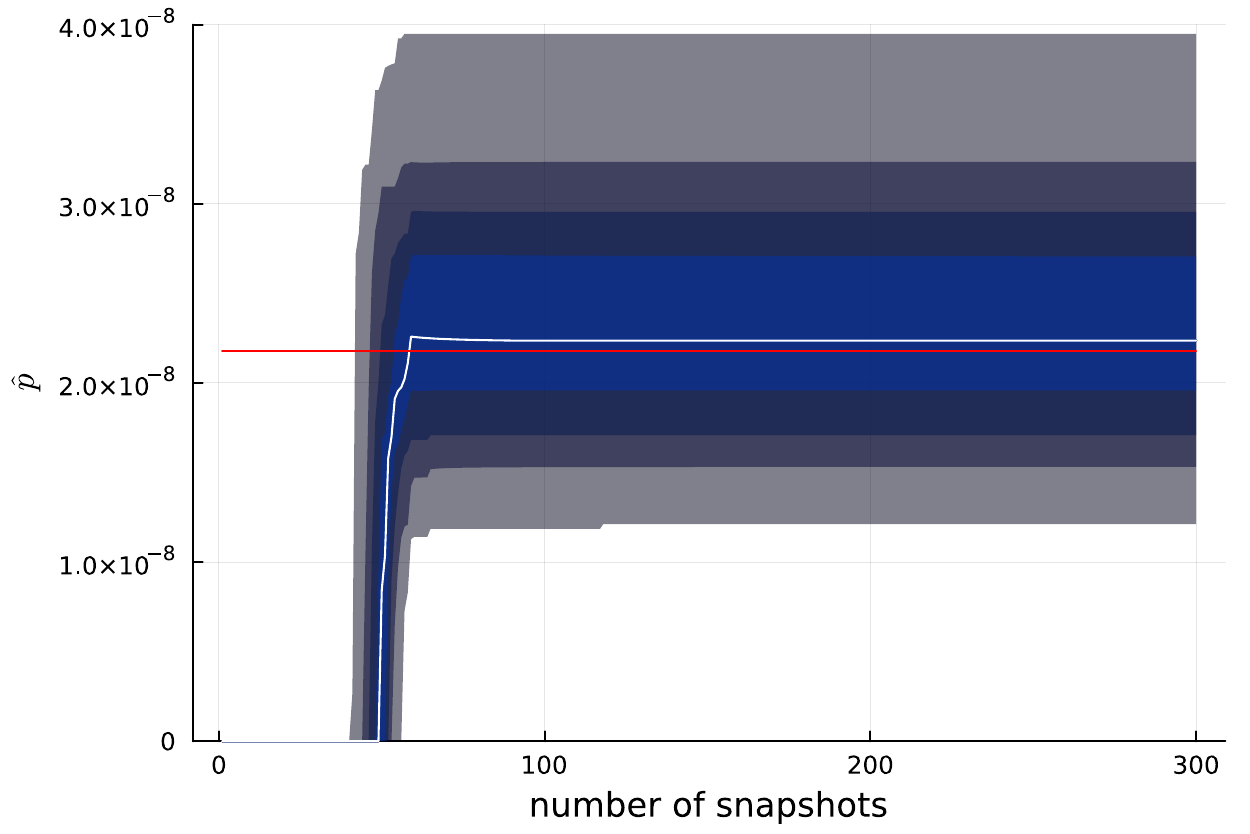}\\
\hspace{-0cm}\includegraphics[width=0.475\textwidth]{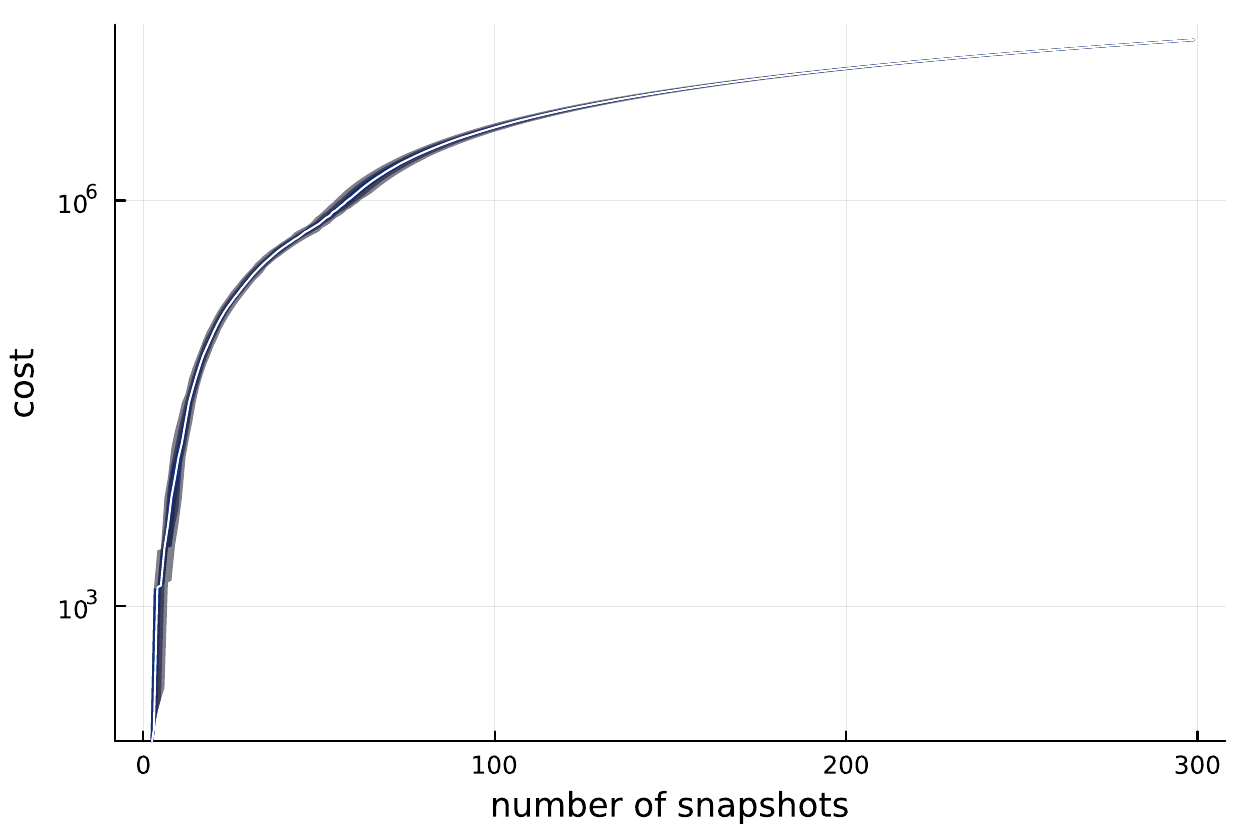}&\hspace{0.25cm}\includegraphics[width=0.475\textwidth]{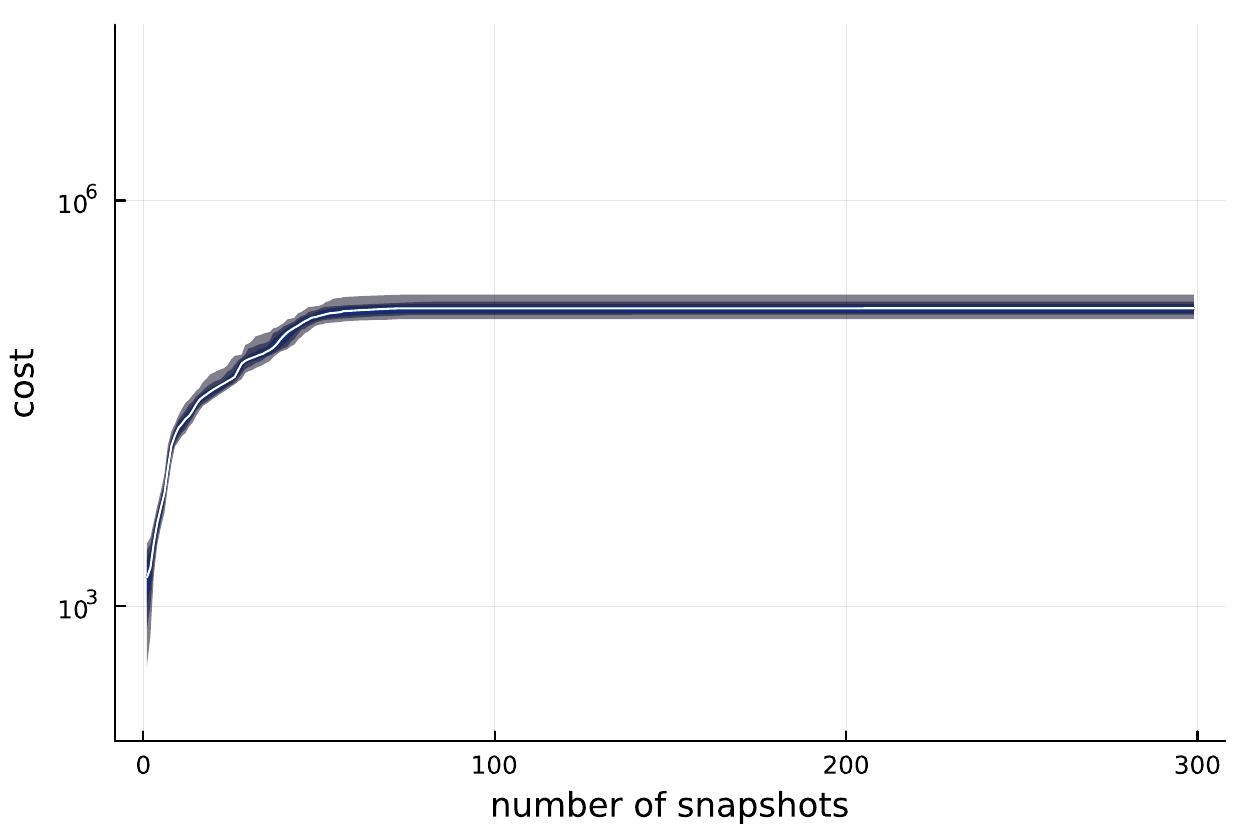}
\end{tabular}
	\caption{{\footnotesize  { {\bf Influence of  bridging (example \#1).}  
	 Median (in white) and  quantiles (in shades of blue-grey) related to the IS estimate \eqref{eq:ISpractice2} (above) and of the cost \eqref{eq:expCost} (above), as a function of the number of snapshots when bridging simulation or not; the red line denotes the true rare event probability.  Number of samples $N=1e3$, entropic criterion set to $\Cte=28 \times e_{\mrm{min}}$,   stopping updates of the reduced score after convergence only in the case of bridging and gain between surrogate and exact is assumed to be $25$. Note that, although imperceptible in the log-log plot at bottom right, the cost increases strictly with the number of snapshots. \label{fig:2tres}}}}
		\end{center}\vspace{-0.cm}
\end{figure}


\begin{figure}[h!]
\begin{center}
\begin{tabular}{cc}
\hspace{-0cm}{\footnotesize Example \#1 }&\hspace{-0.5cm}{\footnotesize Example \#2b ($L^\infty$)}\\
\hspace{-0cm}\includegraphics[width=0.5\textwidth]{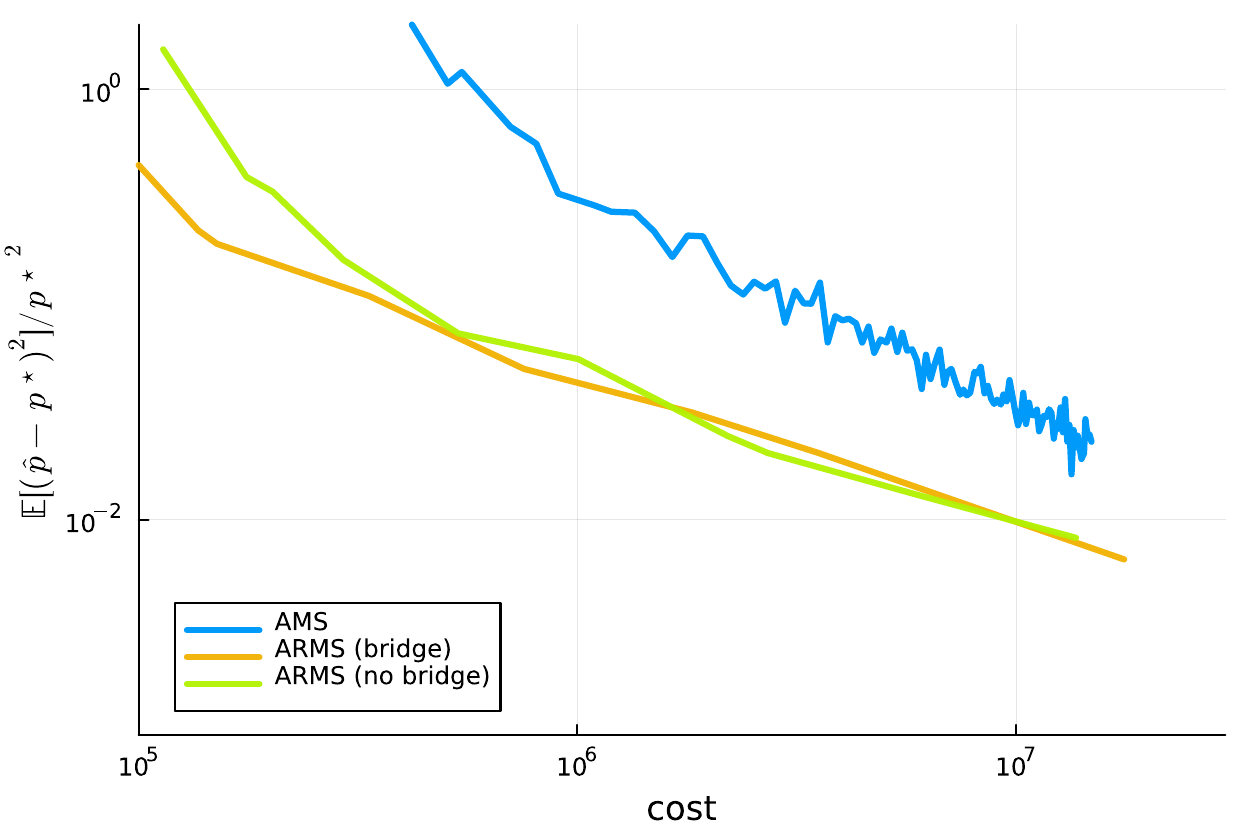}&\hspace{-0.5cm}\includegraphics[width=0.5\textwidth]{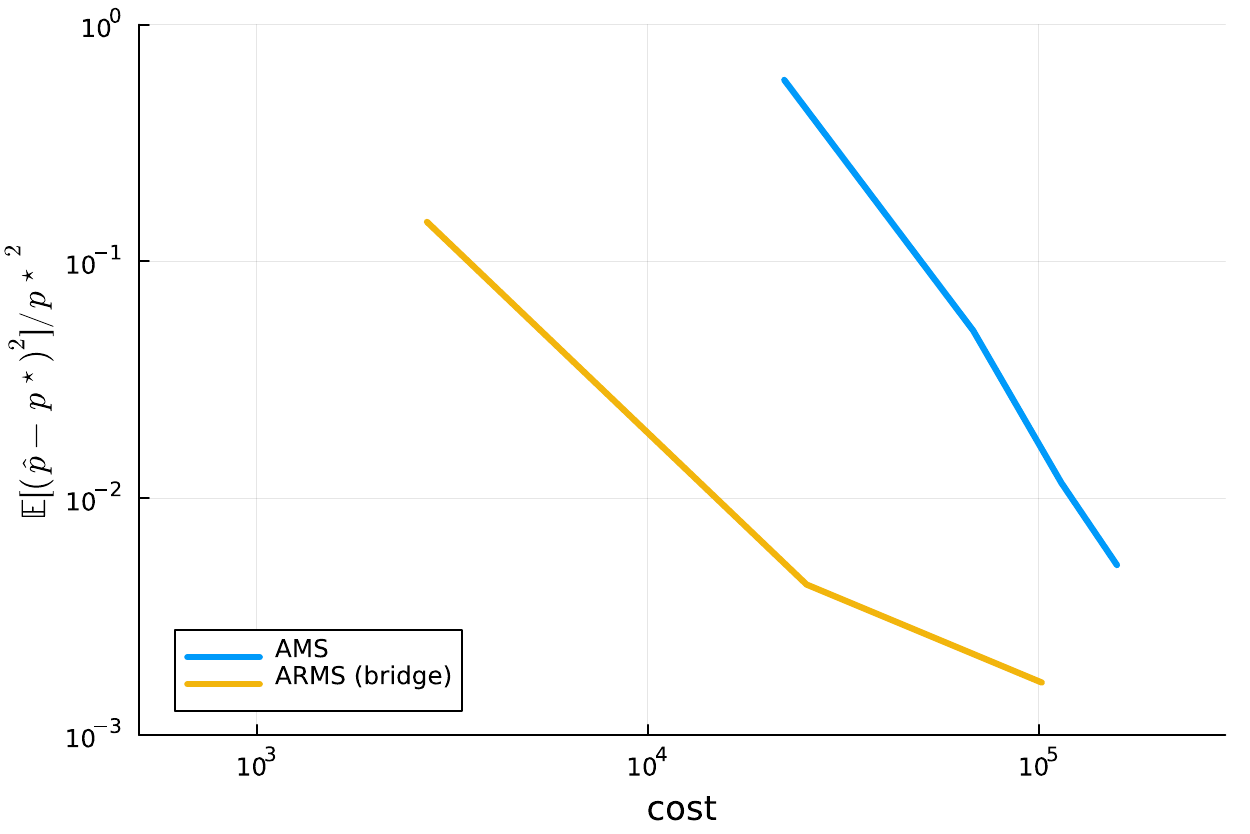}\\
\hspace{-0cm}{\footnotesize Example \#2a ($L^1$) }&\hspace{-0.5cm}{\footnotesize Example \#2c ($L^1$)}\\
\hspace{-0cm}\hspace{-0.5cm}\includegraphics[width=0.5\textwidth]{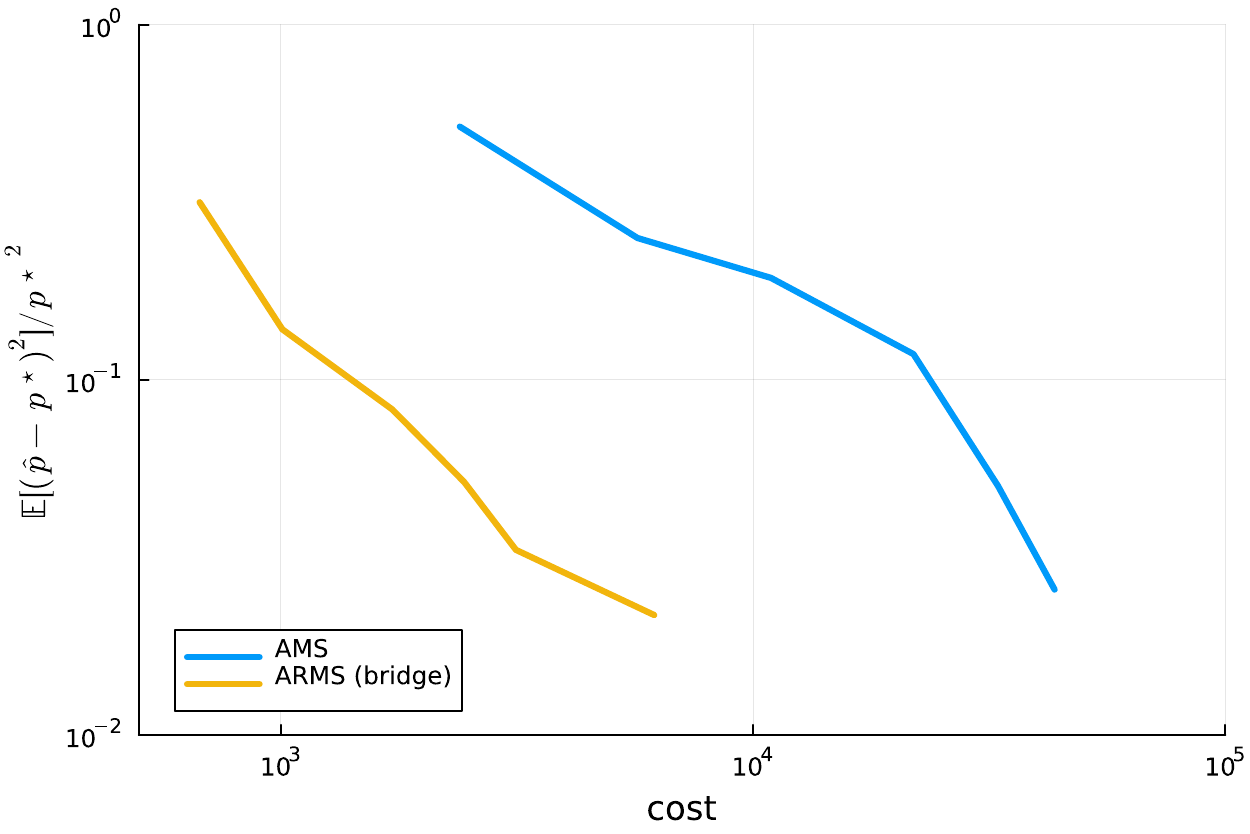}&\hspace{-0.5cm}\includegraphics[width=0.5\textwidth]{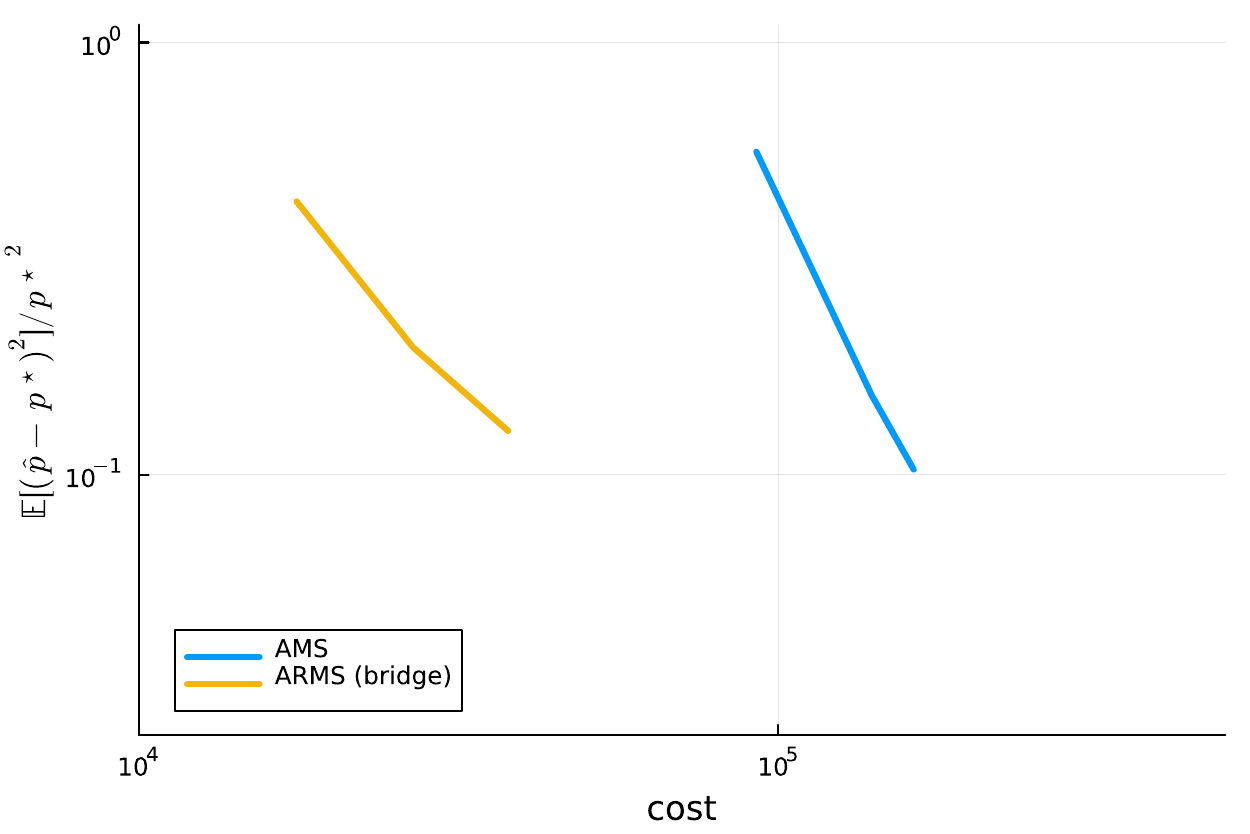}\vspace{-0.35cm}
\end{tabular}
\	\caption{{\footnotesize  { {\bf Performance of ARMS (example \#1 \&2).}   Relative  expected  square error \eqref{eq:expError}  as a function of the expected cost \eqref{eq:expCost} of the IS estimate  \eqref{eq:ISpractice2} at the final iteration (once the snapshots budget $K$ is exhausted),   for a varying sample size $N$. 
Curves can be  compared to standard AMS estimation using the score $\Score$, for a varying sample size $N$.
\label{fig:2}}}}
		\end{center}\vspace{-0.cm}
\end{figure}

\begin{figure}[t!]
\begin{center}
\begin{tabular}{c}
\includegraphics[width=0.5\textwidth]{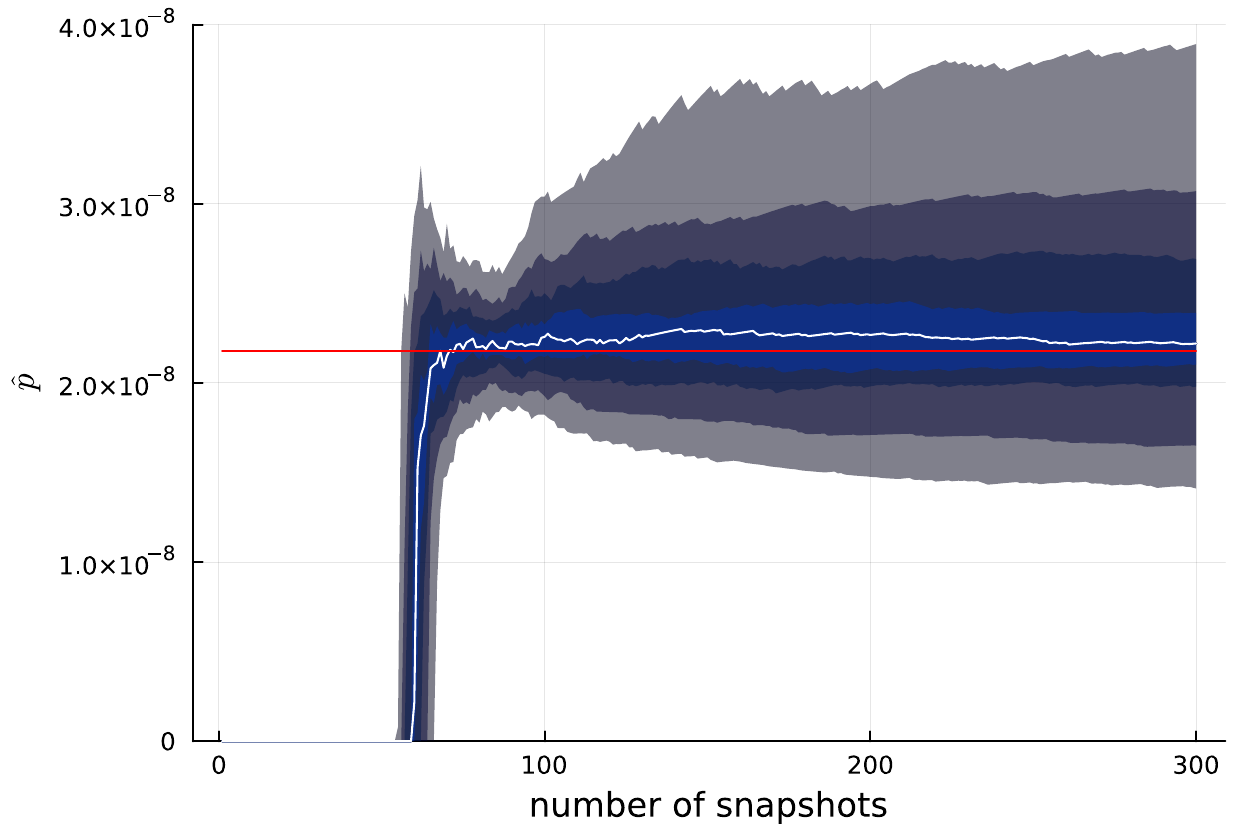}
\end{tabular}
	\caption{{\footnotesize  { {\bf A bad idea: using the stopping criterion  without bridging  (example \#1).}  Median (in white) and  quantiles (in shades of blue-grey) related to the IS estimate \eqref{eq:ISpractice2} as a function of the number of snapshots when stopping  the reduced score updates ($\epsilon=0$) without bridging; the red line denotes the true rare event probability.  Number of samples $N=1e3$, entropic criterion set to $\Cte=28 \times c_{\mrm{min}}$ and gain between surrogate and exact is assumed to be $25$. \label{fig:2quatre}}}}
		\end{center}\vspace{-0.cm}
\end{figure}

\subsection{Results}

We begin by analyzing the results obtained with the toy model of example \#1.  
Figure~\ref{fig:1} evaluates the influence of the  threshold $\Cte$ on the entropic criterion of Section~\ref{sec:prop_temp}. The three quantiles show the distribution of the IS estimate as a function of the number of snapshots for different values of the entropy criterion threshold $\Cte$. As expected, we note (upper left) that a too large threshold produces over 40\% of erroneous IS estimates, associated with zero probability regardless of the number of reduced score updates.   Although the remaining 60\% of estimates nevertheless become accurate after around 60 updates of the reduced score, this demonstrates that increasing the critical level too quickly tends to concentrate particles around unimportant local minima. This results in sampling snapshots and refining the reduced score exclusively in regions of state space unrelated to the rare event of interest.  To avoid this undesirable scenario, it is possible to set the threshold to its minimum value $c_{\min}$ (upper right), which has the disadvantage of delaying the convergence of the IS estimate. Indeed, the plot shows convergence of all estimates towards a non-zero value close to $p^\ast$, but with a considerable delay, since in this case about a hundred more updates of the reduced score are required. A well-tuned value between these extrema (lower left) achieves convergence after about 70 snapshots, with a characterization of the distribution of IS estimates comparable to the safe distribution obtained with the minimum value. 

This trend is confirmed by plotting (lower right) the relative root expected  square error \eqref{eq:expError} as  a function of the expected cost \eqref{eq:expCost}.  
In the case of a not-too-high threshold $c_{\min}$ or $28c_{\min}$, given a certain precision (\ie  after sampling enough snapshots), we observe a gain in expected cost from around a decade onwards, which increases more and more. Clearly, a value of $42c_{\min}$ is too high and do not achieve comparable accuracy, regardless of the number of snapshots. \medskip

Figure~\ref{fig:3} illustrates how the gain in terms of expected cost over accuracy can be dramatic, as the cost of evaluating $\Srom^{(k)}$ becomes negligible compared to $\Score$.     \medskip 

Figure~\ref{fig:2cinq} shows that the AMS estimate \eqref{eq:AMSpractice} significantly underestimates the true probability, demonstrating that relying solely on reduced scores $\Srom^{(k)}$ (here spline approximations) of the true score $\Score$ induces a significant bias.  Indeed, in our specific example, the non-differentiability of $\Score$ at the rare event boundary leads to high-frequency oscillations (which tend to decrease with the number of interpolation points, but never disappear completely) in the spline model. The consequence is that a significant proportion of particles in the rare event set $\{x\in \mathcal{X} :\Score(x) \ge l_{\max}\}$ are to be excluded from the set $\{x\in \mathcal{X} :\Srom^{(k)}(x) \ge l_{\max}\}$, leading to a negative bias.    On the contrary, we  observe that the IS estimate \eqref{eq:ISpractice2} is not biased, independently of the problem of convergence of $\Srom^{(k)}$ to $\Score$. \medskip

The evaluation above focused on the ARMS algorithm in which the simulation is restarted after each reduced score update and in which  the stop criterion introduced in Section~\ref{sec:stop} is never triggered ($\epsilon <0$). Figure~\ref{fig:2tres}  extends the analysis by evaluating the influence of bridging the simulations between two iterations, as presented in Section~\ref{sec:bridge}. The entropic criterion is set to $\Cte=28 \times c_{\mrm{min}}$. 
 As shown in figure~\ref{fig:2tres}, the gain in computational cost brought about by bridging is offset by an increase in variance.  One can remark that, when using the bridging procedure, and for a given sample size, we  cannot expect to increase  the accuracy for the IS estimate  \eqref{eq:ISpractice2} indefinitely, contrary to when  the simulation is restarted  after each reduced score update. Indeed, the estimate depends crucially  on the variance of the normalization constant ${Z_{l^{(k)}}^{N,(k)}}$ which generates a residual error (upper right box of Figure~\ref{fig:2tres}).  In the case of bridging, this residual error will indeed propagate across the iterations according to \eqref{eq:inclusionZRefined} instead of being re-estimated by restarting a fresh simulation at each reduced score update. Nevertheless, the plot of expected squared error versus predicted cost in the upper left-hand panel of figure~\ref{fig:2} shows that, in the case of small sample sizes yielding a relative square error above  20\%,  some saving of order of 2 is achieved by a bridging procedure compared to restarting the simulation, while both procedures are equivalent for larger sample sizes.


On the other hand, we notice that enabling the stopping criterion without bridging, is a bad idea as it induces a spread around the mean of the distribution of the IS estimate as illustrated in  Figure~\ref{fig:2quatre}. Indeed, similarly to the above explanation,  the IS estimate will be affected by the apparent bias of  the estimates ${Z_{l^{(k)}}^{N,(k)}}$ in \eqref{eq:ISpractice2}, all identical for  indices $k$ above some random time (determined by the stopping criterion). The phenomenon does not appear when the stopping criterion is off (Figure~\ref{fig:2tres}) and we restart the simulation as before and calculate a new normalization constant at each iteration. The stopping criterion is, in fact, simply used to marginally reduce the cost in the bridging procedure, without changing the outcome (not shown in plot). \medskip

We then move on to example \#2, the PDE-based scores and their RB approximations. 
Overall, the trends of the algorithm for the RB approximation of a PDE score are very similar to those of the previous example. We thus omit the precise evaluation made for the previous example, and jump directly to the analysis of performances in Figure~\ref{fig:2}. We  use   an initial reduced score built on a $5$-dimensional RB approximation. In the case of examples \#2a and \#2b, we observe that about $10$ to $20$ more snapshots (\ie ARMS iterations) are needed to reach convergence, with a median estimate around $p^\ast$. In the case of example \#2c, about $170$ to $200$ additional snapshots are needed to reach a good accuracy of the reduced model and converge towards $p^\ast$.  Indeed, in the latter situation, the parameter space has a much higher dimensionality (25 dimensions versus 4), and learning the reduced model for parameters in the (initially unknown) region where rare events occur is a difficult task performed progressively over the course of ARMS iterations. 

In all examples, Figure~\ref{fig:2} shows that the proposed ARMS method enables important computational savings as compared to the use of the true model alone. Saving are consistently around a factor $10$ in all cases.
As mentioned in section~\ref{sec:PDERB}, in the case of the $\#2b$ example, ARMS uses error estimates that are not hard bounds for the $L^\infty$ score. Nevertheless, it is surprising that ARMS manages to accurately estimate the desired probability and achieve performances comparable to those obtained with the $L^1$ score.

\section{Conclusions and Perspectives}

This work proposes an adaptive algorithm called ARMS that reduces the cost of AMS simulation when the  score function $\Score$ is very expensive to compute. 

The general idea of ARMS is to  perform iterative importance sampling of a sequence of target distributions of the form $\propto  \one_{{\Score} > l^{(k)}} d\pi$ parametrized   at iteration $k$ by the level  $l^{(k)} \in \mathbb{R}$, using a sequences of proposals  of the form  $\mu_{l^{(k)}}^{(k)} \propto  \one_{{\Srom^{(k)}} > l^{(k)}} d\pi$, much cheaper to simulate with AMS. As the algorithm is iterated, the proposals are adapted by adjusting the level $l^{(k)}$ and refining the approximation of the reduced score $\Srom^{(k)}$ using the history of score evaluations previously calculated for importance sampling. 

As $\Srom^{(k)}$ gains in accuracy and $l^{(k)}$ increases, the reduced score approximation should be refined using samples of the state space in regions closer and closer to the rare event of interest. However, an essential ingredient in achieving this desirable convergence is to determine a criteria for setting the first non-achievable (in the sense of Importance Sampling cost) level $l^{(k)}$ by the accuracy of the current reduced score approximation $\Srom^{(k)}$. In addition, one must ensure that the support of the proposal always includes the final rare event target. Justified by theoretical arguments, these two problems are addressed in ARMS using entropy constraints: $l^{(k)}$ is chosen such that $i)$ the relative entropy between a pessimistic rare event target at level $l^{(k)}$ and the proposal corresponds to a given logarithmic cost of importance sampling; $ii)$ the relative entropy between an optimistic final rare event target and the proposal remains finite. Pessimistic and optimistic rare event targets are constructed using a quantification of the worst-case error associated with the reduced score approximation.

A final point addressed by ARMS is the sampling of the proposal given by a new approximation of the reduced score $\Srom^{(k+1)}$. In particular, the algorithm finds a (usually lower) level $l_{b}(k+1)$ at which the updated proposal is dominated (inclusion in terms of support) by a well-chosen previous (carefully recorded) proposal, so that the updated proposal can be sampled by 'bridging' quickly from the recorded proposal in a single AMS simulation step. 
In addition, the updated proposal at level $l_{b}(k+1)$ must satisfy entropy constraints $i)$ and $ii)$ of the previous paragraph. 

In addition to demonstrating the unbiasedness of our estimator in an idealized setting, we evaluate in our numerical experiments the empirical convergence of the ARMS algorithm on a toy model and a rare-event problem based on a more realistic PDE.  Our study confirms that entropy criteria play a crucial role in ARMS. In particular, as expected, too high a value of the logarithm of the cost in the entropy criterion $i)$ can take ARMS into spurious regions of the state-space remote from the region of interest, resulting in a large variance of the IS estimator. Too small a value gives a more desirable variance but a slower rate of convergence. We show empirically that the approximation procedure using the entropy criterion $iii)$ significantly reduces simulation cost (around a factor $10$) while achieving a comparable squared error.  In the same way in both experiments, we observe a squared error gain of the order of a decade for a given simulation cost. Our numerical experiments show that bridging is a good option for small computational budgets, with a saving of around 2 compared with restarting the simulation at each iteration. However,  for larger computational budgets, both procedures appear to be equivalent in terms of cost multiplied by squared error.

The prospects arising from this work are numerous. These include answering open theoretical questions, including the consistency of ARMS, which is proven in the present paper only in a formal sense. No straightforward answers seem available, as we are confronted with a non-standard framework in which interacting particles in AMS simulations have distributions with varying supports which depend on random approximations of the reduced score. Future research directions also include extending ARMS to smooth proposal distributions, with the advantage of removing nesting and domination constraints  and the possibility of being more flexible in the choice of error estimates that need not have strict error bounds.

\appendix
\section{Sample Size for Importance Sampling}\label{app:cost}

\subsection{Generic quantification rules}
{Importance sampling} of some generic distribution $\eta \propto \gamma$ known up to a normalizing constant $p=\gamma(\one)$ using a proposal distribution $\mu$ is based on the following identity
$$
\gamma(\ph)=p \, \eta(\ph) = \E_\mu \b{f(X) \ph(X)} \qquad \forall \ph:\Space \to \R
$$
where $f(x) = \frac{d\gamma}{d\mu}(x)$ is computable in closed-form. \medskip

Unfortunately, importance sampling becomes quickly useless if the target distribution diverges too widely from the proposal, a typical phenomenon in high dimension. This can be readily seen by computing the relative variance of the importance sampling estimator for $\ph = \one$:
\begin{align}\label{eq:varIsstep}
 \Var_{\mu}(f(X)) / p^2 = \eta( \frac{\d \eta}{\d\mu} )- 1 = \e^{\Ent_2(\eta \mid \mu)}- 1,    
 \end{align}
where we have denoted $\Ent_2(\eta \mid \mu) = \ln \eta( \frac{\d \eta}{\d\mu} ) $ the order $2$ R\'enyi entropy (also known as $\chi^2$-divergence). The latter scales linearly in $d$ if $\mu$ and $\eta$ are a $d$-fold product measure, generating a variance exponential with $d$. \medskip

As is thoroughly argued in~\cite{ChaDia18} (see also~\cite{cerou2022entropy}), the variance ($e^{\Ent_2(\eta \mid \mu)}-1$) is in general a pessimistic quantification of importance sampling, whose cost is better estimated using a $\L^1$-error, or a success probability within a tolerance precision. For a target $\eta$ with proposal $\mu$, it has been proven in ~\cite{ChaDia18} that the cost  of importance sampling -- in terms of sample size $K$-- is roughly given (up to a reasonable tail condition on the log-density $\ln \frac{\d \eta}{\d \mu}$), by the exponential of the relative entropy  between $\eta$ and $\mu$:
$
K \simeq \e^{\Ent(\eta \mid \mu)}.
$\medskip

\begin{Rem}\label{rem:Jensen}
Jensen inequality $$\Ent_2(\eta \mid \mu) = \ln \eta( \frac{\eta}{\mu}) \geq   \eta( \ln\frac{\eta}{\mu})=\Ent(\eta \mid \mu),$$   shows that  indeed, variance is a more pessimistic estimator of the required sample size for importance sampling.
\end{Rem}

\subsection{Particular case of interest}
We will particularize the above discussion to the family of distributions studied in our paper: distributions of propositions of the form $ \mu \propto \one_{B} d\pi$ and a target $\eta \propto \one_{A} d \pi$ such that $A \subseteq B$. 

In accordance with the importance sampling identity, the normalization $p$ is estimated without bias using the random variable $f(X) = \pi(B) \one_{A} (X)$ where $X \sim \mu$. 

In this remarkable case, it turns out that the relative variance and the exponential of the relative entropy reduces to the same quantity. In other words, we reach the limit of Jensen's inequality in the Remark~\ref{rem:Jensen} 
 $$\Ent_2(\eta \mid \mu) =\Ent(\eta \mid \mu)=\ln\frac{\pi(B)}{\pi(A)}.$$ 
Therefore, in our study, we will refer indifferently to the exponential of the relative entropy $e^{\Ent(\eta \mid \mu)}$ or to the relative variance $e^{\Ent_2(\eta \mid \mu)}$ to designate the {\it importance sampling cost} of the target $\eta$ by the proposal~$\mu$.

\section{Consistency of the Estimator}\label{app:martingal}
This section analyses the idealized algorithm obtained with $N = +\infty$ many particles (clones) in ARMS;

For simplicity (but the general case is similar), we study  the estimator \eqref{eq:estim2} in the case of a test function  $\ph = \one$. We thus consider the rare event probability estimator:
$ \hat p_l^H  = \hat{\gamma}^{H}_l(\one).$\medskip

\begin{Pro}
 The  estimator $\hat{p}_l^{H} $ is unbiased, \ie 
$
\E \, \hat{p}_l^{H}  = p_l^\ast,
$
and  its relative variance is given by
\begin{equation}\label{eq:varIS}
\Var\b{ \hat p_l ^{H}}/{p^\ast_l}^2 =\frac{1}{H^2} \sum_{k=K_{j_0}+1}^{K_{j_0}+H}\left(\frac{\mathbb{E}({Z_{l^{(k)}}^{(k)}} )}{p^\ast_l}-1\right),  
\end{equation}
where  the random variable ${Z_{l^{(k)}}^{(k)}}$ is a function of the past snapshots.
\end{Pro}\medskip

\proof{
We first show that for any test function $\varphi$
the random sequence
$
H \mapsto M^\varphi_H \eqdef H \, (\hat \gamma_l^{H}(\varphi)- \gamma^\ast_l(\varphi)),
$
is a martingale with respect to the filtration generated by the snapshots. Indeed, defining this filtration by $\mathcal{F}_H=\sigma(X_1,\ldots,X_{K_{j_0}+1+H})$,
it follows from the definition \eqref{eq:estim2} that
\begin{align*}
\mathbb{E}[  M^\varphi_H | \mathcal{F}_{H-1}]&=\mathbb{E}[  \sum_{k=K_{j_0}+1}^{K_{j_0}+H} f_l^{(k)}(X_{k+1}) \varphi({X_{k+1}}) - H\gamma^\ast_l (\varphi)|  \mathcal{F}_{H-1}],\\
&=M^\varphi_{H-1}+\mathbb{E}[ f_l^{({K_{j_0}+H})}(X_{{K_{j_0}+H}+1}) \varphi({X_{{K_{j_0}+H}+1}}) - \gamma^\ast_l(\varphi) |  \mathcal{F}_{H-1} ],\\
&=M^\varphi_{H-1}+\pi(\varphi \one_{\Score> l })- \gamma^\ast_l(\varphi)=M^\varphi_{H-1}.
\end{align*}

Now, since $
M^\varphi_H $
is a martingale, the estimator is unbiased $\E \, \hat{\gamma}_l^{H} (\varphi) = \gamma_l^\ast (\varphi),$ and its variance is given by its quadratic variations given by the sum over $k$ of the local variances. In particular, using \eqref{eq:varIsstep}, the variance of the rare event probability estimator is then 
\begin{eqnarray*}
\Var\b{ \hat p_l ^{H}}/{p^\ast_l}^2
&=  \frac{1}{H^2} \sum_{k=K_{j_0}+1}^{K_{j_0}+H} \left( \mathbb{E}(\e^{\Ent_2\p{\eta_l^\ast \mid \mu^{(k)}_{l^{(k)}}}})-1\right),
\end{eqnarray*}
which for $\eta_l^\ast \propto \one_{\Score > l} d\pi$ and $ \mu^{(k)}_{l^{(k)}} \propto  \one_{\Srom^{(k)} > l^{(k)}} d\pi$ yields~\eqref{eq:varIS}.$\quad \square$

}

\bibliographystyle{plain}
\bibliography{mathias}

\end{document}

%% file: macros.tex
\newcommand{\Space}{\mathcal{X}}
\newcommand{\mref}{\pi}
\newcommand{\Score}{S^{\ast}}
\newcommand{\Srom}{S}
\newcommand{\Err}{E}
\newcommand{\Cte}{c}

\newcommand{\mrm}[1]{\mathrm{#1}}

\newcommand{\tit}[1]{\textit{#1}}

\newcommand{\Ent}{\mathrm{Ent}}

\newcommand{\ie}{\textit{i.e.}, }

\DeclareMathOperator{\Var}{\mathbb{V}ar}

\newcommand{\bigO}{\mathcal{O}}

\newcommand{\E}{\mathbb{E}}

\newcommand{\N}{\mathbb{N}}
\newcommand{\R}{\mathbb{R}}

\renewcommand{\L}{\mathbb{L}}

\newcommand{\ph}{\varphi}

\newcommand{\e}{\mathrm{e}}
\renewcommand{\d}{\mathrm{d}}

\newcommand{\eqdef}{ \dps \mathop{=}^{{\rm def}} }
\newcommand{\one}{ \, {\rm l} \hspace{-.7 mm} {\rm l}}
\newcommand{\dps}{\displaystyle}

\newcommand{\abs}[1]{\left | #1\right |}
\newcommand{\set}[1]{\left\{#1\right\}}
\newcommand{\p}[1]{ \left(#1\right) }
\renewcommand{\b}[1]{\left [ #1\right ]}

\newtheorem{The}{Theorem}[section]

\newtheorem{Pro}[The]{Proposition}

\newtheorem{Rem}[The]{Remark}

\numberwithin{equation}{section}